\documentclass[11pt]{article}
\newtheorem{remark}{Remark}

\usepackage{epsfig}
\usepackage{color}
\usepackage{amsmath}
\usepackage{setspace}
\usepackage{array}
\usepackage{color}
\usepackage{amssymb}
\usepackage{amsbsy}
\usepackage{multirow}
\usepackage{multicol}
\usepackage{lineno}
\usepackage{listings}
\usepackage{url}
\usepackage{float}
\usepackage{graphicx}
\usepackage{pdflscape}
\usepackage{soul}
\usepackage[normalem]{ulem}
\usepackage{lineno}

\newcommand{\bl}[1]{{\color{black} #1}}

\title{\textbf{Optimal Crossover Designs for Generalized Linear Models}}

\author{Jeevan Jankar\\Department of Statistics\\
	University of Georgia\\Athens, GA 30602\\
	\and Abhyuday Mandal \\ {Email: \url{amandal@stat.uga.edu}} \\Department of Statistics\\University of Georgia\\Athens, GA 30602 \and Jie Yang\\Department of Mathematics, Statistics, and Computer Science\\ University of Illinois at Chicago\\ Chicago, IL 60607 }
\date{\today}

\begin{document}

	\maketitle
	
\textit{Abstract:} 
We identify locally $D$-optimal crossover designs for generalized linear models. We use generalized estimating equations to estimate the model parameters along with their variances. To capture the dependency among the observations coming from the same subject, we propose six different correlation structures. We identify the optimal allocations of units for different sequences of treatments. For two-treatment crossover designs, we show via simulations that the optimal allocations are reasonably robust to different choices of the correlation structures. We discuss a real example of multiple treatment crossover experiments using Latin square designs. Using a simulation study, we show that a two-stage design with our locally $D$-optimal design at the second stage is more efficient than the uniform design, especially when the responses from the same subject are correlated. 
	
	\textit{Key words and phrases:} Approximate Designs, $D$-Optimality, Compound Symmetric Correlation, AR(1) Correlation Structure, Generalized Estimating Equations, Two-Stage Design.

	%%%%%%%%%%%%%%%%%%%%%%%%%%%%%%%%%%%%%%%%%%%%%%%%%%%%%%%%%%%%%%%%%%%%%%%%%%%%%%%%%%%%%
	%                                                                                   %
	%                                                                                   %
	%                                                                                   %
	%                                                                                   %
	%                                                                                   %
	%                        SECTION 1: INTRODUCTION                                    %
	%                                                                                   %
	%                                                                                   %
	%                                                                                   %
	%                                                                                   %
	%                                                                                   %
	%                                                                                   %
	%%%%%%%%%%%%%%%%%%%%%%%%%%%%%%%%%%%%%%%%%%%%%%%%%%%%%%%%%%%%%%%%%%%%%%%%%%%%%%%%%%%%%
	
	\section{Introduction}\label{intro}
	
	Pharmaceutical companies frequently conduct clinical trials where the outcome is either success or failure of a particular therapy. Crossover designs, also known as repeated measurements designs or change-over designs, have been used extensively in pharmaceutical research.  There is a rich literature on optimal crossover designs when the response can be adequately modeled by normal distributions. However, for a binary outcome, where the response needs to be described using generalized linear models (GLMs), limited results are known. Consequently, these trials are usually designed using the guidelines of traditional crossover designs obtained using the theory of linear models. However, these designs can be quite inefficient for GLMs. Our goal is to bridge this gap in the literature and determine efficient designs specifically for crossover experiments with responses under univariate GLMs, including binary, binomial, Poisson, Gamma, Inverse Gaussian responses, etc. 
	
	\medskip\noindent Among different types of experiments that are available for treatment comparisons with multiple periods, the crossover designs are among the most important ones. In these experiments, every subject is exposed to a sequence of treatments over different time periods, i.e., subjects crossover from one treatment to another. One of the most important aspects of crossover designs is that we can get the same number of observations as other designs but with less number of subjects. This is an important consideration since human participants are often scarce in clinical trials. The order in which treatments are applied to subjects is known as a {\it sequence} and the time at which these sequences are applied is known as a {\it period}. In most of the cases, the main aim of such experiments is to compare $t$ treatments over $p$ periods. In each period, each subject receives a treatment, and the corresponding response is recorded. In different periods, a subject may receive different treatments, but treatment may also be repeated on the same subject. Naturally, crossover designs also provide within-subject information about treatment differences. 
	
	\medskip\noindent Most of the research in the crossover design literature dealt with continuous response variables (see, for example, Kershner and Federer (1981), Laska and Meisner (1985), Matthews (1987), Carriere and Huang (2000), and the references therein). The problem of determining optimal crossover designs for continuous responses has been studied extensively (see, for example, Bose and Dey (2009), for a review of results).    For examples of practical cases where the responses are discrete in nature, such as binary responses, one may refer to Jones and Kenward (2014) and Senn (2003).

	\medskip\noindent Among many fixed effects models proposed in the literature, the following linear model is used extensively to formulate crossover designs. 
	\begin{eqnarray}\label{eq:LM}
	Y_{ij} = \lambda + \beta_{i} + \alpha_{j} + \tau_{d(i,j)} + \rho_{d(i-1,j)} + \epsilon_{ij} ,
	\end{eqnarray}
	where $Y_{ij}$ is the observation from the $j$th subject in the $i$th time period, with $i=1,\ldots,p$ and $j=1,\ldots,n$. Here $d(i,j)$ stands for the treatment assignment to the $j$th subject at time period $i$ and $\lambda, \beta_{i}, \alpha_{j}, \tau_{d(i,j)}, \rho_{d(i-1,j)}$ are the corresponding overall mean, the $i$th period effect, the $j$th subject effect, the direct treatment effect and the carryover treatment effect respectively. Here $\epsilon_{ij}$'s are the uncorrelated error terms which follow a normal distribution with zero mean and constant variance. Model~(\ref{eq:LM}) is sometimes referred to as the traditional model due to its extensive use in the literature.  
	
	\medskip\noindent As all the effects are fixed, for the linear model~(\ref{eq:LM}), the Fisher information matrix is independent of model parameters. Various optimality criteria such as $A$-, $D$-, $E$-optimality depend on this information matrix (see, for example, Pukelsheim (1993)). Numerous results corresponding to the optimality of crossover designs for linear models are available in the literature. Hedayat and Afsarinejad (1978), Cheng and Wu (1980) and Kunert (1984b) studied the optimality of balanced, uniform designs. Cheng and Wu (1980) formulated theorems for optimality of strongly balanced design. Kunert (1983) produced results for optimality of designs which are neither balanced nor strongly balanced. Dey et al. (1983) were among the first ones to provide results for optimality of designs when $p$ $\leq$ $t$. Considering arbitrary $p$ and $t$ with both $p \leq t$ and $p \geq t$, Kushner (1997b) obtained conditions for universal optimality through approximate theory. Such results cannot be readily extended for binary responses since the Fisher information matrix for GLMs depends on the model parameters (McCullagh and Nelder  (1989), Stufken and Yang (2012)). In this paper, we focus on local optimality to circumvent this problem (Khuri et al. (2006)).
	
	\medskip\noindent{This paper is organized as follows. We describe a preliminary setup of a model for crossover designs for GLMs in Section~\ref{PreSetup} and then discuss generalized estimating equations in Section~\ref{GEE}. We propose different correlation structures in Section~\ref{DiffCorrStr} and formulate locally optimal crossover designs along with an algorithm for obtaining such designs, in Section~\ref{LocOptDes}. In Sections~\ref{OptDesTwoTrt} we provide examples of optimal design for two-treatment crossover trials. We calculate optimal designs for examples with binary response in Section~\ref{Binary} and for example with Poisson response in Section~\ref{Poisson}. In Section~\ref{LatSqDes}, we provide examples of optimal designs for multi-treatment crossover trials, where we use Latin square design. Sensitivity study and Relative $D$-efficiency are presented in Section~\ref{RelD_eff}. Simulation studies are presented in Section~\ref{SimStd}. The paper concludes with comments in Section~\ref{DisFW}.} Some technical details and additional results are presented in Appendix and Supplementary Materials.

	%%%%%%%%%%%%%%%%%%%%%%%%%%%%%%%%%%%%%%%%%%%%%%%%%%%%%%%%%%%%%%%%%%%%%%%%%%%%%%%%%%%%%
	%                                                                                   %
	%                                                                                   %
	%                                                                                   %
	%                                                                                   %
	%                                                                                   %
	%                        SECTION 3: Crossover Design for GLM                        %
	%                                                                                   %
	%                                                                                   %
	%                                                                                   %
	%                                                                                   %
	%                                                                                   %
	%                                                                                   %
	%%%%%%%%%%%%%%%%%%%%%%%%%%%%%%%%%%%%%%%%%%%%%%%%%%%%%%%%%%%%%%%%%%%%%%%%%%%%%%%%%%%%%
	
	\section{Crossover Designs for GLM}\label{CrossoverDGLM}
	
	Although there is a rich literature on optimal crossover designs for linear models, the results on crossover designs under generalized linear models (GLMs) is meagre. Before identifying optimal crossover design we first formally introduce the GLM and the associated optimal crossover designs.

	%%%%%%%%%%%%%%%%%%%%%%%%%%%%%%%%%%%%%%%%%%%%%%%%%%%%%%%%%%%%%%%%%%%%%%%%%%%%%%%%%%%%%                                                          %                                                                                   %                        
	%                                                                                   %
	%                                                                                   %
	%                                                                                   %
	%                        SUB-SECTION 1: Preliminary Setup                           %
	%                                                                                   %
	%                                                                                   %
	%                                                                                   %
	%                                                                                   %
	%%%%%%%%%%%%%%%%%%%%%%%%%%%%%%%%%%%%%%%%%%%%%%%%%%%%%%%%%%%%%%%%%%%%%%%%%%%%%%%%%%%%%
	
	\subsection{Preliminary Setup}\label{PreSetup}
	
	We consider a crossover trial with $t$ treatments, $n$ subjects, and $p$ periods. The responses obtained from these $n$ subjects are denoted as $Y_{1} , \ldots , Y_{n}$, where the response from the $jth$ subject is $ Y_{j} = (Y_{1j} , \ldots , Y_{pj})^\prime $. As discussed above we use a generalized linear model (GLM) to describe the marginal distribution of $Y_{ij}$ as in Liang and Zeger (1986). Let $\mu_{ij}$ denote the mean of a binary response $Y_{ij}$. To fix ideas, first we consider the logistic regression, which models the marginal mean $\mu_{ij}$ for crossover trial as 
	\begin{eqnarray}\label{logitmodel}
	\textrm{logit}(\mu_{ij}) = \textrm{log}\left(\frac{\mu_{ij}}{1-\mu_{ij}}\right) = \eta_{ij} = \lambda + \beta_{i} + \tau_{d(i,j)} + \rho_{d(i-1,j)} ,
	\end{eqnarray} 
	where $ i = 1,\ldots,p ; j = 1,\ldots,n $; $\lambda$ is the overall mean, $\beta_{i}$ represents the effect of the $ith$ period, $\tau_{s}$ is the direct effect due to treatment $s$ and $\rho_{s}$ is the carryover effect due to treatment $s$, where $s = 1,\ldots,t$.
	
	\begin{remark}
		\medskip\noindent Unlike model~(\ref{eq:LM}), model~(\ref{logitmodel}) does not contain a subject effect term $\alpha_{j}$. Note that the response here is described by a GLM, where the Fisher information matrix depends on model parameters. In this paper, we consider the local optimality approach of Chernoff (1953), in which the parameters are replaced by assumed values. In the linear model, the subject effect can be estimated from the data, but for our local optimality approach for the GLM, an educated guess for the subject effect is needed. It would be reasonable to guess the fixed treatment effects from prior knowledge, while from a design point of view the subject effect if included, have to be treated as random. Instead of incorporating a random effects term, in this paper, the mean response is modeled through the logit link function in equation~(\ref{logitmodel}) with an extra assumption that the responses from a particular subject are mutually correlated, while the responses from different subjects are uncorrelated. In the case of generalized linear models, only the mean response is modeled through the link function, and hence we are free to choose a variance-covariance matrix as long as that is positive definite. So, in this paper, we use this opportunity of choosing the covariance matrix and capture the subject effect by putting different meaningful structures on this matrix and studying the robustness of the design. In this way, we can exclude a random subject effect from the model and calculate optimal designs more easily. 
	\end{remark}
	
	\medskip\noindent {As the main interest is in estimating the treatment effects and variance of its estimator, carryover effects are treated as nuisance parameters. To ensure estimability of the model parameters, we set the baseline constraints as $\beta_{1} = \tau_{1} = \rho_{1} = 0$. Consider $ \beta = ( \beta_{2},\ldots,\beta_{p} )^{\prime} $ , $ \tau = ( \tau_{2},\ldots,\tau_{t} )^{\prime} $ and $ \rho = ( \rho_{2},\ldots,\rho_{t} )^{\prime} $, which define the parameter vector $ \theta = ( \lambda, \beta, \tau, \rho )^{\prime}$. Then the linear predictor corresponding to the $jth$ subject, $ \eta_{j} = ( \eta_{1j},\ldots,\eta_{pj} )^{\prime} $ can be written as
		\begin{eqnarray*} 
			\eta_{j} &=& X_{j}\theta.
		\end{eqnarray*}
	\noindent The corresponding design matrix $X_{j}$ can be written as $X_{j} = \left[ 1_{p}, P_{j}, T_{j}, F_{j} \right ]$, where $ P_{j}$ is $ p \times (p-1) $ such that $P_{j} = [0_{(p-1)1},I_{p-1}]'$; where $T_{j}$ is a $ p \times (t-1) $ matrix with its $(i,s)th$ entry equal to 1 if subject $j$ receives the direct effect of the treatment $s$ in the $ith$ period and zero otherwise; where $F_{j}$ is a $ p \times (t-1) $ matrix with its $(i,s)th$ entry equal to 1 if subject $j$ receives the carryover effect of the treatment $s$ in the $ith$ period and zero otherwise, where columns of $T_{j}$ and $F_{j}$ are indexed by $2,\ldots,t$.}
	
	\medskip\noindent If the number of subjects is fixed to $n$ and the number of periods is $p$, then we determine the proportion of subjects assigned to a particular treatment sequence. As the number of periods is fixed to $p$, each treatment sequence will be of length $p$ and a typical sequence can be written as $\omega = (t_{1},\ldots,t_{p})^{\prime}$ where $t_{i}\in\{1,\ldots,t\}$. Now, let $\Omega$ be the set of all such sequences and $n_{\omega}$ denote the number of subjects assigned to sequence $\omega$. Then, the total number of subjects $n$ can be written as $n = \Sigma_{\omega\in\Omega}n_{\omega} ,n_{\omega} \geq 0$. A crossover design $\zeta$ in approximate theory is specified by the set $ \{ p_{\omega}, \omega\in\Omega \} $, where $ p_{\omega} = n_{\omega}/n$ is the proportion of subjects assigned to treatment sequence $\omega$. Such a crossover design $\zeta$ can be denoted as follows:
	\[
	\zeta = \left\{ \begin{array} { l l l l }
	{ \omega _ { 1 } } & { \omega _ { 2 } } & { \ldots } & { \omega _ { k } } \\
	{ p _ { \omega _ { 1 } } } & { p _ { \omega _ { 2 } } } & { \ldots } & { p _ { \omega _ { k } } } \end{array} \right\}
	\]
	where $k$ is the number of treatment sequences involved, such that $\sum _ { i = 1 } ^ { k } p _ { \omega _ { i } } = 1 , \text { for } i = 1 , \ldots , k$. From the definitions of matrices $T_{j}$ and $ F_{j}$ it can be noted that they depend only on the treatments sequence $\omega$ that subject $j$ receives. So it can be inferred that $T_{j} = T_{\omega}$ and $F_{j} = F_{\omega}$. This implies, $X_{j} = X_{\omega}$ as $P_{j} = [0_{(p-1)1},I_{p-1}]'$.

	%%%%%%%%%%%%%%%%%%%%%%%%%%%%%%%%%%%%%%%%%%%%%%%%%%%%%%%%%%%%%%%%%%%%%%%%%%%%%%%%%%%%%                                                          %                                                                                   %                        
	%                                                                                   %
	%                                                                                   %
	%                                                                                   %
	%                        SUB-SECTION 2: Generalized Estimating Equations            %
	%                                                                                   %
	%                                                                                   %
	%                                                                                   %
	%                                                                                   %
	%%%%%%%%%%%%%%%%%%%%%%%%%%%%%%%%%%%%%%%%%%%%%%%%%%%%%%%%%%%%%%%%%%%%%%%%%%%%%%%%%%%%%
	
	\subsection{Generalized Estimating Equations}\label{GEE}	
	
	Generalized estimating equations are quasi-likelihood equations which allow us to estimate quasi-likelihood estimators. In this paper, instead of using maximum likelihood estimation (MLE) or ordinary least squares (OLS) to estimate the parameters we use quasi-likelihood estimation. Earlier we made one important assumption in crossover trials that observations from each subject are mutually correlated while the observations from different subjects are uncorrelated. This dependency between repeated observations from a subject is modeled using what is called ``working correlation'' matrix $C$. If $C$ is the true correlation matrix of $Y_{j}$, then from the definition of covariance we can write
	\begin{eqnarray*} 
		Cov(Y_{j}) &=& D_{j}^{1/2}CD_{j}^{1/2},
	\end{eqnarray*}
	where $D_{j} = diag\Big(\mu_{1j}(1-\mu_{1j}),\ldots,\mu_{pj}(1-\mu_{pj})\Big)$. Let us denote $	Cov(Y_{j})$ by $W_{j}$. In Zeger $et$ $al$. (1988, equation (3.1)) it has been shown that for repeated measurement model, the generalized estimating equations (GEE) are defined to be 
	\begin{eqnarray*} 
		\sum _ { j = 1 } ^ { n } \frac { \partial \mu _ { j } ^ { \prime } } { \partial \theta } W _ { j } ^ { - 1 } \left({ Y } _ { j } - \mu _ { j } \right) = 0
	\end{eqnarray*}
	where $\mu _ { j } = \left( \mu _ { 1 j } , \ldots , \mu _ { p j } \right) ^ { \prime }$ and the asymptotic variance for the GEE estimator $\hat{\theta}$ (see Zeger $et$ $al$., 1988, equation (3.2)) is 
	\begin{eqnarray}\label{Var}
	{\rm Var} (\hat{\theta}) &=& \left[ \sum _ { j = 1 } ^ { n } \frac { \partial \mu _ { j } ^ { \prime } } { \partial \theta } W _ { j } ^ { - 1 } \frac { \partial \mu _ { j } } { \partial \theta } \right] ^ { - 1 }
	\end{eqnarray}
	where $W_{j} = Cov(Y_{j})$. \bl{As mentioned by Singh and Mukhopadhyay (2016) in the paper (Zeger $et$ $al$., 1988, equation (3.2)) it has also been shown that if the true correlation structure varies from ``working correlation'' structure, then ${\rm Var} (\hat{\theta})$ is given by the sandwich formula 
		\begin{eqnarray*} 
			{\rm Var} (\hat{\theta}) &=& U^{-1}VU^{-1},
		\end{eqnarray*}
		where the $U$ and $V$ in above equation are as follows: 
		\begin{eqnarray}\label{UV}
		U = \sum_{\omega \epsilon \Omega} np_{\omega} \frac{\partial \mu_{\omega}^{\prime}}{\partial \theta} W_{\omega}^{-1} \frac{\partial \mu_{\omega}}{\partial \theta} &,&
		V = \sum_{\omega \epsilon \Omega} np_{\omega} \frac{\partial \mu_{\omega}^{\prime}}{\partial \theta} W_{\omega}^{-1} Cov(Y_{\omega}) W_{\omega}^{-1} \frac{\partial \mu_{\omega}}{\partial \theta}.
		\end{eqnarray}
	So it is expected that the effect of variance misspecification on the locally optimal designs will be minimal. Table A1 presented in the Appendix confirms this.}
	
	\medskip\noindent From above equations~(\ref{Var}) and~(\ref{UV}), it can be seen that if the true correlation of $Y_j$ is equal to $C$, then ${\rm Var} (\hat{\theta}) = U^{-1}$.\iffalse 
	\begin{eqnarray}\label{Uinv}
	{\rm Var} (\hat{\theta}) = U^{-1} = \left[\sum_{\omega \epsilon \Omega} np_{\omega} \frac{\partial \mu_{\omega}^{\prime}}{\partial \theta} W_{\omega}^{-1} \frac{\partial \mu_{\omega}}{\partial \theta}\right]^{-1}.
	\end{eqnarray} \fi 
	We have considered carryover effects to be nuisance parameters as the main interest usually lies in estimating the direct treatment effect contrasts. So, instead of working with the full variance-covariance matrix of parameter estimator $\hat\theta$ we concentrate only on the variance of the estimator of treatment effect ${\rm Var}(\hat\tau)$ where
	\begin{eqnarray}\label{ObjectiveFunction}
	{\rm Var}(\hat\tau) &=& H{\rm Var}(\hat\theta)H^{\prime},
	\end{eqnarray}
	$H$ is a $(t-1)\times m$ matrix given by $[0_{(t-1)1},0_{(t-1)(p-1)},I_{t-1},0_{(t-1)(t-1)}]$ where $m=p+2t-2$ is the total number of parameters in $\theta$ and $0_{(t-1)(p-1)}$ is a $(t-1)\times(p-1)$ matrix of zeros.
	
	\medskip\noindent We calculate optimal proportions such that the variances of estimators of treatment effect is minimized. In this paper we focus on $D$-optimality and use the determinant of ${\rm Var}(\hat\tau)$ as our objective function. Note that other optimality criteria such as $A$-,$E$-optimality can be applied similarly. Then an optimal design $\zeta^{*}$ minimizes the determinant of ${\rm Var}(\hat\tau)$ in equation~(\ref{ObjectiveFunction}) with respect to $p_{\omega}$ such that $\sum _ { w \in \Omega } p _ { w } = 1$. \bl{For illustration, we give an explicit expression of the information matrix and present the associated calculations for a crossover design in the Supplementary Materials.}

	%%%%%%%%%%%%%%%%%%%%%%%%%%%%%%%%%%%%%%%%%%%%%%%%%%%%%%%%%%%%%%%%%%%%%%%%%%%%%%%%%%%%%                                                          %                                                                                   %                        
	%                                                                                   %
	%                                                                                   %
	%                                                                                   %
	%                        SUB-SECTION 3: Proposed Correlation Structure              %
	%                                                                                   %
	%                                                                                   %
	%                                                                                   %
	%                                                                                   %
	%%%%%%%%%%%%%%%%%%%%%%%%%%%%%%%%%%%%%%%%%%%%%%%%%%%%%%%%%%%%%%%%%%%%%%%%%%%%%%%%%%%%%
	
	\subsection{Proposed Correlation Structures}\label{DiffCorrStr}
	
	As mentioned in the above section, to calculate the variance matrix of parameter estimates, a predefined working correlation structure for the responses is needed. Any correlation structure can be assumed for the responses, but if the design is not robust, then the optimal proportions will vary as the correlation structure varies. So, to check the robustness of design and to make the design more practically acceptable, optimal proportions using different correlation structures are calculated. For the design in equation (2) with two treatments $A$ and $B$, six different types of correlation structures are proposed, and optimal proportions are calculated. Out of these six correlation structures, the correlation matrices defined by the first three correlation structures are fixed and do not depend on treatment sequence whereas the correlation matrices of the fourth, fifth and sixth types depend on treatment sequences and vary along with treatment sequences. 
	
	\medskip\noindent \bl{The first correlation structure is a compound symmetric correlation structure, i.e., 
		$$Corr(1) = (1-\rho)I_{p} + \rho J_{p},$$
	where $I_p$ is the identity matrix of order $p$, and $J_p$ is a $p\times p$ matrix with all elements unity.
		
	\medskip\noindent The second correlation structure is the AR(1) correlation structure, i.e.,
		$$Corr(2) = \Big(\rho ^ { \left| i - i ^ { \prime } \right| }\Big),$$
	so that the correlation between responses decreases as the time gap between responses increases.
		
	\medskip\noindent The third correlation structure is as follows:
		\begin{eqnarray*}
			Corr(3) = 
			\left(\begin{array}{rrrrrrr}
				1          & \rho       & 0   & \ldots & 0 & 0 & 0 \\
				\rho       & 1          & \rho       & \ldots & 0 & 0 & 0 \\
				\vdots     &      &   & \vdots &       & & \vdots     \\
				0 & 0 & 0 & \ldots & \rho & 1          & \rho       \\
				0 & 0 & 0 & \ldots & 0 & \rho       & 1          \\
			\end{array}\right).
		\end{eqnarray*}
	For each correlation structure different correlation matrices using different $\rho$ values are considered.}
	
	\medskip\noindent To understand the other three correlation structures, we denote the correlation coefficient between the response  when a subject receives treatment $A$ first and the response when the same subject receives treatment $B$ afterwards as $\rho_{AB}$ and, $\rho_{BA}$ when the subject receives $B$ first and $A$ afterwards. Note that in general $\rho_{AB}$ is not necessarily the same as $\rho_{BA}$. In a similar manner we define $\rho_{AA}$ and $\rho_{BB}$. {To define the fourth type of correlation structure we will use the same structure as $Corr(3)$ but with different values of correlation coefficient for different treatment sequences. For fourth type of correlation we use $\rho_{AB} = 0.2, \rho_{BA} = 0.5$ and $\rho_{AA} = 0.1, \rho_{BB} = 0.3$.}
	
	\medskip\noindent To define fifth and sixth type of correlation structures, we use AR(1) correlation structure with correlation coefficient depending on treatment sequence. For the fifth type, we use the same values for $\rho_{AB}$ and $\rho_{BA}$ and for the sixth type of correlation structure we use different values for $\rho_{AB}$ and $\rho_{BA}$. For both fifth and sixth type of correlation structure we keep $\rho_{AA} = \rho_{BB}$. These values might vary from example to example and would depend on what treatments $A$ and $B$ are. As the entries of the correlation matrix depend on which treatment the subject receives in a particular period, these correlation matrices are different for different treatment sequences. Here, our aim is to see how optimal proportions vary as we vary values of $\rho_{AB}$ and $\rho_{BA}$.
	
	\medskip\noindent As an illustration, we consider $p = 2$ with treatment sequences ${ AB, BA}$. Then the third type correlation matrices for both treatment sequences $AB$ and $BA$ will have same structure as $Corr(1)$. The fourth, fifth and sixth type correlation matrices will have same structure as follows with different $\rho$ values,
	
	\medskip
	\begin{scriptsize}
	\begin{minipage}{0.45\textwidth}
		\[
		Corr(4/5/6)_{AB} = \left(\begin{array}{rr}
		1   & \rho_{AB} \\
		\rho_{AB} & 1   \\
		\end{array}\right),
		\]
	\end{minipage}
	\hspace{0.01\textwidth}
	\begin{minipage}{0.45\textwidth}
		\[
		Corr(4/5/6)_{BA} = \left(\begin{array}{rr}
		1   & \rho_{BA} \\
		\rho_{BA} & 1   \\
		\end{array}\right).
		\]
	\end{minipage}
	\end{scriptsize}
	
	\medskip\noindent For $p = 3$ case we consider an example with treatment sequences ${ ABB, BAA}$. The fourth type of correlation matrix will have values as mentioned above. The fifth type correlation matrices for both treatment sequences $ABB$ and $BAA$ will be the same if in treatment sequences, $A$ and $B$ are interchangeable and $\rho_{AB} = \rho_{BA}$ along with $\rho_{AA} = \rho_{BB}$. The sixth type correlation matrices for both treatment sequences $ABB$ and $BAA$ will be different as  $\rho_{AB}$ and $\rho_{BA}$ are different. We get
	
	\begin{scriptsize}
	\begin{minipage}{0.45\textwidth}
		\[
		Corr(4)_{ABB} = \left(\begin{array}{rrr}
		1             & \rho_{AB} & 0  \\
		\rho_{AB}     & 1         & \rho_{BB}      \\
		0             & \rho_{BB} & 1              \\
		\end{array}\right),
		\]
	\end{minipage}
	\hspace{0.01\textwidth}
	\begin{minipage}{0.45\textwidth}
		\[
		Corr(4)_{BAA} = \left(\begin{array}{rrr}
		1             & \rho_{BA} & 0 \\
		\rho_{BA}     & 1         & \rho_{AA}     \\
		0             & \rho_{AA} & 1             \\
		\end{array}\right),
		\]
	\end{minipage}
	\end{scriptsize}
	
	and
	
	\medskip
	\begin{scriptsize}
	\[
	Corr(5)_{ABB} = Corr(5)_{BAA} = \left(\begin{array}{rrr}
	1             & \rho_{AB} & \rho_{AB}^{2} \\
	\rho_{AB}     & 1         & \rho_{BB}     \\
	\rho_{AB}^{2} & \rho_{BB} & 1             \\
	\end{array}\right),
	\]
	\end{scriptsize}
	
	and
	
	\medskip
	\begin{scriptsize}
	\begin{minipage}{0.45\textwidth}
		\[
		Corr(6)_{ABB} = \left(\begin{array}{rrr}
		1             & \rho_{AB} & \rho_{AB}^{2}  \\
		\rho_{AB}     & 1         & \rho_{BB}      \\
		\rho_{AB}^{2} & \rho_{BB} & 1              \\
		\end{array}\right),
		\]
	\end{minipage}
	\hspace{0.01\textwidth}
	\begin{minipage}{0.45\textwidth}
		\[
		Corr(6)_{BAA} = \left(\begin{array}{rrr}
		1             & \rho_{BA} & \rho_{BA}^{2} \\
		\rho_{BA}     & 1         & \rho_{AA}     \\
		\rho_{BA}^{2} & \rho_{AA} & 1             \\
		\end{array}\right).
		\]
	\end{minipage}
	\end{scriptsize}
	
	\medskip\noindent Same as the above two cases, for $p = 4$ case we consider an example with treatment sequences ${ AABB, BBAA}$. The fourth type of correlation matrix will be as given below. The fifth type of correlation matrices for both treatment sequences $AABB$ and $BBAA$ will be same because in treatment sequences $A$, $B$ are interchangeable and $\rho_{AA} = \rho_{BB}$ and $\rho_{AB} = \rho_{BA}$. Sixth type of correlation matrices for both treatment sequences $ABB$ and $BAA$ will be different as  $\rho_{AB}$ and $\rho_{BA}$ are different.  We get
	
	\medskip
	\begin{scriptsize}
	\begin{tabular}{lr}
		\begin{minipage}{0.45\textwidth}
			\[
			Corr(4)_{AABB} = \left(\begin{array}{rrrr}
			1             & \rho_{AA}     & 0             & 0 \\
			\rho_{AA}     & 1             & \rho_{AB}     & 0 \\
			0             & \rho_{AB}     & 1             & \rho_{BB} \\
			0             & 0             & \rho_{BB}     &  1        \\
			\end{array}\right),
			\]
		\end{minipage}
		&
		\begin{minipage}{0.45\textwidth}
			\[
			\hspace{.1in}
			Corr(4)_{BBAA} = \left(\begin{array}{rrrr}
			1             & \rho_{BB}     & 0             & 0 \\
			\rho_{BB}     & 1             & \rho_{BA}     & 0 \\
			0             & \rho_{BA}     & 1             & \rho_{AA}     \\
			0             & 0             & \rho_{AA}     &  1            \\
			\end{array}\right),
			\]
		\end{minipage}
	\end{tabular}
	\end{scriptsize}
	
	and
	
	\medskip
	\begin{scriptsize}
		\[
		Corr(5)_{AABB} = Corr(5)_{BBAA} = \left(\begin{array}{rrrr}
		1             & \rho_{BB}     & \rho_{BA}^{2} & \rho_{BA}^{3} \\
		\rho_{BB}     & 1             & \rho_{BA}     & \rho_{BB}^{2} \\
		\rho_{BA}^{2} & \rho_{BA}     & 1             & \rho_{BB}     \\
		\rho_{BA}^{3} & \rho_{BA}^{2} & \rho_{BB}     &  1            \\
		\end{array}\right),
		\]
	\end{scriptsize}
	
	and
	
	\medskip
	\begin{scriptsize}
	\begin{tabular}{lr}
		\begin{minipage}{0.45\textwidth}
			\[
			Corr(6)_{AABB} = \left(\begin{array}{rrrr}
			1             & \rho_{AA}     & \rho_{AB}^{2} & \rho_{AB}^{3} \\
			\rho_{AA}     & 1             & \rho_{AB}     & \rho_{AB}^{2} \\
			\rho_{AB}^{2} & \rho_{AB}     & 1             & \rho_{BB}     \\
			\rho_{AB}^{3} & \rho_{AB}^{2} & \rho_{BB}     &  1            \\
			\end{array}\right),
			\]
		\end{minipage}
		&
		\begin{minipage}{0.45\textwidth}
			\[
			\hspace{.1in}
			Corr(6)_{BBAA} = \left(\begin{array}{rrrr}
			1             & \rho_{BB}     & \rho_{BA}^{2} & \rho_{BA}^{3} \\
			\rho_{BB}     & 1             & \rho_{BA}     & \rho_{BA}^{2} \\
			\rho_{BA}^{2} & \rho_{BA}     & 1             & \rho_{AA}     \\
			\rho_{BA}^{3} & \rho_{BA}^{2} & \rho_{AA}     &  1            \\
			\end{array}\right).
			\]
		\end{minipage}
	\end{tabular}
	\end{scriptsize}
	
	\medskip\noindent For $p=4$ case, we discuss another interesting example with four treatments $A$, $B$, $C$ and $D$. The set of treatment sequences for this example is $\Omega = \{ABCD$,$BDAC$, $CADB$,$DCBA\}.$  This experiment will be discussed in detail later in Section~\ref{OptMultTrt}. Note that the treatment sequences are given by a Latin square design and the treatments are interchangeable.  
	\begin{center}
		\begin{tabular}{ cccc } 
			%\hline
			$A$ & $B$ & $C$ & $D$ \\
			%\hline 
			$B$ & $D$ & $A$ & $C$ \\
			%\hline
			$C$ & $A$ & $D$ & $B$ \\ 
			%\hline
			$D$ & $C$ & $B$ & $A$ \\
			%\hline
		\end{tabular}
	\end{center}

	\medskip\noindent For this example above six different types of correlation matrices are considered. The first three correlation matrices will be the same as above with $\rho = 0.3$, $\rho = 0.2$ and $\rho = 0.1$ respectively. The fourth type correlation structure will be defined in similar manner as discussed above. The fifth type correlation matrix is defined using AR(1) correlation structure with  $\rho_{AB} = \rho_{AC} = \rho_{AD} = \rho_{BA} = \rho_{CA} = \rho_{DA} = 0.4$, $\rho_{BC} = \rho_{BD} = \rho_{CB} = \rho_{DB} = 0.3$ and $\rho_{CD} = \rho_{DC} = 0.2$. For fourth type and sixth type of correlation matrix, $\rho_{AB} = \rho_{AC} = \rho_{AD} $ is taken to be 0.4. In a similar manner $\rho_{BA} = \rho_{BC} = \rho_{BD}$ is taken to be 0.3 and $\rho_{CA} = \rho_{CB} = \rho_{CD}$ is taken to be 0.2 and $\rho_{DA} = \rho_{DB} = \rho_{DC}$ taken to be 0.1. As the entries of the correlation matrix depend on which treatment the subject receives in a particular period, these correlation matrices are different for different treatment sequences and are listed as follows:
	
		\medskip\noindent 
		\begin{scriptsize}
			\begin{tabular}{lc}
				\begin{minipage}{0.45\textwidth}	
					\[
					Corr(4)_{ABCD} = \left(\begin{array}{rrrr}
					1             & \rho_{AB}     & 0             & 0 \\
					\rho_{AB}     & 1             & \rho_{BC}     & 0 \\
					0             & \rho_{BC}     & 1             & \rho_{CD}     \\
					0             & 0             & \rho_{CD}     & 1             \\
					\end{array}\right),
					\]
				\end{minipage}
				&
				\begin{minipage}{0.45\textwidth}
					\[
					Corr_{BDAC} = \left(\begin{array}{rrrr}
					1             & \rho_{BD}     & 0             & 0 \\
					\rho_{BD}     & 1             & \rho_{DA}     & 0 \\
					0             & \rho_{DA}     & 1             & \rho_{AC}     \\
					0             & 0             & \rho_{AC}     & 1             \\
					\end{array}\right),
					\]
				\end{minipage}
		\end{tabular}
		
		\medskip\noindent
			\begin{tabular}{lc}
				\begin{minipage}{0.45\textwidth}	
					\[
					Corr(4)_{CADB} = \left(\begin{array}{rrrr}
					1             & \rho_{CA}     & 0             & 0 \\
					\rho_{CA}     & 1             & \rho_{AD}     & 0 \\
					0             & \rho_{AD}     & 1             & \rho_{DB}     \\
					0             & 0             & \rho_{DB}     & 1             \\
					\end{array}\right),
					\]
				\end{minipage}
				&
				\begin{minipage}{0.45\textwidth}
					\[
					Corr(4)_{DCBA} = \left(\begin{array}{rrrr}
					1             & \rho_{DC}     & 0             & 0 \\
					\rho_{DC}     & 1             & \rho_{CB}     & 0 \\
					0             & \rho_{CB}     & 1             & \rho_{BA}     \\
					0             & 0             & \rho_{BA}     & 1             \\
					\end{array}\right),
					\]
				\end{minipage}
			\end{tabular}
		\end{scriptsize}

	and
	
		\medskip\noindent
		\begin{scriptsize}
			\begin{tabular}{lc}
				\begin{minipage}{0.45\textwidth}	
					\[
					Corr(5/6)_{ABCD} = \left(\begin{array}{rrrr}
					1             & \rho_{AB}     & \rho_{AC}^{2} & \rho_{AD}^{3} \\
					\rho_{AB}     & 1             & \rho_{BC}     & \rho_{BD}^{2} \\
					\rho_{AC}^{2} & \rho_{BC}     & 1             & \rho_{CD}     \\
					\rho_{AD}^{3} & \rho_{BD}^{2} & \rho_{CD}     & 1             \\
					\end{array}\right),
					\]
				\end{minipage}
			&
				\begin{minipage}{0.45\textwidth}
					\[
					Corr(5/6)_{BDAC} = \left(\begin{array}{rrrr}
					1             & \rho_{BD}     & \rho_{BA}^{2} & \rho_{BC}^{3} \\
					\rho_{BD}     & 1             & \rho_{DA}     & \rho_{DC}^{2} \\
					\rho_{BA}^{2} & \rho_{DA}     & 1             & \rho_{AC}     \\
					\rho_{BC}^{3} & \rho_{DC}^{2} & \rho_{AC}     & 1             \\
					\end{array}\right),
					\]
				\end{minipage}
			\end{tabular}
		
		\medskip\noindent
		\begin{tabular}{lc}
				\begin{minipage}{0.45\textwidth}	
					\[
					Corr(5/6)_{CADB} = \left(\begin{array}{rrrr}
					1             & \rho_{CA}     & \rho_{CD}^{2} & \rho_{CB}^{3} \\
					\rho_{CA}     & 1             & \rho_{AD}     & \rho_{AB}^{2} \\
					\rho_{CD}^{2} & \rho_{AD}     & 1             & \rho_{DB}     \\
					\rho_{CB}^{3} & \rho_{AB}^{2} & \rho_{DB}     & 1             \\
					\end{array}\right),
					\]
				\end{minipage}
			&
				\begin{minipage}{0.45\textwidth}
					\[
					Corr(5/6)_{DCBA} = \left(\begin{array}{rrrr}
					1             & \rho_{DC}     & \rho_{DB}^{2} & \rho_{DA}^{3} \\
					\rho_{DC}     & 1             & \rho_{CB}     & \rho_{CA}^{2} \\
					\rho_{DB}^{2} & \rho_{CB}     & 1             & \rho_{BA}     \\
					\rho_{DA}^{3} & \rho_{CA}^{2} & \rho_{BA}     & 1             \\
					\end{array}\right).
					\]
				\end{minipage}
			\end{tabular}
		\end{scriptsize}
	
	\bigskip\noindent In the above, we only specified the  forms of correlation structures. Note that for this particular example, the form of $Corr(5)$ is the same as that of $Corr(6)$ since the treatment sequences are obtained using a Latin square design. In Section~\ref{OptMultTrt}, we will consider the above six types of correlation structures and calculate the corresponding optimal proportions. We will also perform a simulation analysis using this example. For simulation analysis, AR(1) correlation structure will be considered with different $\rho$ values.  We have performed robustness in the Appendix~\ref{Appendix} and provided explicit expressions on how to obtain objective function in Supplementary Section~\ref{ObjFn}.

	%%%%%%%%%%%%%%%%%%%%%%%%%%%%%%%%%%%%%%%%%%%%%%%%%%%%%%%%%%%%%%%%%%%%%%%%%%%%%%%%%%%%%                                      %                                                                                   %                        
	%                                                                                   %
	%                                                                                   %
	%                                                                                   %
	%           SUB-SECTION 4: Algorithm for Locally Optimal Crossover Trials           %
	%                                                                                   %
	%                                                                                   %
	%                                                                                   %
	%                                                                                   %
	%%%%%%%%%%%%%%%%%%%%%%%%%%%%%%%%%%%%%%%%%%%%%%%%%%%%%%%%%%%%%%%%%%%%%%%%%%%%%%%%%%%%%
	
	\subsection{Algorithm for Locally Optimal Crossover Trials}\label{LocOptDes}

	In this section, we propose an algorithm to find locally optimal designs for crossover trials. Assumed values of the model parameters are obtained from some prior knowledge or pilot studies. To identify the locally optimal crossover design, the major challenge is in minimizing the objective function.  The complexity of the objective function increases with the increase of $t$, $p$ and $k$. We use the {\tt solnp} function in R for numerical optimization.

	\medskip \noindent
	\vspace{8pt}
	\hrule
	\vspace{8pt}
	\textbf{Algorithm : } Pseudo-code for finding locally optimal crossover designs.
	\vspace{8pt}
	\hrule
	\vspace{8pt}
	\noindent Given assumed values of the parameters, construct the design matrix, correlation matrix, and the parameter vector.
	
	\noindent \textbf{for}\vspace{-.05in}
	\begin{itemize}
		\item[] Each subject in each period \vspace{-.01in}
		\item[] Calculate the mean of the response \vspace{-.02in}
	\end{itemize}
	\textbf{end}
	
	\noindent \textbf{for}\vspace{-.05in}
	\begin{itemize}
		\item[] Each treatment sequence \vspace{-.01in}
		\item[] Calculate the covariance matrix using the correlation matrix \vspace{-.01in}
		\item[] Diagonal entries of covariance matrix are variances of observations \vspace{-.01in}
		\item[] Variance depends on the distribution of the response \vspace{-.01in}
		\item[] Calculate the inverse of covariance matrix \vspace{-.02in}
	\end{itemize}
	\textbf{end}
	
	\noindent \textbf{for}\vspace{-.05in}
	\begin{itemize}
		\item[] Each treatment sequence \vspace{-.01in}
		\item[] Calculate the corresponding derivative matrix \vspace{-.01in}
		\item[] Using calculated matrices and variables corresponding to each treatment sequence, compute the variance matrix of parameter estimates \vspace{-.01in}
		\item[] Calculate variance matrix of treatment effects. Its determinant is the required objective function \vspace{-.02in}
	\end{itemize} 
	\textbf{end}
	
	\noindent \textbf{function}\vspace{-.05in}
	\begin{itemize}
		\item[] Define the objective function along with the constraints, i.e., sum of proportions is equal to one \vspace{-.02in}
	\end{itemize}	
	\textbf{end}
	
	\textbf{solnp} Using this constraint optimization function calculate optimal proportions 
	
	\vspace{6pt}
	\hrule
	\vspace{6pt}

	%%%%%%%%%%%%%%%%%%%%%%%%%%%%%%%%%%%%%%%%%%%%%%%%%%%%%%%%%%%%%%%%%%%%%%%%%%%%%%%%%%%%%
	%                                                                                   %
	%                                                                                   %
	%                                                                                   %
	%                                                                                   %
	%                                                                                   %
	%          SECTION 3: Optimal Design for two-treatment crossover trails             %
	%                                                                                   %
	%                                                                                   %
	%                                                                                   %
	%                                                                                   %
	%                                                                                   %
	%                                                                                   %
	%%%%%%%%%%%%%%%%%%%%%%%%%%%%%%%%%%%%%%%%%%%%%%%%%%%%%%%%%%%%%%%%%%%%%%%%%%%%%%%%%%%%%
	
	\section{Optimal Designs for Two-treatment Crossover Trials }\label{OptDesTwoTrt}
	
	The crossover designs for which we will calculate the optimal proportions are similar to those discussed by Laska and Meisner (1985) and Carriere and Huang (2000). Optimal proportions are listed below for $p = 2, 3, 4$ for binary response and for $p = 2$ for poisson response under two sets of parameter estimates. In this section, we consider only two treatments $A$ and $B$. Considering our baseline constraint to be $\tau_{A} = \rho_{A} = 0$ and $\beta_{1} = 0$ we only have $p+2$ parameters in vector $\theta$. So, when there are only two treatments involved in the crossover trial, the parameter vector $\theta$ is $[\lambda,\beta_{2},\ldots,\beta_{p},\tau_{2},\rho_{2}]$.
	
	\medskip \noindent Optimal proportions for different crossover designs are calculated with each of the six different correlation structures mention above. For each correlation matrix that we consider, an optimal design $\zeta^{*}$ is the one minimizing the determinant of ${\rm Var}(\hat\tau)$ in equation~(\ref{ObjectiveFunction}) with respect to $p_{\omega}$ such that $\sum _ { w \in \Omega } p _ { w } = 1$.
	
	\medskip \noindent We use different colors to represent different correlation structures. The color scheme that we use is as follows:
	
		\begin{center}		
			\begin{tabular}{ |lc| }
				\hline
				&\\
				Correlation Structure & Color \\
				\hline
				&\\		      
				Corr(1) $(1-\rho)I_{p} + \rho J_{p}$ with $\rho = 0.1$ &\includegraphics[width=.2in,height = 0.02\textheight]{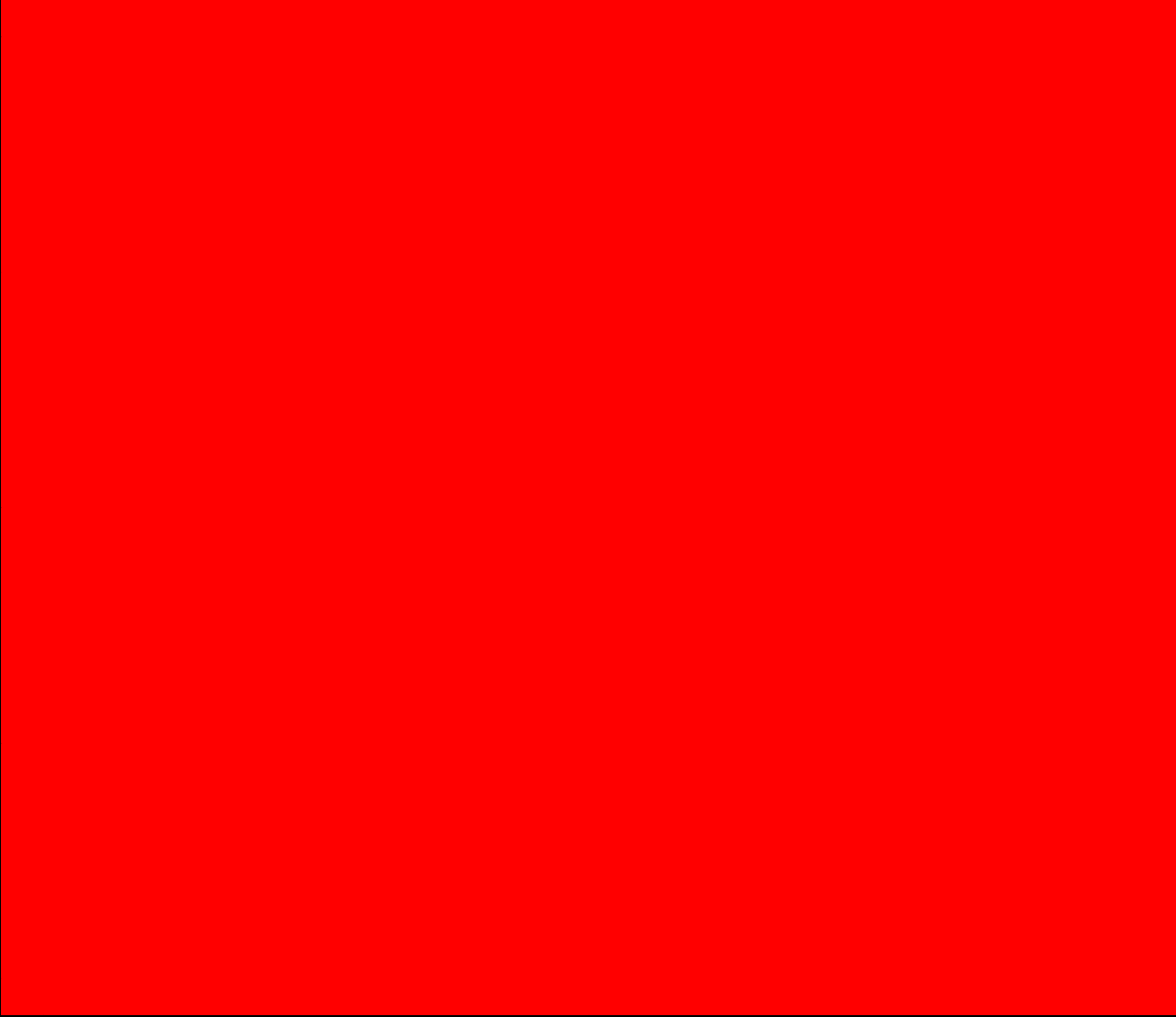} \\
				Corr(2) $\rho ^ { \left| i - i ^ { \prime } \right| } , i \neq i ^ { \prime }$ with $\rho = 0.1$ &\includegraphics[width=.2in,height = 0.02\textheight]{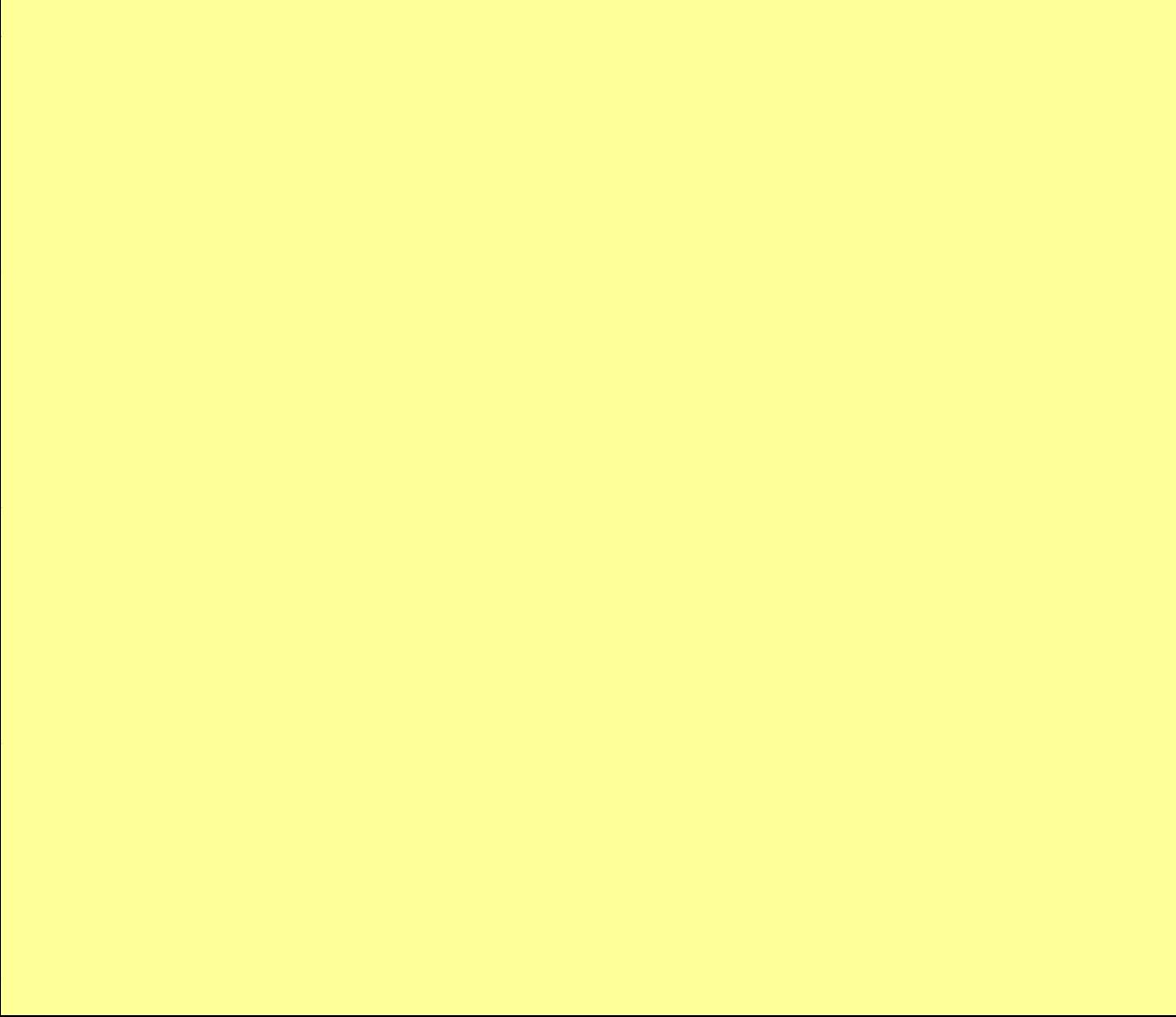} \\
				Corr(3) with $\rho = 0.1$  &\includegraphics[width=.2in ,height = 0.02\textheight]{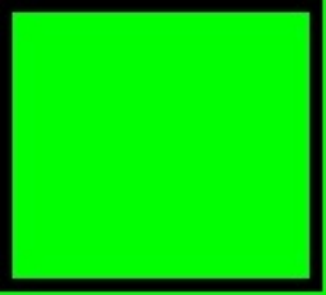} \\
				Corr(4) with $\rho_{AB} = 0.2 , \rho_{BA} = 0.5$ &\includegraphics[width=.2in,height = 0.02\textheight]{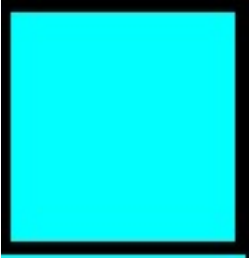} \\
				Corr(5) with $\rho_{AB} = \rho_{BA} = 0.4$ &\includegraphics[width=.2in,height = 0.02\textheight]{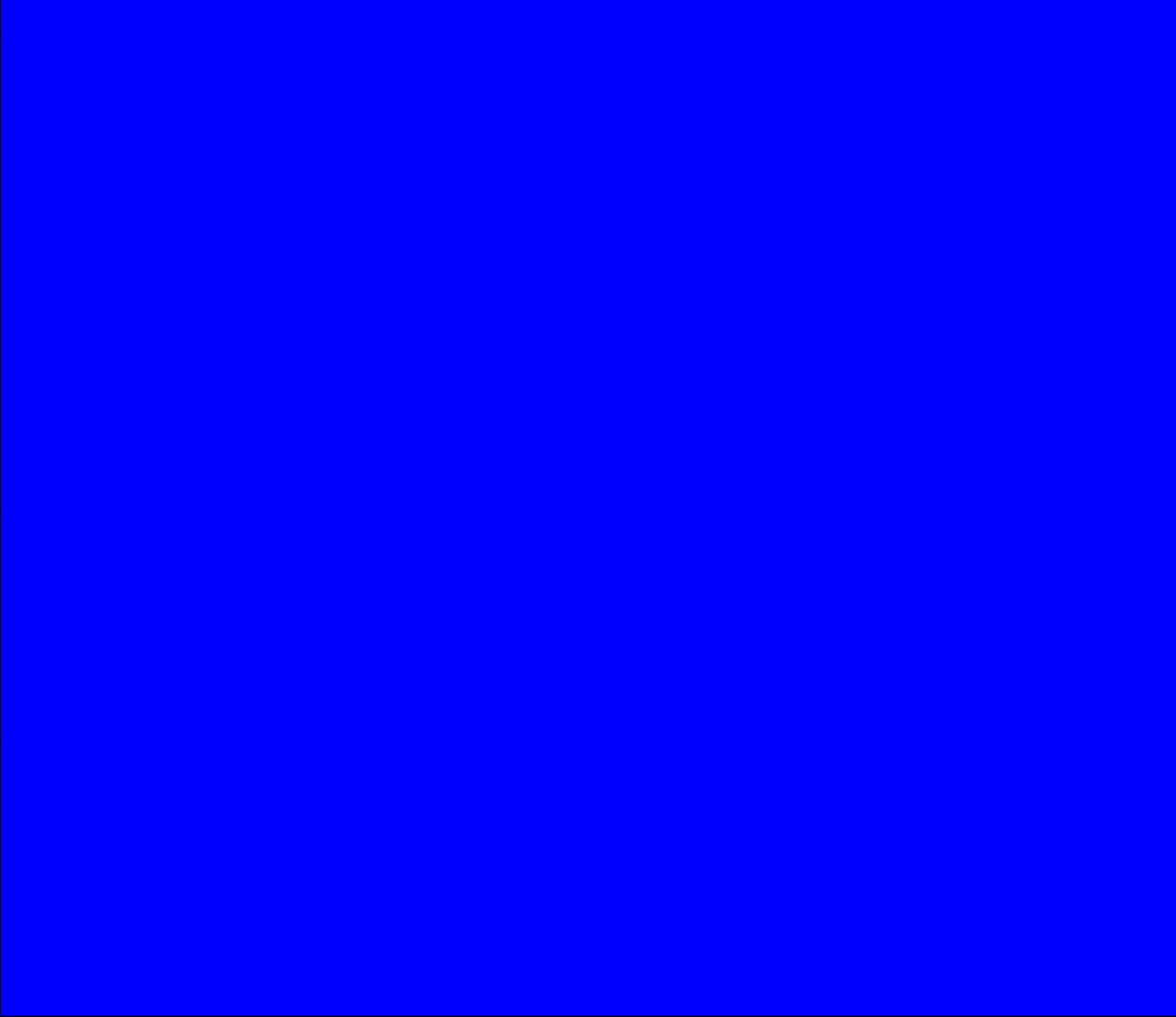} \\
				Corr(6) with $\rho_{AB} = 0.4 , \rho_{BA} = 0.3$ &\includegraphics[width=.2in,height = 0.02\textheight]{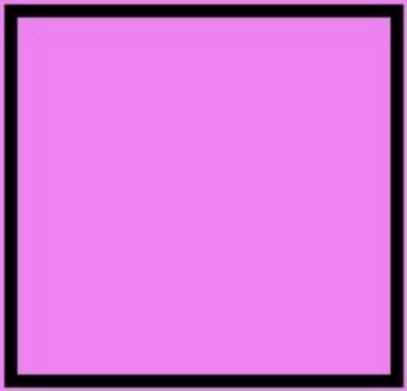} \\
				\hline
			\end{tabular}
		\end{center}

	%%%%%%%%%%%%%%%%%%%%%%%%%%%%%%%%%%%%%%%%%%%%%%%%%%%%%%%%%%%%%%%%%%%%%%%%%%%%%%%%%%%%%                                                         
	%                                                                                   %                        
	%                                                                                   %
	%                                                                                   %
	%                                                                                   %
	%                 SUB-SECTION 1: Optimal Designs when response is Binary            %
	%                                                                                   %
	%                                                                                   %
	%                                                                                   %
	%                                                                                   %
	%%%%%%%%%%%%%%%%%%%%%%%%%%%%%%%%%%%%%%%%%%%%%%%%%%%%%%%%%%%%%%%%%%%%%%%%%%%%%%%%%%%%%
	
	\subsection{Optimal Designs for Binary Response}\label{Binary}
	
	In case of binary response we calculate locally optimal designs under model~(\ref{logitmodel}) for different crossover designs.

	%%%%%%%%%%%%%%%%%%%%%%%%%%%%%%%%%%%%%%%%%%%%%%%%%%%%%%%%%%%%%%%%%%%%%%%%%%%%%%%%%%%%%                                                          
	%                                                                                   %                        
	%                                                                                   %
	%                                                                                   %
	%                                                                                   %
	%                                       p = 2                                       %
	%                                                                                   %
	%                                                                                   %
	%                                                                                   %
	%                                                                                   %
	%%%%%%%%%%%%%%%%%%%%%%%%%%%%%%%%%%%%%%%%%%%%%%%%%%%%%%%%%%%%%%%%%%%%%%%%%%%%%%%%%%%%%
	
\begin{center}
		\begin{figure}[h]
			\centering
			\begin{tabular}{cc}
				\includegraphics[scale=.5]{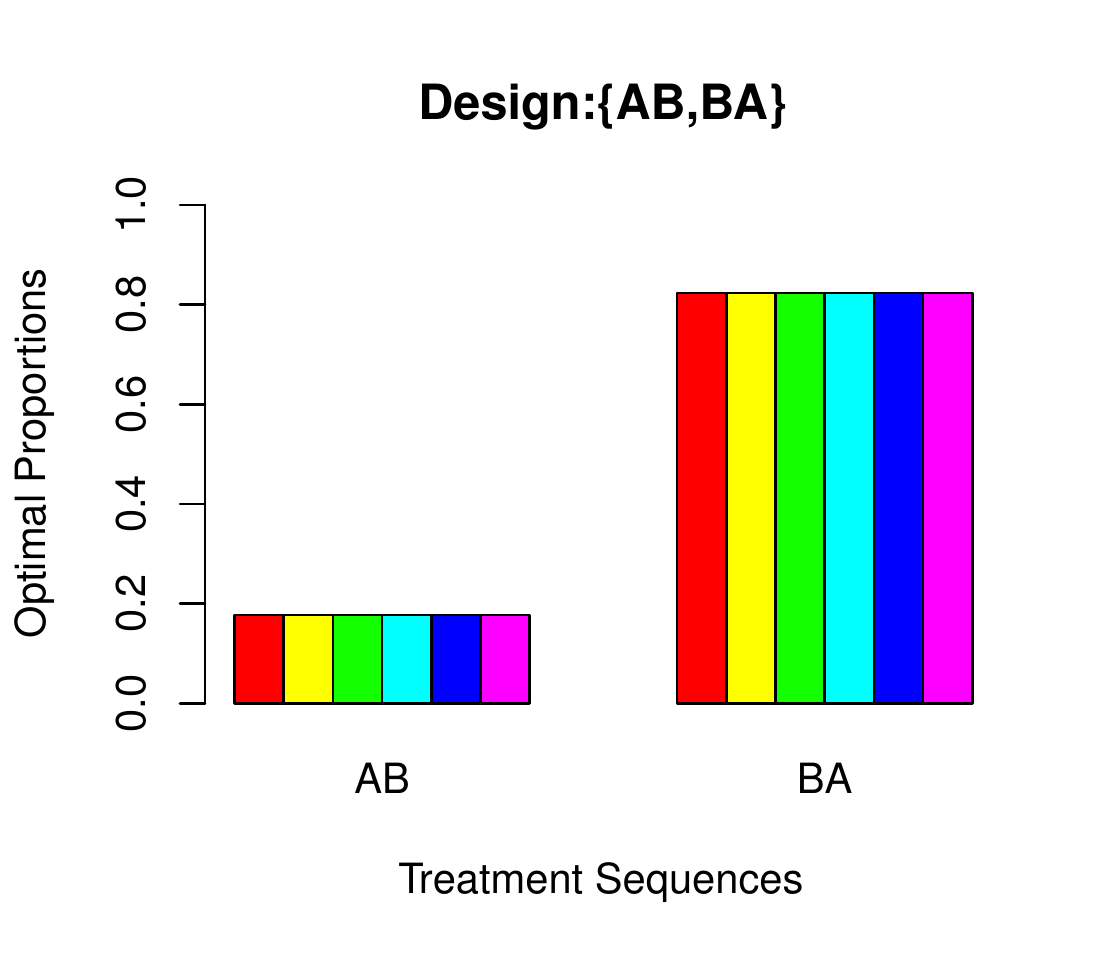} & \includegraphics[scale=.5]{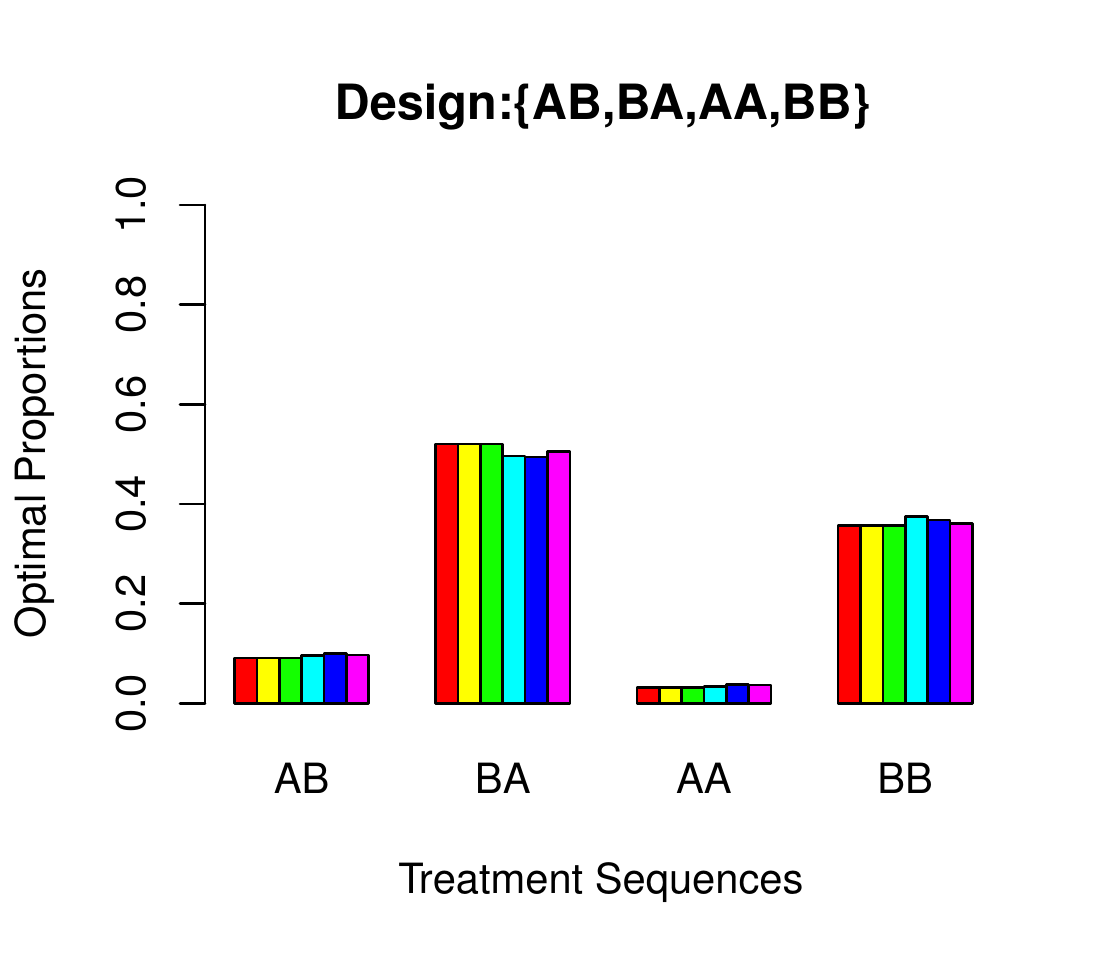}\\
			\end{tabular}
			\caption{Optimal proportions for $p=2$ case under $\theta_1$.}
			\label{Fig:OP2_1}
		\end{figure}
	\end{center}

\begin{center}
		\begin{figure}[h]
			\centering 
			\begin{tabular}{cc}
				\includegraphics[scale=.65]{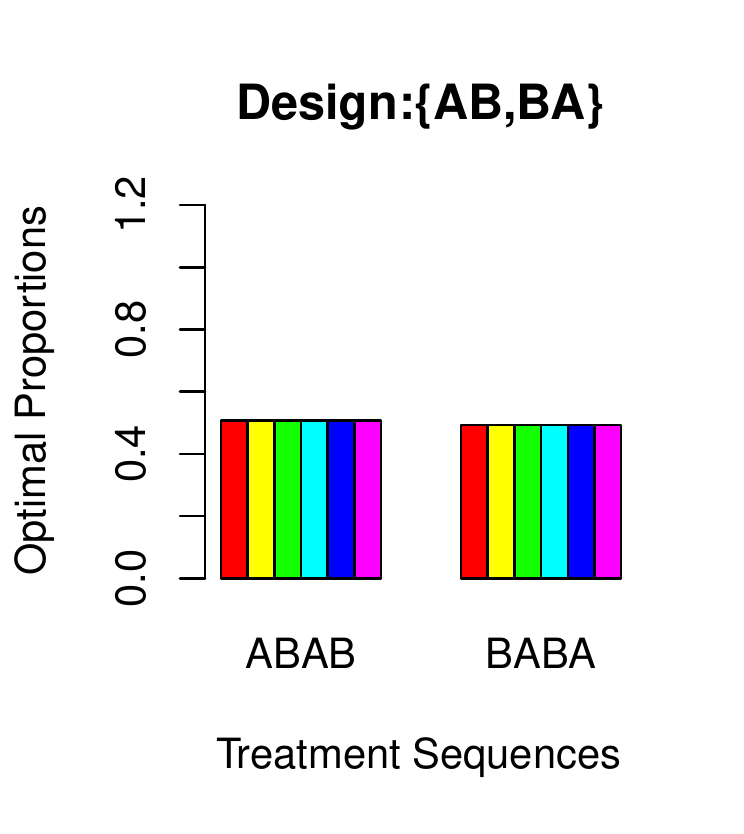} & \includegraphics[scale=.65]{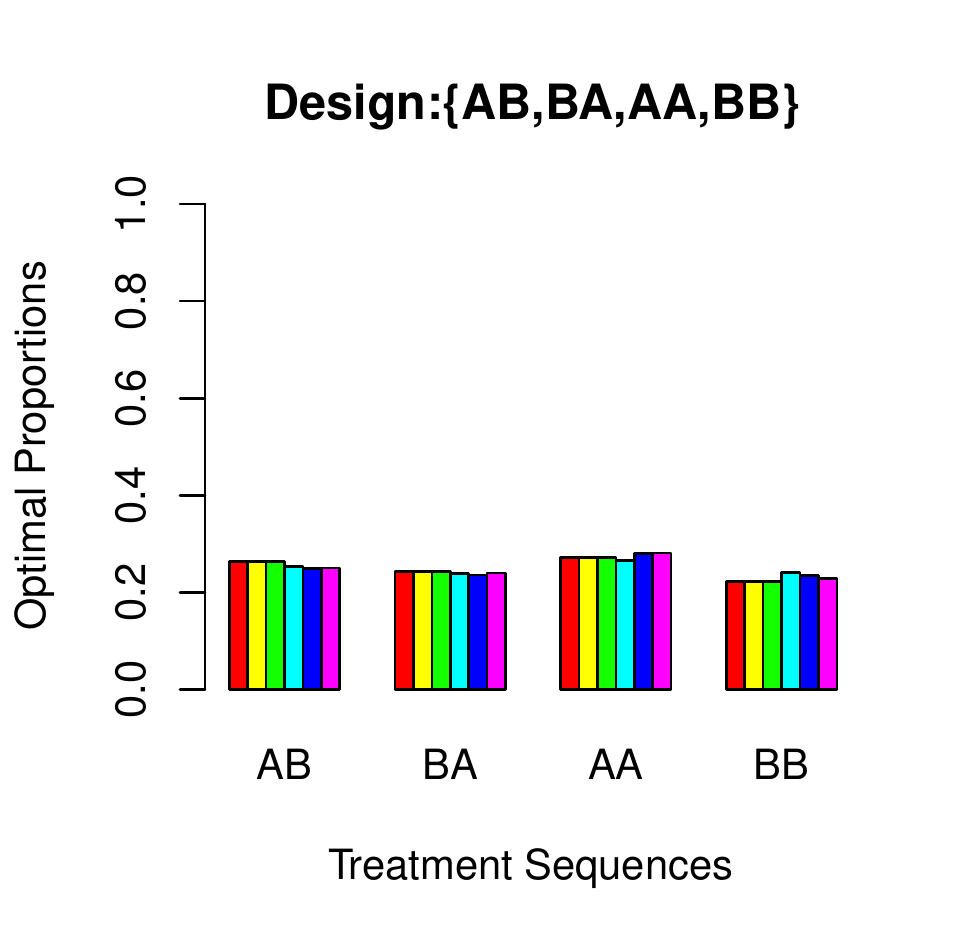}\\
			\end{tabular}
			\caption{Optimal proportions for $p=2$ case under $\theta_2$.}
			\label{Fig:OP2_2}
		\end{figure}
	\end{center}

	\medskip \noindent We first consider the local optimality approach, for $p = 2$ case. \bl{For illustration purpose, we assume that the parameter values are $ \theta_1 = $ $[\lambda, \beta_{2}, \tau_{B}, \rho_{B}] =$ $ [0.5, -1.0, 4.0,$ $-2.0]$ which gives us non-uniform optimal allocations and $ \theta_2 =$ $[\lambda,\beta_{2},\tau_{B},\rho_{B}] =$ $ [0.5,0.06,-0.35,$ $0.73]$ which gives us approximately uniform allocations.} Note that we need to know the parameter values before calculating the optimal proportions. If the initial guess for the model parameters changes, the obtained optimal proportions will change as well. For different correlation structures, the optimal designs (proportions) are stated in Table \ref{Tab1:OP2Case}. The same information is presented in Figure~\ref{Fig:OP2_1} and Figure~\ref{Fig:OP2_2} as well.
	
	\begin{table}[h]\caption{Optimal proportions for $p = 2$ case.}\label{Tab1:OP2Case}
		\scriptsize
		\begin{center}		
			\begin{tabular}{ |l|l|c|c| }
				\hline
				&&&\\
				Design Points & Corr & {Optimal proportions under $\theta_1$} & {Optimal proportions under $\theta_2$} \\
				\hline
				&&&\\		
				& Corr(1) & $\{ 0.1770, 0.8230\}$ & $\{ 0.5070, 0.4930\}$\\
				& Corr(2) & $\{ 0.1770, 0.8230\}$ & $\{ 0.5070, 0.4930\}$\\
				$\{ AB, BA \}$ & Corr(3) & $\{ 0.1770, 0.8230\}$ & $\{ 0.5070, 0.4930\}$\\
				& Corr(4) & $\{ 0.1770, 0.8230\}$ & $\{ 0.5070, 0.4930\}$\\
				& Corr(5) & $\{ 0.1770, 0.8230\}$ & $\{ 0.5070, 0.4930\}$\\
				& Corr(6) & $\{ 0.1770, 0.8230\}$ & $\{ 0.5070, 0.4930\}$\\
				\hline
				&&&\\	
				& Corr(1) & $\{ 0.0908, 0.5207, 0.0315, 0.3570\}$ & $\{ 0.2633, 0.2425, 0.2722, 0.2220\}$ \\
				& Corr(2) & $\{ 0.0908, 0.5207, 0.0315, 0.3570\}$ & $\{ 0.2633, 0.2425, 0.2722, 0.2220\}$ \\
				$\{ AB, BA,$ & Corr(3) & $\{ 0.0908, 0.5207, 0.0315, 0.3570\}$ & $\{ 0.2633, 0.2425, 0.2722, 0.2220\}$ \\
				$ AA, BB \}$ & Corr(4) & $\{ 0.0957, 0.4960, 0.0338, 0.3745 \}$ & $\{ 0.2534, 0.2393, 0.2661, 0.2412 \}$ \\
				& Corr(5) & $\{ 0.1002, 0.4941, 0.0379, 0.3678\}$ & $\{ 0.2496, 0.2359, 0.2801, 0.2344\}$ \\
				& Corr(6) & $\{ 0.0972, 0.5050, 0.0367, 0.3611\}$ & $\{ 0.2502, 0.2400, 0.2808, 0.2290\}$ \\
				\hline
			\end{tabular}
		\end{center}
	\end{table}

	\medskip \noindent It can be seen from the graphs in Figure~\ref{Fig:OP2_1} and Figure~\ref{Fig:OP2_2} that in case of $p = 2$ the optimal proportions do not vary when correlation structure changes both under $\theta_2$ and $\theta_1$. Uniform designs (same proportions for each sequence) are often used in practice. It is clear that those uniform designs are sub-optimal under $\theta_1.$

	%%%%%%%%%%%%%%%%%%%%%%%%%%%%%%%%%%%%%%%%%%%%%%%%%%%%%%%%%%%%%%%%%%%%%%%%%%%%%%%%%%%%%                                                          
	%                                                                                   %                        
	%                                                                                   %
	%                                                                                   %
	%                                                                                   %
	%                                       p = 3                                       %
	%                                                                                   %
	%                                                                                   %
	%                                                                                   %
	%                                                                                   %
	%%%%%%%%%%%%%%%%%%%%%%%%%%%%%%%%%%%%%%%%%%%%%%%%%%%%%%%%%%%%%%%%%%%%%%%%%%%%%%%%%%%%%
	
\begin{center}
		\begin{figure}[H]
			\centering
			\begin{tabular}{ccc}
				\includegraphics[scale=.45]{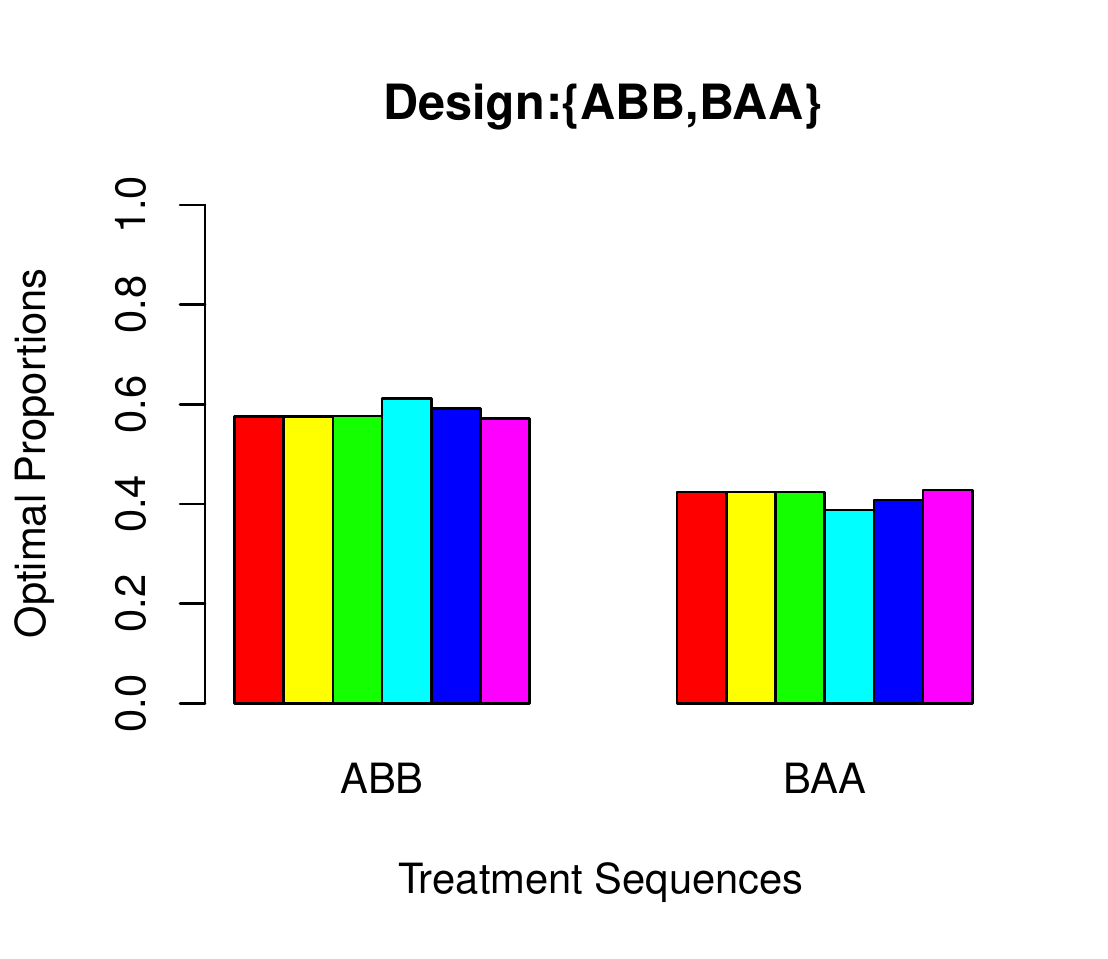} & \includegraphics[scale=.45]{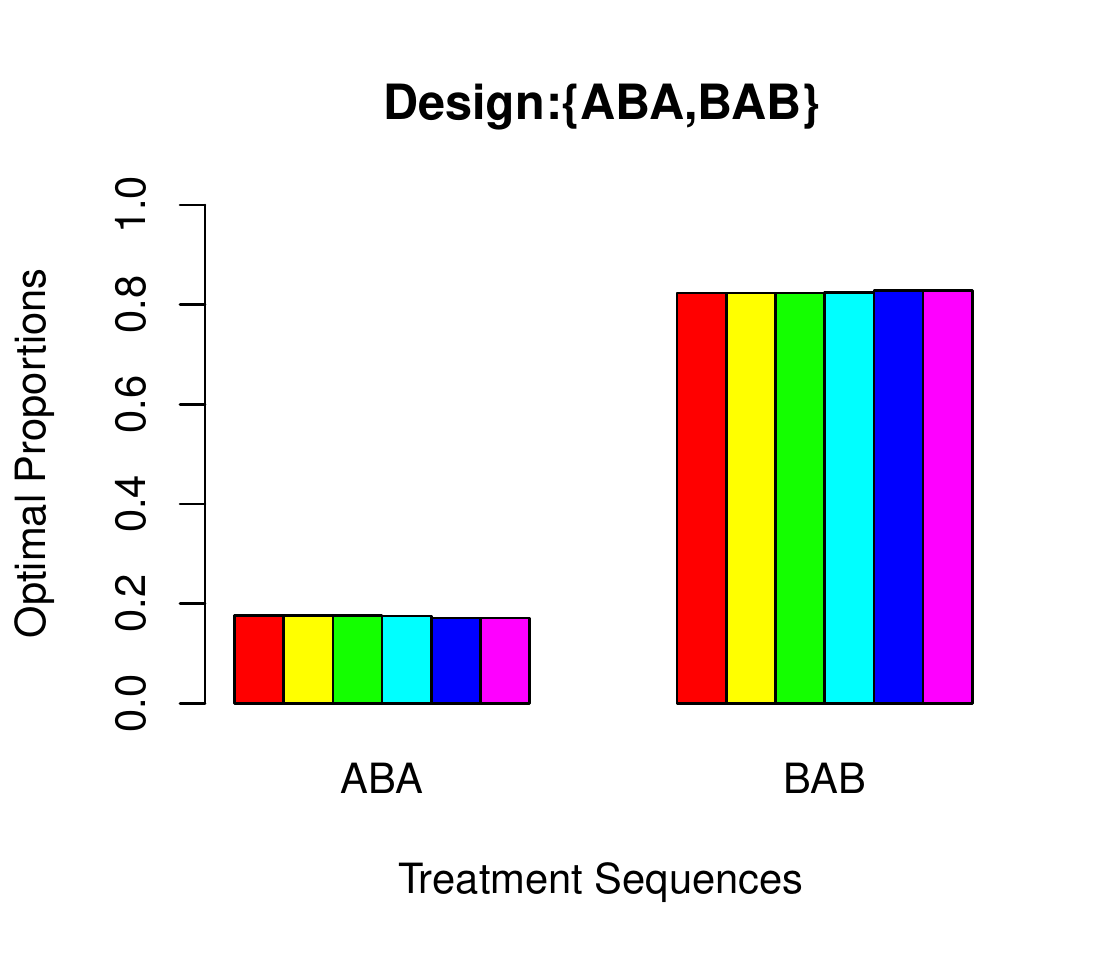}
				& \multicolumn{1}{c}{\includegraphics[scale=.45]{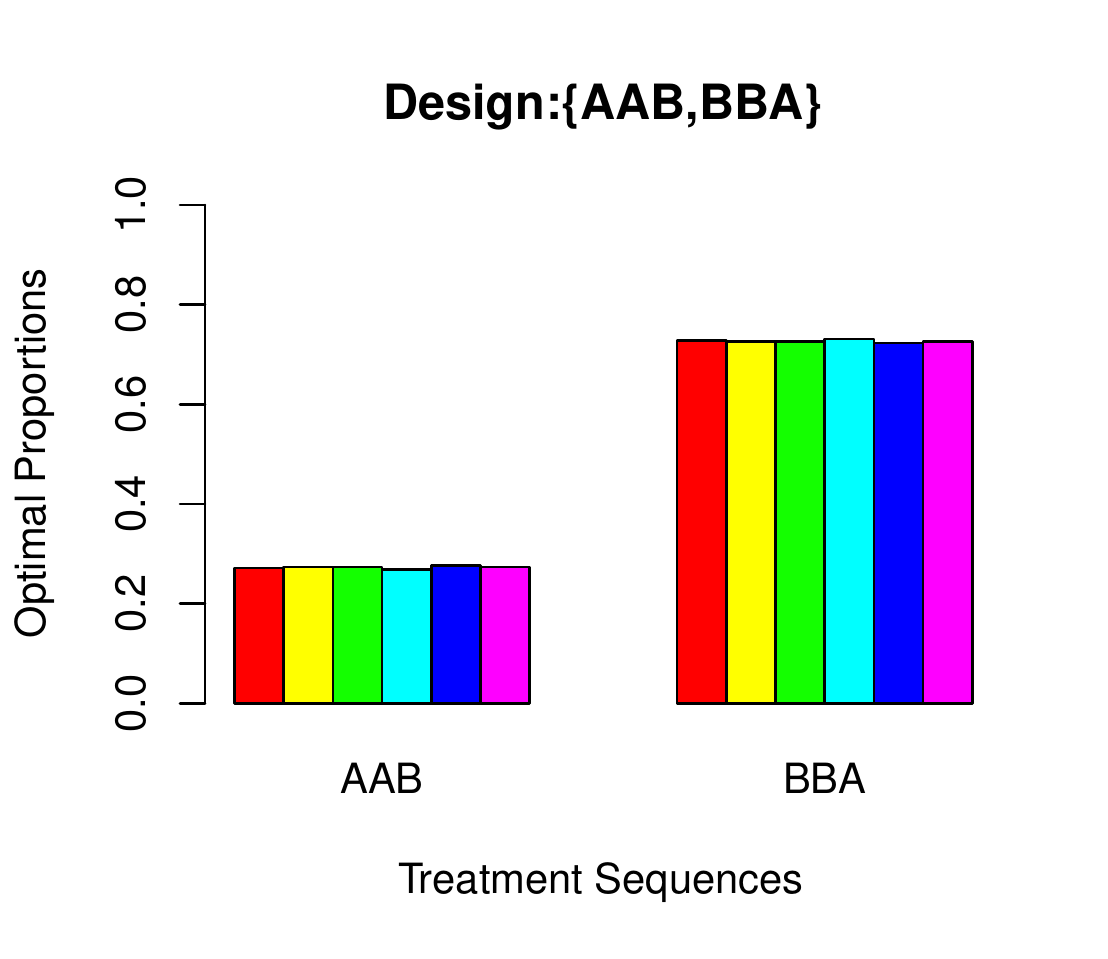}} \\
%			\end{tabular}
%			\caption{Optimal proportions for $p=3$ case with two-treatment sequences under $\theta_1$}
%			\label{Fig:OP3_21}
%		\end{figure}
%	\end{center}

%	\begin{center}
%		\begin{figure}
%			\centering
%			\begin{tabular}{ccc}
				\includegraphics[scale=.5]{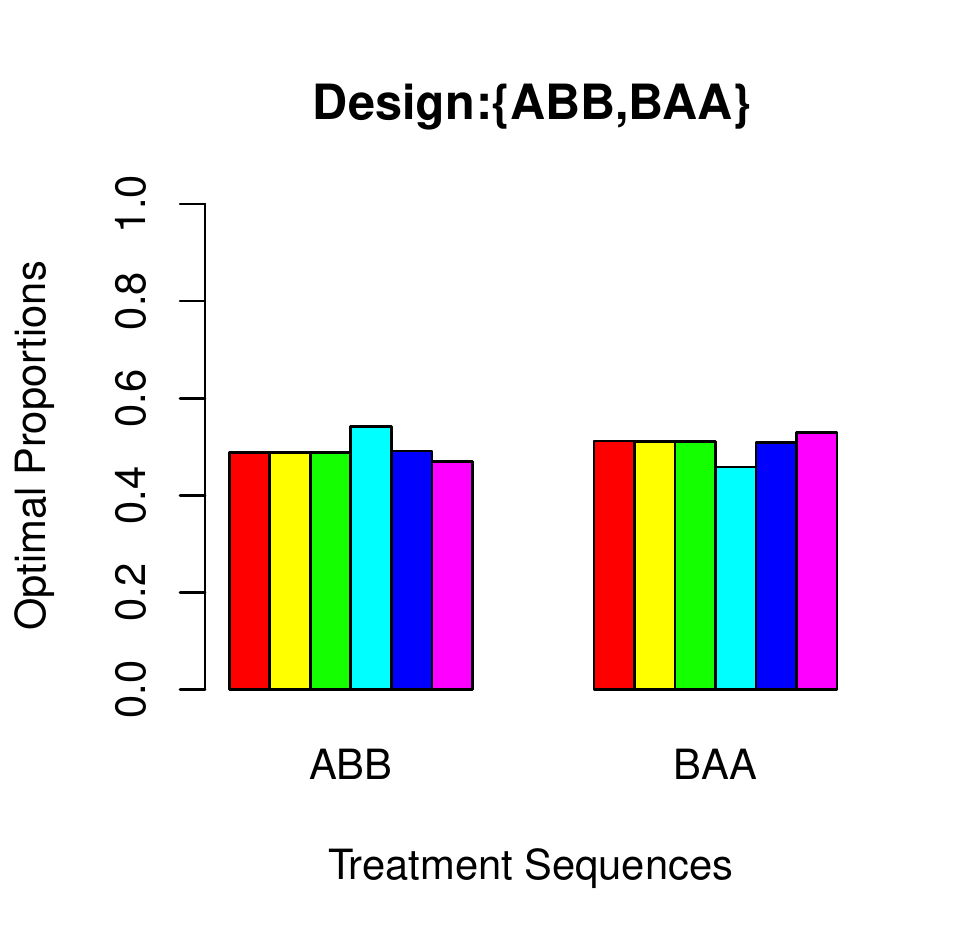} & \includegraphics[scale=.5]{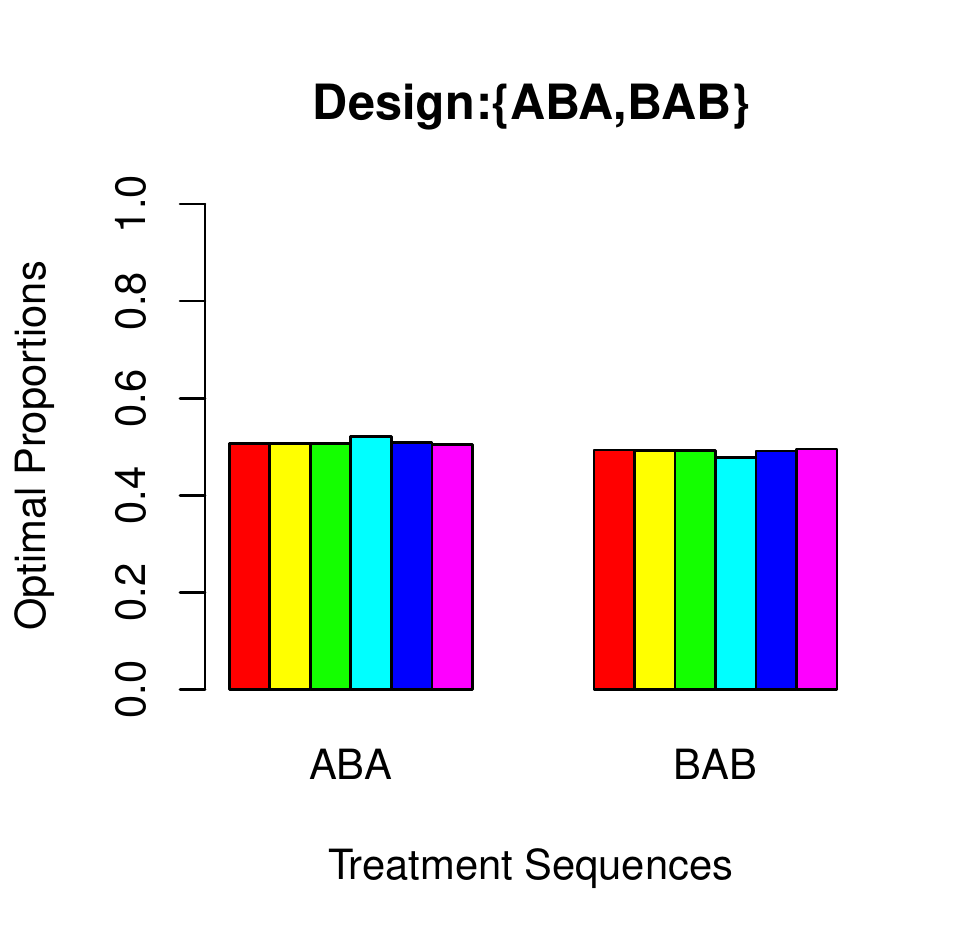}
				& \multicolumn{1}{c}{\includegraphics[scale=.5]{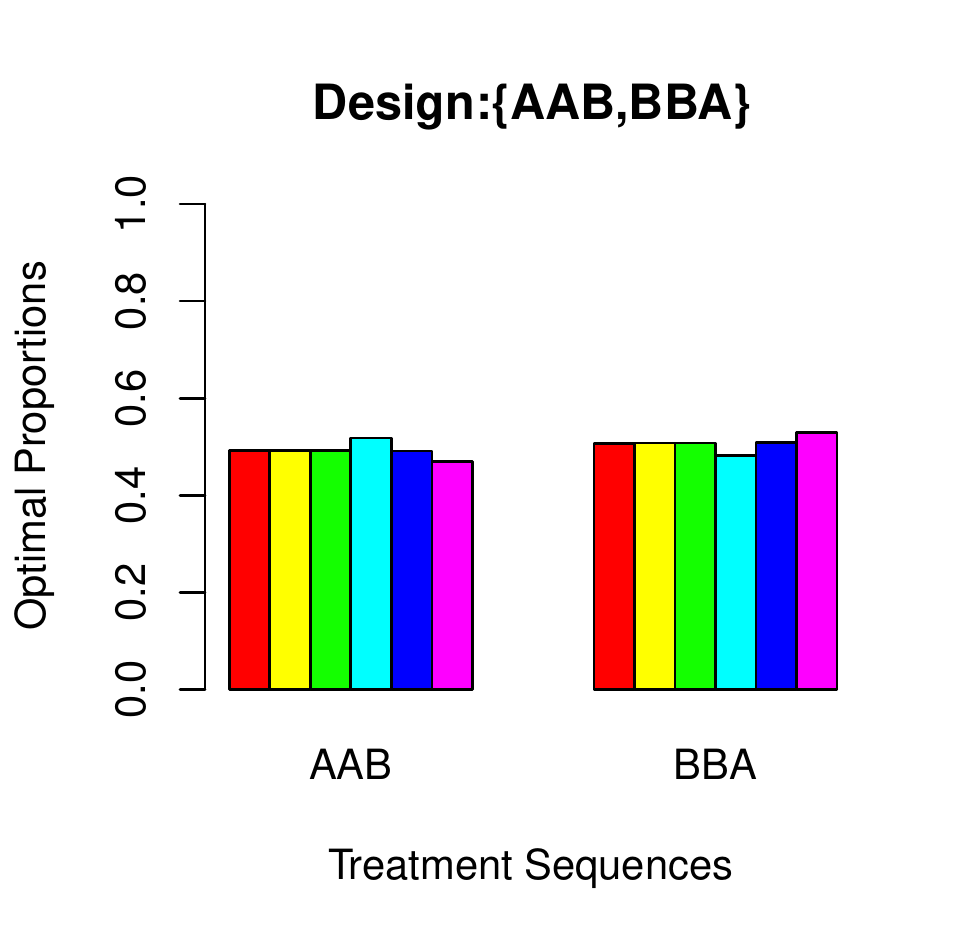}}
			\end{tabular}
			\caption{Optimal proportions for $p=3$ case with two-treatment sequences under $\theta_1$ and $\theta_2$ respectively.}
			\label{Fig:OP3_212}
		\end{figure}
	\end{center}

\begin{center}
		\begin{figure}[H]
			\centering
			\begin{tabular}{ccc}
				\includegraphics[scale=.45]{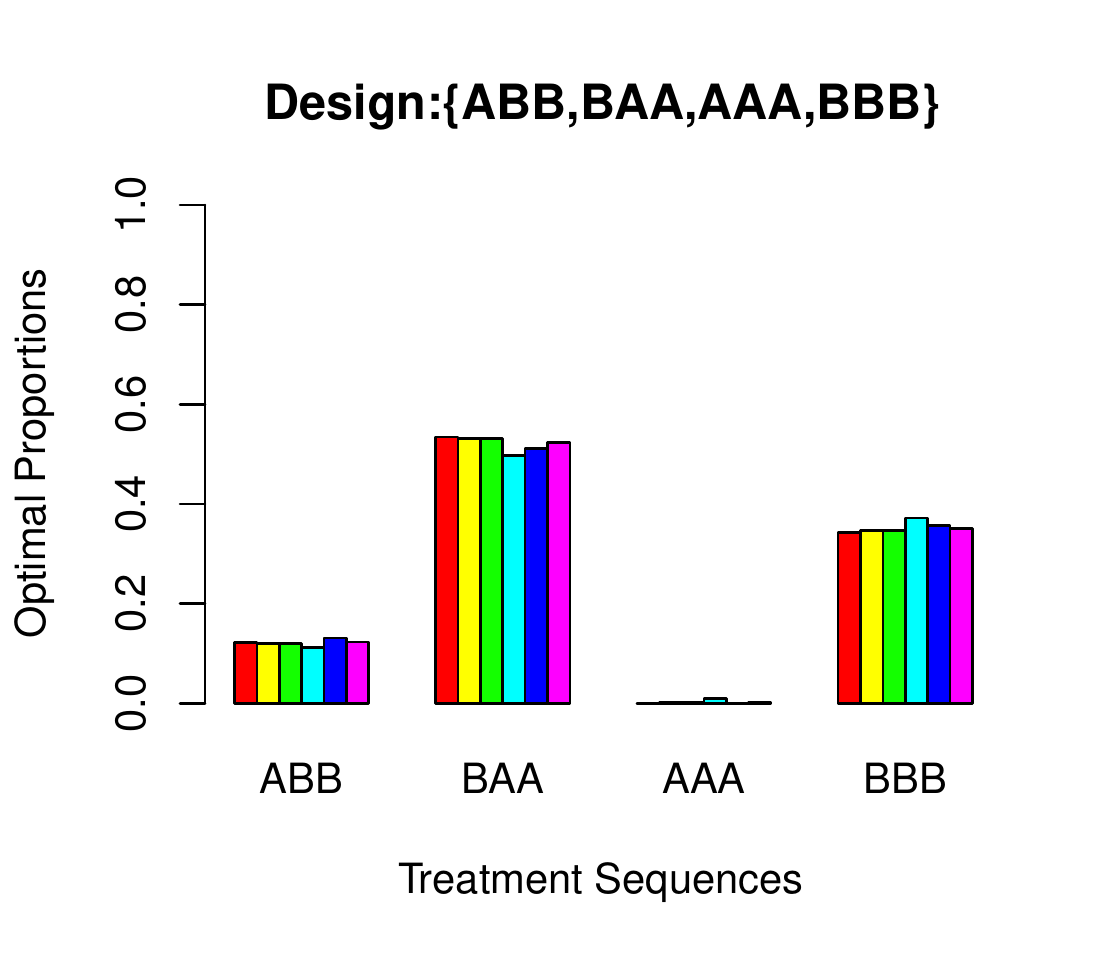} & \includegraphics[scale=.45]{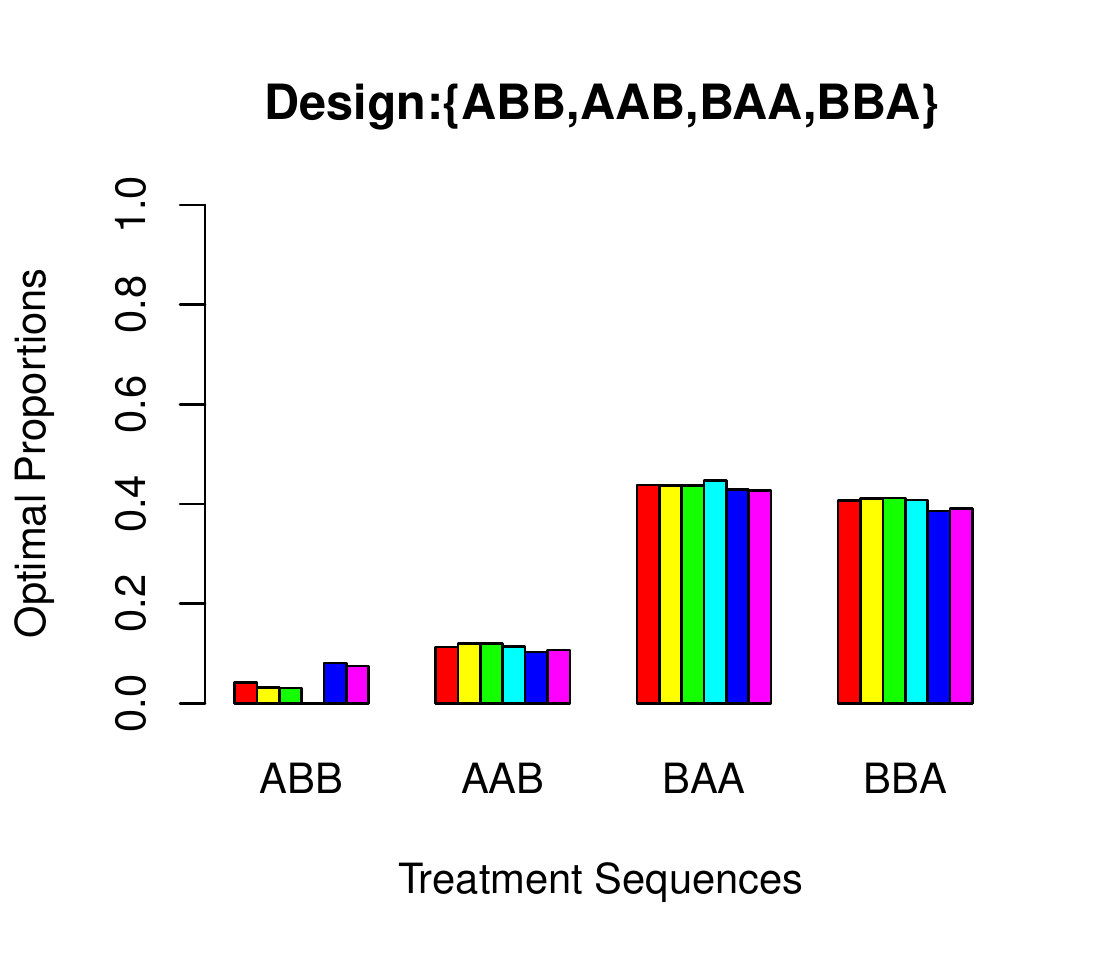} &
				\multicolumn{1}{c}{\includegraphics[scale=.45]{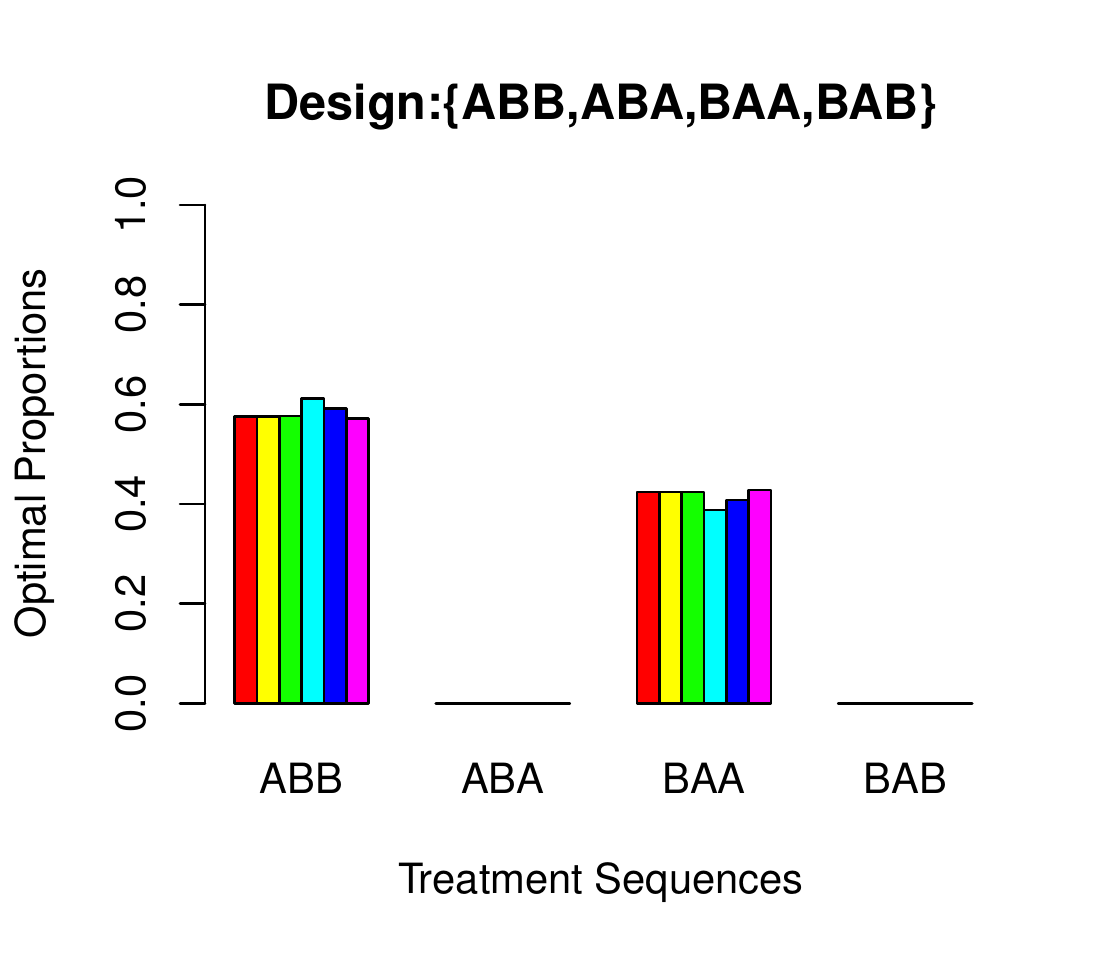}}\\
%			\end{tabular}
%			\caption{Optimal proportions for $p=3$ case with four-treatment sequences under $\theta_1$.}
%			\label{Fig:OP3_41}
%		\end{figure}
%	\end{center}	
	
%	\begin{center}
%		\begin{figure}[h]
%			\centering
%			\begin{tabular}{ccc}
				\includegraphics[scale=.5]{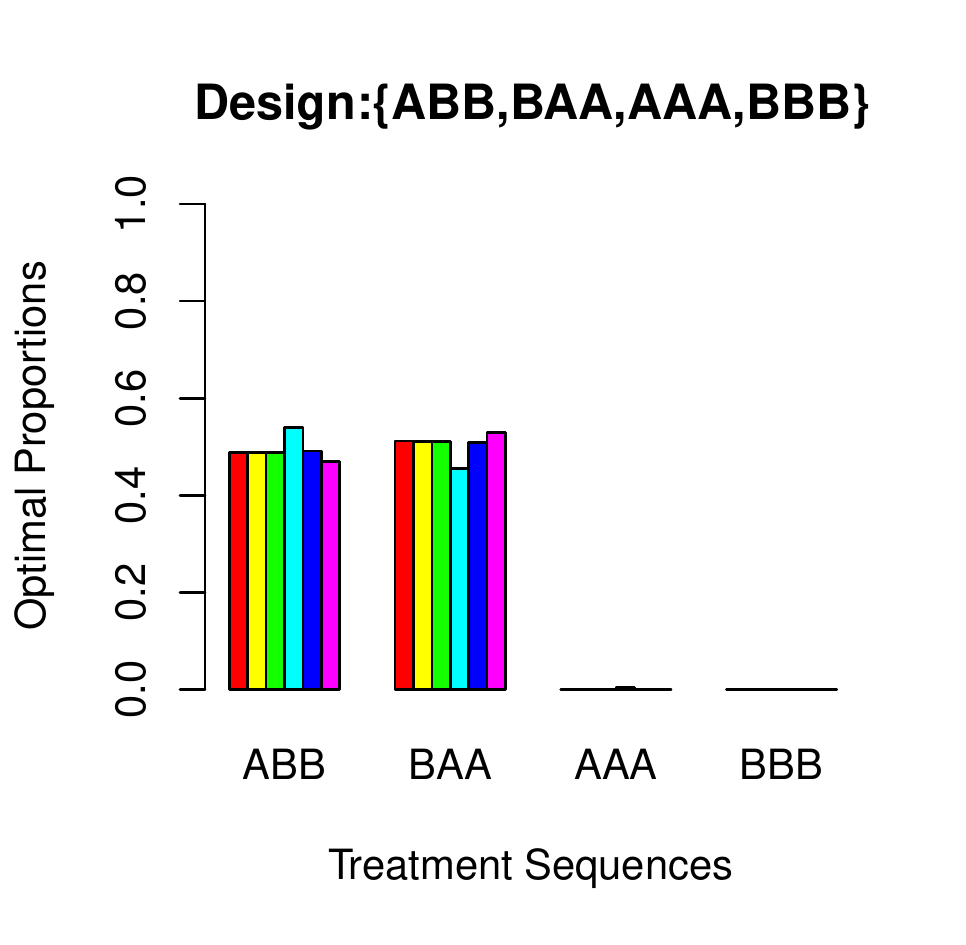} & \includegraphics[scale=.5]{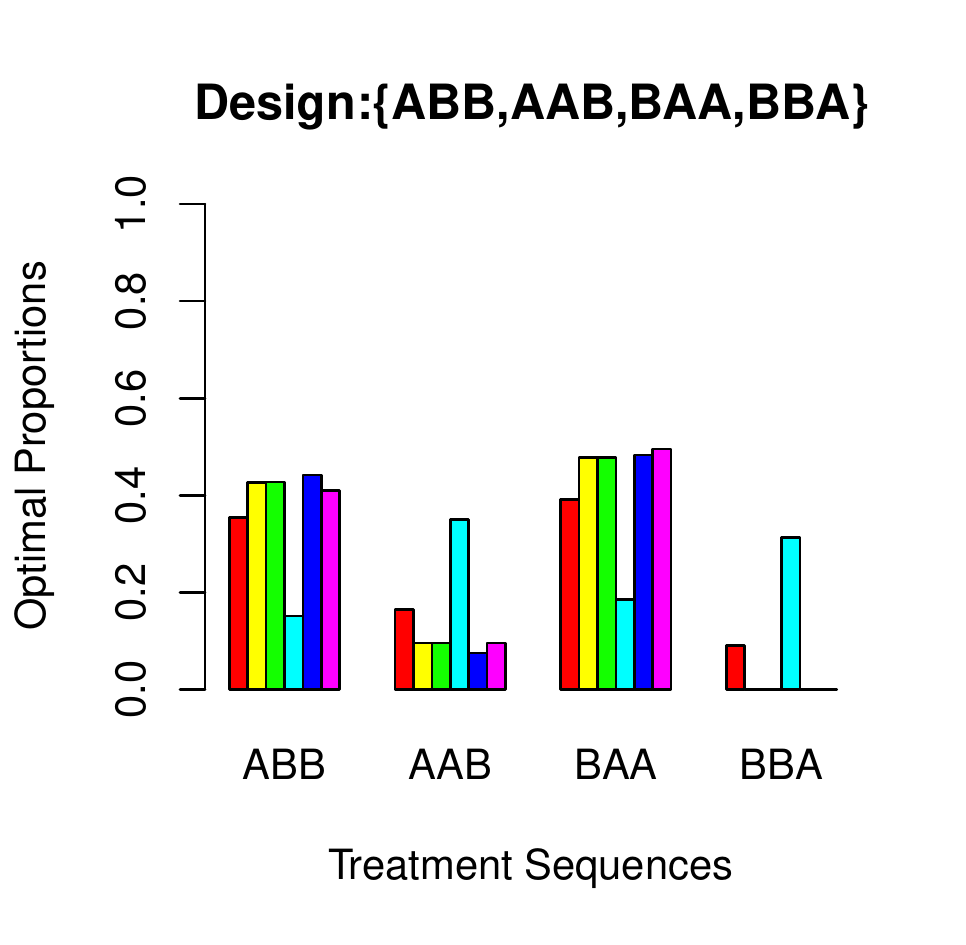} &
				\multicolumn{1}{c}{\includegraphics[scale=.5]{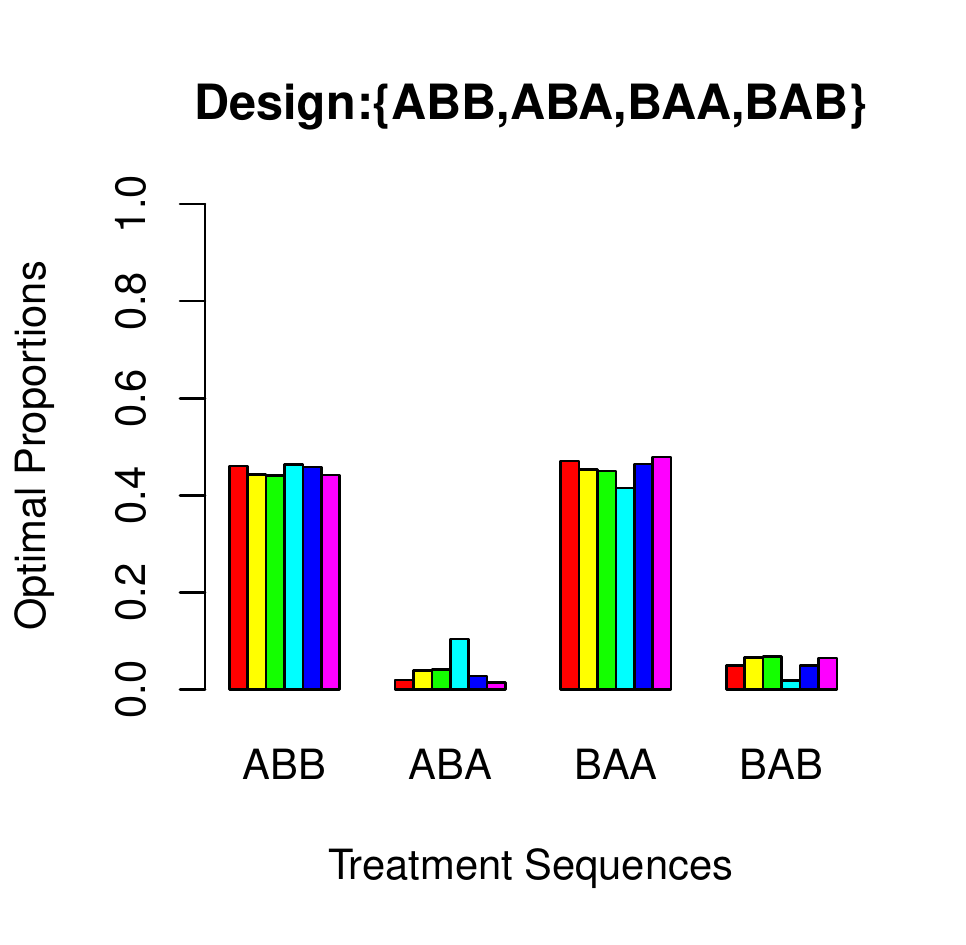}}
			\end{tabular}
			\caption{Optimal proportions for $p=3$ case with four-treatment sequences under $\theta_1$ and $\theta_2$ respectively.}
			\label{Fig:OP3_412}
		\end{figure}
	\end{center}

	\medskip 
	\noindent 
	For $p = 3$ case, \bl{as before suppose our guess for the parameter values are $\theta_1$ $= [\lambda$, $\beta_{2}$, $\beta_{3}$, $\tau_{B}$, $\rho_{B}] =$ $[0.5,-1.0,2.0,4.0,-2.0]$ which gives us non-uniform optimal allocations and $\theta_2$ $= [\lambda$ ,$\beta_{2}$, $\beta_{3}$ ,$\tau_{B}$, $\rho_{B}] =$ $[0.5,0.06,-0.53,-0.35 ,0.73]$ which gives us approximately uniform optimal allocations.} The designs are presented in Table~\ref{Tab2:OP32Case}, Figure~\ref{Fig:OP3_212} for the first example, and in Table~\ref{Tab3:OP34Case}, Figure~\ref{Fig:OP3_412} for the second example.  It can be seen that in case of $p = 3$ also the optimal proportions do not vary much when correlation structure changes under both $\theta_1$ and $\theta_2$. Similar to $p=2$ case it is clear from above table that uniform designs are sub-optimal for $p=3$ case with two- and four-treatment sequences under $\theta_1.$

	\begin{table}[h]
		\caption{Optimal proportions for $p = 3$ case for designs with two treatment sequences.}\label{Tab2:OP32Case}
		\scriptsize
		\begin{center}
			\begin{tabular}{ |l|l|p{2.2cm}|p{2.2cm}| }
				\hline
				&&&\\
				Design Points & Corr & {Optimal \hspace{.3in} proportions under $\theta_1$} & {Optimal \hspace{.3in} proportions under $\theta_2$} \\
				\hline
				&&&\\
				& Corr(1) & $\{ 0.5756, 0.4244\}$ & $\{ 0.4880, 0.5120\}$ \\
				& Corr(2) & $\{ 0.5761, 0.4239\}$ & $\{ 0.4887, 0.5113\}$ \\
				$\{ ABB, BAA \}$ & Corr(3) & $\{ 0.5762, 0.4238\}$ & $\{ 0.4888, 0.5112\}$ \\
				& Corr(4) & $\{ 0.6120, 0.3880\}$ & $\{ 0.5416, 0.4584\}$ \\
				& Corr(5) & $\{ 0.5921, 0.4079\}$ & $\{ 0.4917, 0.5083\}$ \\
				& Corr(6) & $\{ 0.5721, 0.4279\}$ & $\{ 0.4700, 0.5300\}$ \\
				\hline
				& Corr(1) & $\{ 0.1768, 0.8232\}$ & $\{ 0.5070, 0.4930\}$ \\
				& Corr(2) & $\{ 0.1766, 0.8234\}$ & $\{ 0.5072, 0.4928\}$ \\
				$\{ ABA, BAB \}$	 & Corr(3) & $\{ 0.1766, 0.8234\}$ & $\{ 0.5072, 0.4928\}$ \\
				& Corr(4) & $\{ 0.1756, 0.8244\}$ & $\{ 0.5217, 0.4783\}$ \\
				& Corr(5) & $\{ 0.1714, 0.8286\}$ & $\{ 0.5088, 0.4912\}$ \\
				& Corr(6) & $\{ 0.1715, 0.8285\}$ & $\{ 0.5043, 0.4957\}$ \\
				\hline
				& Corr(1) & $\{ 0.2713, 0.7287\}$ & $\{ 0.4927, 0.5073\}$ \\
				& Corr(2) & $\{ 0.2738, 0.7262\}$ & $\{ 0.4926, 0.5074\}$ \\
				$\{ AAB, BBA \}$ & Corr(3) & $\{ 0.2740, 0.7260\}$ & $\{ 0.4926, 0.5074\}$ \\
				& Corr(4) & $\{ 0.2685, 0.7315\}$ & $\{ 0.5181, 0.4819\}$ \\
				& Corr(5) & $\{ 0.2771, 0.7229\}$ & $\{ 0.4911, 0.5089\}$ \\
				& Corr(6) & $\{ 0.2740, 0.7260\}$ & $\{ 0.4702, 0.5298\}$ \\
				\hline
			\end{tabular}
		\end{center}
	\end{table}

	\begin{table}[H]\caption{Optimal proportions for $p = 3$ case for designs with four treatment sequences.}\label{Tab3:OP34Case}
		\scriptsize	
		\begin{center}	
			\begin{tabular}{ |p{1cm}|l|c|c| }
				\hline
				&&&\\
				Design Points & Corr & {Optimal proportions under $\theta_1$} & {Optimal proportions under $\theta_2$} \\
				\hline
				&&&\\
				& Corr(1) & $\{ 0.1222, 0.5344, 0.0000, 0.3434\}$ & $\{ 0.4880, 0.5120, 0.0000, 0.0000\}$ \\
				$\{ ABB,$ & Corr(2) & $\{ 0.1199, 0.5316, 0.0022, 0.3463\}$ & $\{ 0.4887, 0.5113, 0.0000, 0.0000\}$ \\
				$BAA,$ & Corr(3) & $\{ 0.1197, 0.5312, 0.0025, 0.3466\}$ & $\{ 0.4888, 0.5112, 0.0000, 0.0000\}$ \\
				$AAA,$ & Corr(4) & $\{ 0.1115, 0.4975, 0.0100, 0.3720\}$ & $\{ 0.5398, 0.4556, 0.0046, 0.0000\}$ \\
				$BBB\}$& Corr(5) & $\{ 0.1313, 0.5113, 0.0000, 0.3574\}$ & $\{ 0.4917, 0.5083, 0.0000, 0.0000\}$ \\
				& Corr(6) & $\{ 0.1233, 0.5236, 0.0018, 0.3513\}$ & $\{ 0.4700, 0.5300, 0.0000, 0.0000\}$ \\
				\hline
				& Corr(1) & $\{ 0.0413, 0.1130, 0.4384, 0.4073\}$ & $\{ 0.3544, 0.1646, 0.3908, 0.0902\}$ \\
				$\{ ABB,$ & Corr(2) & $\{ 0.0316, 0.1196, 0.4373, 0.4115\}$ & $\{ 0.4266, 0.0957, 0.4777, 0.0000\}$ \\
				$AAB,$ & Corr(3) & $\{ 0.0304, 0.1204, 0.4371, 0.4121\}$ &  $\{ 0.4271, 0.0953, 0.4776, 0.0000\}$\\
				$BAA,$ & Corr(4) & $\{ 0.0005, 0.1440, 0.4471, 0.4084\}$ & $\{ 0.1512, 0.3503, 0.1854, 0.3131 \}$ \\
				$BBA\}$& Corr(5) & $\{ 0.0811, 0.1033, 0.4297, 0.3858\}$ &  $\{ 0.4420, 0.0747, 0.4833, 0.0000\}$ \\
				& Corr(6) & $\{ 0.0749, 0.1070, 0.4270, 0.3911\}$ & $\{ 0.4094, 0.0955, 0.4951, 0.0000\}$ \\
				\hline
				& Corr(1) & $\{ 0.5755, 0.0000, 0.4244, 0.0000\}$ & $\{ 0.4606, 0.0194, 0.4710, 0.0490\}$ \\
				$\{ ABB,$ & Corr(2) & $\{ 0.5761, 0.0000, 0.4239, 0.0000\}$ & $\{ 0.4430, 0.0391, 0.4526, 0.0653\}$ \\
				$ABA,$ & Corr(3) & $\{ 0.5762, 0.0000, 0.4238, 0.0000\}$ & $\{ 0.4408, 0.0415, 0.4504, 0.0673\}$ \\
				$BAA,$ & Corr(4) & $\{ 0.6120, 0.0000, 0.3880, 0.0000 \}$ & $\{ 0.4634, 0.1036, 0.4152, 0.0178 \}$ \\
				$BAB\}$& Corr(5) & $\{ 0.5921, 0.0000, 0.4079, 0.0000\}$ & $\{ 0.4582, 0.0280, 0.4642, 0.0496\}$ \\
				& Corr(6) & $\{ 0.5721, 0.0000, 0.4279, 0.0000\}$ & $\{ 0.4420, 0.0142, 0.4787, 0.0651\}$ \\
				\hline	
			\end{tabular}
		\end{center}
	\end{table}

	\noindent Figure~\ref{Fig:OP3_412} shows that, unlike the previous examples, here under $\theta_1$ the optimal proportions vary a little for different correlation structures. Also, as before, not only the uniform design is sub-optimal here, the first and third designs have optimal allocations very low for some sequences. Also it can be observed from Figure~\ref{Fig:OP3_412} that under $\theta_2$ for different correlation structures some of the optimal proportions are zero for all the three designs. Hence under $\theta_2$ these designs fail to have uniform allocations.
	
	%\begin{landscape}
	%\end{landscape}

	%\begin{landscape}
	%\end{landscape}	
	
	%%%%%%%%%%%%%%%%%%%%%%%%%%%%%%%%%%%%%%%%%%%%%%%%%%%%%%%%%%%%%%%%%%%%%%%%%%%%%%%%%%%%%                                                          
	%                                                                                   %                        
	%                                                                                   %
	%                                                                                   %
	%                                                                                   %
	%                                       p = 4                                       %
	%                                                                                   %
	%                                                                                   %
	%                                                                                   %
	%                                                                                   %
	%%%%%%%%%%%%%%%%%%%%%%%%%%%%%%%%%%%%%%%%%%%%%%%%%%%%%%%%%%%%%%%%%%%%%%%%%%%%%%%%%%%%%
	
	\medskip\noindent For $p = 4$ case, \bl{in a similar way, we calculate locally optimal designs with nominal parameter values as $\theta_1$ $= [\lambda$, $\beta_{2}$, $\beta_{3}$, $\beta_{4}$, $\tau_{B}$, $\rho_{B}] =$ $[0.5, -1.0, 2.0, -1.5, 4.0, -2.0]$ which gives us non-uniform allocations and  $\theta_2$ $= [\lambda$, $\beta_{2}$, $\beta_{3}$, $\beta_{4}$, $\tau_{B}$, $\rho_{B}] =$ $[0.5$, $0.06$, $-0.53$, $-0.6$, $-0.35$, $0.73]$ which gives us approximately uniform allocations.} From Table~\ref{Tab4:OP4Case} and Figure~\ref{Fig:OP4_12} it is clear that similar to $p=2$ and $p=3$ cases the uniform designs are sub-optimal for $p=4$ case under $\theta_1$. 
	
	\begin{table}[H]\caption{Optimal proportions for $p = 4$ case.}\label{Tab4:OP4Case}
		\scriptsize	
		\begin{center}	
			\begin{tabular}{ |l|l|p{2.2cm}|p{2.2cm}| }
				\hline
				&&&\\
				Design Points & Corr & {Optimal \hspace{.3in} proportions under $\theta_1$} & {Optimal \hspace{.3in} proportions under $\theta_2$} \\
				\hline
				&&&\\
				& Corr(1) & $\{ 0.2723, 0.7277\}$ & $\{ 0.4953, 0.5047\}$ \\
				& Corr(2) & $\{ 0.2743, 0.7257\}$ & $\{ 0.4949, 0.5051\}$ \\
				$\{ AABB, BBAA \}$ & Corr(3) & $\{ 0.2744, 0.7256\}$ & $\{ 0.4949, 0.5051\}$ \\
				& Corr(4) & $\{ 0.2690, 0.7310 \}$ & $\{ 0.5244, 0.4756 \}$ \\
				& Corr(5) & $\{ 0.2772, 0.7228\}$ & $\{ 0.4937, 0.5063\}$ \\
				& Corr(6) & $\{ 0.2745, 0.7255\}$ & $\{ 0.4700, 0.5300\}$ \\
				\hline
				& Corr(1) & $\{ 0.6075, 0.3925\}$ & $\{ 0.4992, 0.5008\}$ \\
				& Corr(2) & $\{ 0.6045, 0.3955\}$ & $\{ 0.4998, 0.5002\}$ \\
				$\{ ABBA, BAAB \}$ & Corr(3) & $\{ 0.6042, 0.3958\}$ & $\{ 0.4998, 0.5002\}$ \\
				& Corr(4) & $\{ 0.5815, 0.4185\}$ & $\{ 0.4927, 0.5073\}$ \\
				& Corr(5) & $\{ 0.6444, 0.3556\}$ & $\{ 0.5021, 0.4979\}$ \\
				& Corr(6) & $\{ 0.6419, 0.3581\}$ & $\{ 0.5007, 0.4993\}$ \\
				\hline
				& Corr(1) & $\{ 0.1763, 0.8237\}$ & $\{ 0.5071, 0.4929\}$ \\
				& Corr(2) & $\{ 0.1767, 0.8233\}$ & $\{ 0.5071, 0.4929\}$ \\
				$\{ ABAB, BABA \}$	   & Corr(3) & $\{ 0.1767, 0.8233\}$ & $\{ 0.5071, 0.4929\}$ \\
				& Corr(4) & $\{ 0.1722, 0.8278\}$ & $\{ 0.5086, 0.4914\}$ \\
				& Corr(6) & $\{ 0.1714, 0.8286\}$ & $\{ 0.5031, 0.4969\}$ \\
				\hline
				
			\end{tabular}
		\end{center}
	\end{table}

	\medskip\noindent In most cases we may not have a clear idea about true correlation structure for responses and hence we choose an working correlation structure. The results in this section show that no matter what correlation structure we choose or what parameter estimates we choose, the proposed design gives almost similar optimal proportions in each case, which suggests that optimal designs are robust.
	
	%%%%%%%%%%%%%%%%%%%%%%%%%%%%%%%%%%%%%%%%%%%%%%%%%%%%%%%%%%%%%%%%%%%%%%%%%%%%%%%%%%%%%                                                          
	%                                                                                   %                        
	%                                                                                   %
	%                                                                                   %
	%                                                                                   %
	%                 SUB-SECTION 2: Optimal Designs when response is Poisson           %
	%                                                                                   %
	%                                                                                   %
	%                                                                                   %
	%                                                                                   %
	%%%%%%%%%%%%%%%%%%%%%%%%%%%%%%%%%%%%%%%%%%%%%%%%%%%%%%%%%%%%%%%%%%%%%%%%%%%%%%%%%%%%%
	
	\subsection{Optimal Designs for Poisson Response}\label{Poisson}
	
	\bl{In the case of Poisson response we calculate locally optimal design for following example under the model,
		\begin{eqnarray}\label{poisson}
		\textrm{log}(\mu_{ij}) = \eta_{ij} = \lambda + \beta_{i} + \tau_{d(i,j)} + \rho_{d(i-1,j)} ,
		\end{eqnarray}  
		where notations have the same meaning as in equation~(\ref{logitmodel}).
		
		\medskip \noindent We consider an example described in Layard and Arvesen (1978). In a crossover clinical trial to test a standard anti-nausea treatment (drug $A$) against a proposed treatment (drug $B$), twenty subjects were tested, ten for each order of administration. The response variable is the number of episodes of nausea suffered by a patient during the first two hours after cancer chemotherapy, and for a given patient is approximately Poisson distributed. The data are given in Table~\ref{Tab:PoissonData}.

	%\begin{landscape}
	\begin{center}
		\begin{figure}[H]
			\centering
			\begin{tabular}{ccc}
				\includegraphics[scale=.45]{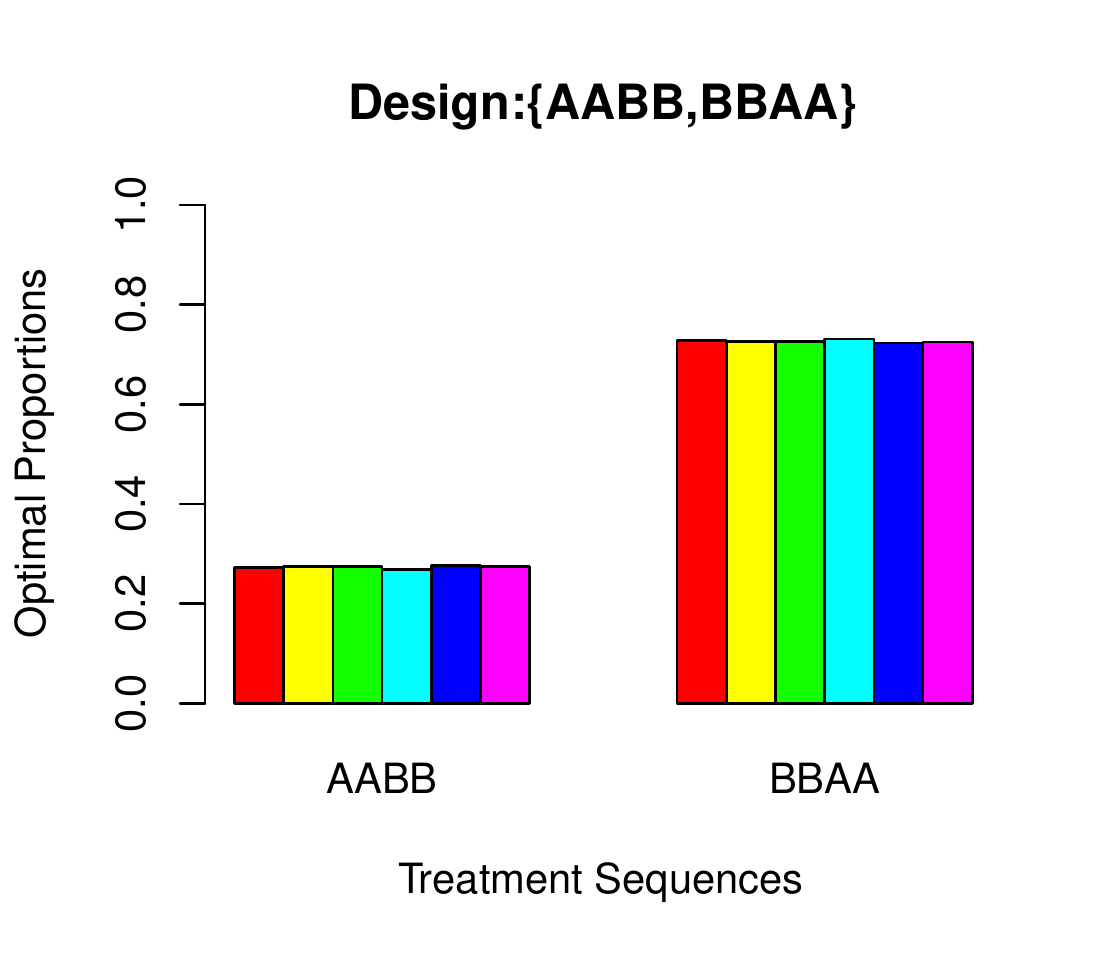} & \includegraphics[scale=.45]{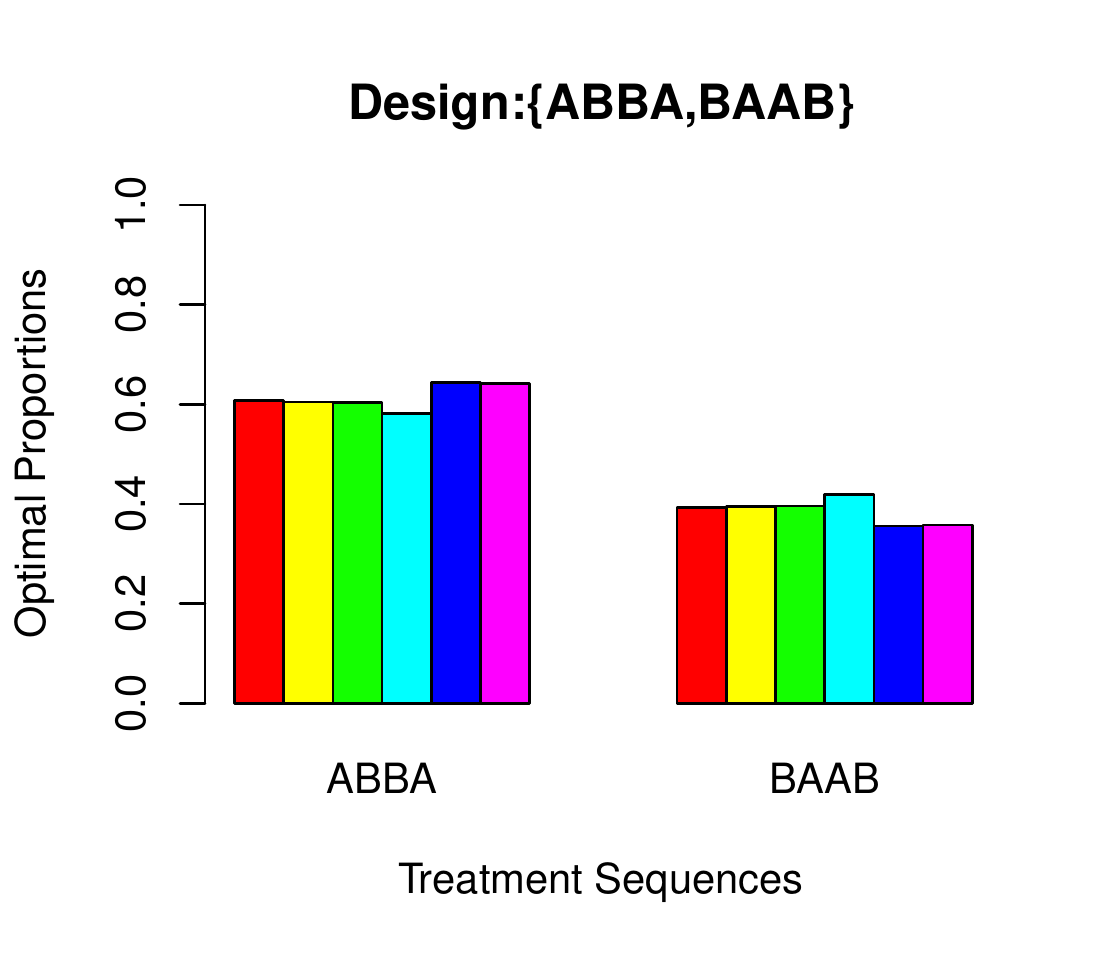} & 
				\multicolumn{1}{c}{\includegraphics[scale=.45]{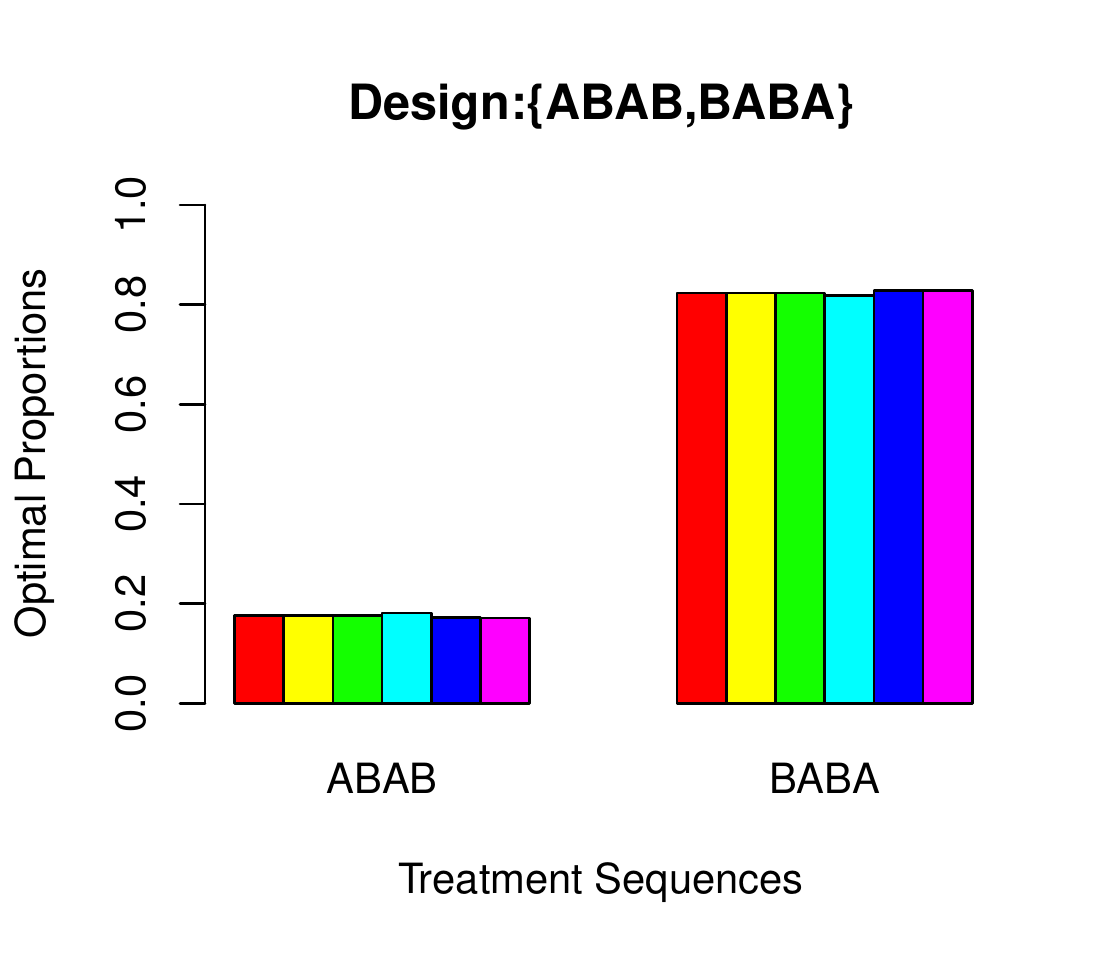}} \\
%			\end{tabular}
%			\caption{Optimal proportions for $p=4$ case under $\theta_1$}
%			\label{Fig:OP4_1}
%		\end{figure}
%	\end{center}
	
%	\begin{center}
%		\begin{figure}[H]
%			\centering
%			\begin{tabular}{ccc}
				\includegraphics[scale=.5]{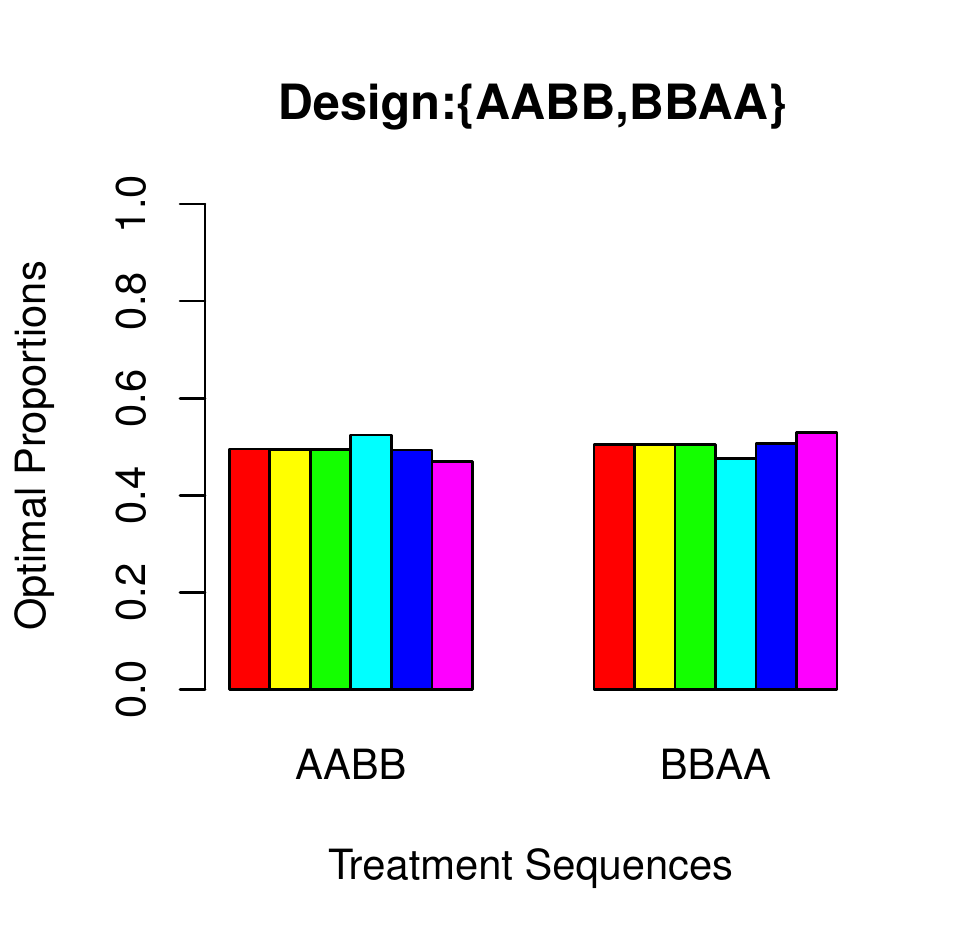} & \includegraphics[scale=.5]{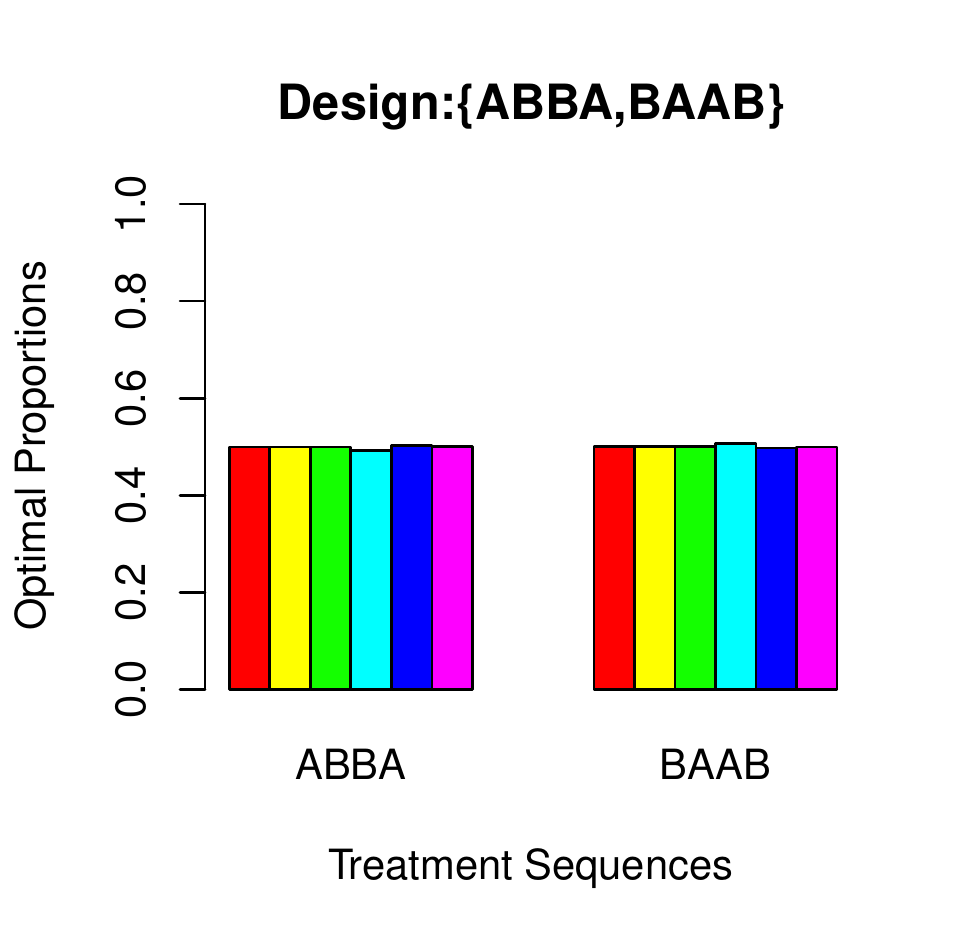} &
				\multicolumn{1}{c}{\includegraphics[scale=.5]{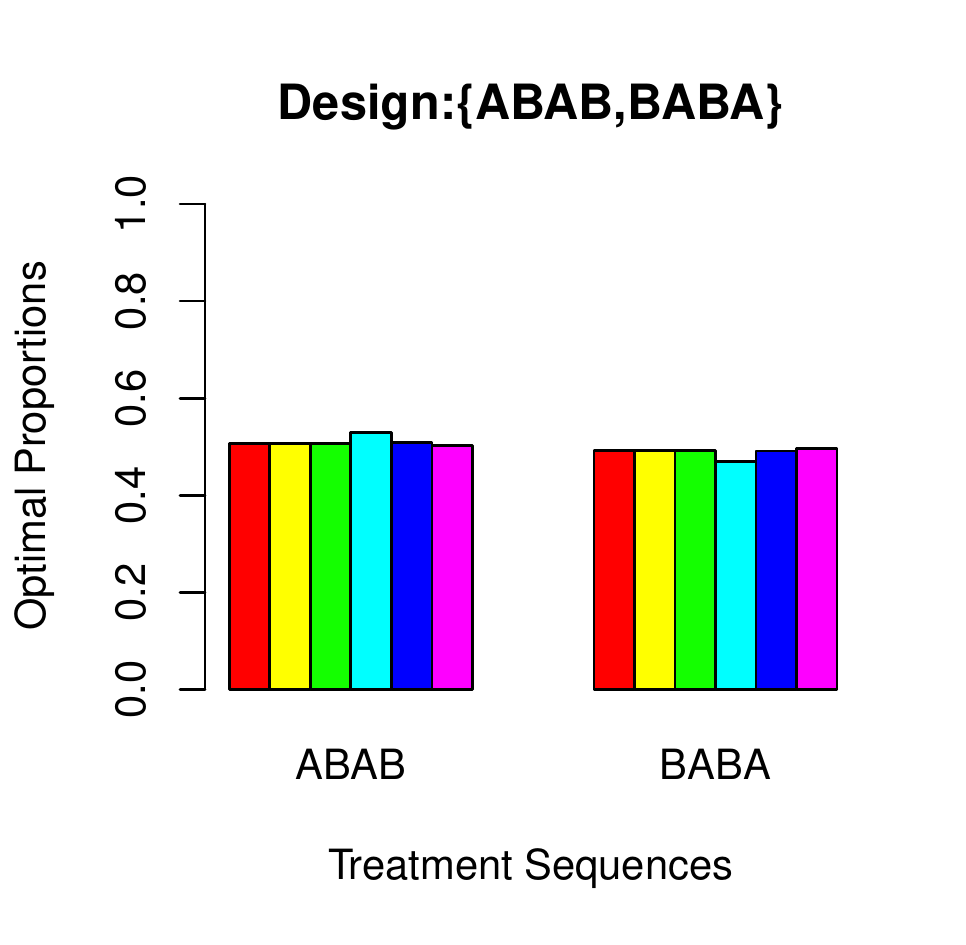}} \\
			\end{tabular}
			\caption{Optimal proportions for $p=4$ case under $\theta_1$ and $\theta_2$ respectively.}
			\label{Fig:OP4_12}
		\end{figure}
	\end{center}
	%\end{landscape} 

	\begin{table}
	\caption{Poisson data from anti-nausea experiment.}
		\label{Tab:PoissonData}
	\begin{center}
		\begin{tabular}{|c|cc|cc| }%|c|p{0.8cm}p{0.8cm}|p{0.8cm}p{0.8cm}|
			\hline 
			&            \multicolumn{4}{p{0.8cm}|}{}   \\[-5pt] 
			Subject &  \multicolumn{2}{c|}{$(A,B)$ Group} &  \multicolumn{2}{c|}{$(B,A)$ Group} \\
			\cline{2-5} 
			&&&&\\[-5pt] 
			&Drug $A$&Drug $B$&Drug $A$&Drug $B$\\
			%&            &            &        &    \\ 
			\hline
			%&	   &       &       &      \\ 
			1  & 1 & 1 & 1 & 0\\
			2  & 0 & 0 & 1 & 0\\
			3  & 1 & 0 & 2 & 0\\
			4  & 1 & 1 & 1 & 0\\
			5  & 1 & 1 & 0 & 1\\
			6  & 0 & 1 & 1 & 1\\
			7  & 2 & 1 & 1 & 0\\
			8  & 1 & 0 & 2 & 0\\
			9  & 0 & 0 & 3 & 1\\
			10 & 1 & 0 & 0 & 0\\
			\hline
			%&	   &       &       &      \\ 
			Total & 8 & 5 & 12 & 3\\
			\hline
		\end{tabular}
	\end{center}
	\end{table}
	
	\medskip \noindent We calculate optimal designs using two values of parameter estimates. $\theta_1$ $=$ $[ 0.2$, $0.34$, $-1.60$, $-1.65]$ represents those parameter estimates that give us non-uniform designs and $\theta_2 = [-0.223$, $-0.875$, $0.405$, $-0.105]$ corresponds to parameter estimates guessed from the data presented in Table~\ref{Tab:PoissonCase} below.
	
	\begin{table}[H]
	\caption{Optimal proportions for anti-nausea experiment.}
	\label{Tab:PoissonCase}
	\medskip
		\begin{center}		
			\begin{tabular}{ |c|lc| }
				\hline
				&&  \\	
				Design Points & Correlation Structure & Optimal Design: $\theta_1$ \\
				&&  \\	
				\hline
				&&  \\
				& Corr(1) $(1-\rho)I_{p} + \rho J_{p}$ with $\rho = 0.1$                                            & $\{0.3632 , 0.6368\}$ \\
				& Corr(2) $\rho ^ { \left| i - i ^ { \prime } \right| } , i \neq i ^ { \prime }$ with $\rho = 0.1$  & $\{0.3632 , 0.6368\}$ \\
				$\{ AB, BA \}$ & Corr(3) with $\rho = 0.1$                                                          & $\{0.3632 , 0.6368\}$ \\
				& Corr(4) with $\rho_{AB} = 0.2 , \rho_{BA} = 0.5$                                                  & $\{0.3632 , 0.6368\}$ \\
				& Corr(5) with $\rho_{AB} = \rho_{BA} = 0.4$                                                        & $\{0.3632 , 0.6368\}$ \\
				& Corr(6) with $\rho_{AB} = 0.4 , \rho_{BA} = 0.3$                                                  & $\{0.3632 , 0.6368\}$ \\
				&&  \\	
				\hline
				&&  \\	
				Design Points & Correlation Structure & Optimal Design: $\theta_2$ \\
				&&  \\
				\hline
				&&  \\		
				& Corr(1) $(1-\rho)I_{p} + \rho J_{p}$ with $\rho = 0.1$                                            & $\{0.5505 , 0.4495\}$ \\
				& Corr(2) $\rho ^ { \left| i - i ^ { \prime } \right| } , i \neq i ^ { \prime }$ with $\rho = 0.1$  & $\{0.5505 , 0.4495\}$ \\
				$\{ AB, BA \}$ & Corr(3) with $\rho = 0.1$                                                          & $\{0.5505 , 0.4495\}$ \\
				& Corr(4) with $\rho_{AB} = 0.2 , \rho_{BA} = 0.5$                                                  & $\{0.5505 , 0.4495\}$ \\
				& Corr(5) with $\rho_{AB} = \rho_{BA} = 0.4$                                                        & $\{0.5505 , 0.4495\}$ \\
				& Corr(6) with $\rho_{AB} = 0.4 , \rho_{BA} = 0.3$                                                  & $\{0.5505 , 0.4495\}$ \\
				&&  \\	
				\hline		
			\end{tabular}
		\end{center}
	\end{table}
		
		\medskip \noindent It can be noted from the above table that when responses are Poisson in nature the optimal proportions do not vary much when correlation structure changes under both $\theta_1$ and $\theta_2$. This suggests us that even when responses are Poisson in nature the proposed design gives almost similar optimal proportions for different choice of correlation matrices. Hence, obtained optimal designs are robust.} 
	
	%%%%%%%%%%%%%%%%%%%%%%%%%%%%%%%%%%%%%%%%%%%%%%%%%%%%%%%%%%%%%%%%%%%%%%%%%%%%%%%%%%%%%
	%                                                                                   %
	%                                                                                   %
	%                                                                                   %
	%                                                                                   %
	%                                                                                   %
	%          SECTION 4: Optimal Design for multiple-treatment crossover trails        %
	%                                                                                   %
	%                                                                                   %
	%                                                                                   %
	%                                                                                   %
	%                                                                                   %
	%                                                                                   %
	%%%%%%%%%%%%%%%%%%%%%%%%%%%%%%%%%%%%%%%%%%%%%%%%%%%%%%%%%%%%%%%%%%%%%%%%%%%%%%%%%%%%%
	
	\section{Optimal Design for Multiple-treatment Crossover Trials }\label{OptMultTrt}
	
	So far we have considered crossover designs with two treatments only. In this section, we extend our study for multiple treatments. This is motivated by a four-period four treatment trial which was first given in Kenward and Jones (1992) and later discussed as Example 6.1 in their book (Kenward and Jones, 2014), \textit{Design and Analysis for Crossover Trials}.
	
	\medskip\noindent
	%%%%%%%%%%%%%%%%%%%%%%%%%%%%%%%%%%%%%%%%%%%%%%%%%%%%%%%%%%%%%%%%%%%%%%%%%%%%%%%%%%%%%                                                          
	%                                                                                   %                        
	%                                                                                   %
	%                                                                                   %
	%                                                                                   %
	%              SUB-SECTION 1: Latin Square Design and Optimal Proportions           %
	%                                                                                   %
	%                                                                                   %
	%                                                                                   %
	%                                                                                   %
	%%%%%%%%%%%%%%%%%%%%%%%%%%%%%%%%%%%%%%%%%%%%%%%%%%%%%%%%%%%%%%%%%%%%%%%%%%%%%%%%%%%%%
	
	\subsection{Latin Square Design and Optimal Proportions }\label{LatSqDes}
	
	In this example, binary responses for four-period crossover trial were obtained. There were four treatments and treatment sequences were allocated at random to eighty different subjects at four different periods. At the end of each period, efficacy measurement of each subject was recorded as success or failure, which resulted in joint outcome at the end of the trial as shown in Table~\ref{Tab5:LSD_data}. The dataset contains four different treatment sequences which were decided before the trial $\Omega = \{ABCD$, $BDAC$, $CADB$, $DCBA\}$ along with the joint outcome of four different periods from the same subject according to a particular treatment sequence. The numbers below each sequence denote how many subjects received that particular treatment sequence, and the particular response was recorded.
	
	\begin{table}[h]\caption{Binary data from a four-period crossover trial.}\label{Tab5:LSD_data}
		\small	
		\begin{center}
			\begin{tabular}{|c|c|c|c|c|}
				\hline
				Joint Outcome   &   \multicolumn{4}{c|}{Frequency of Outcome} \\
				(1=Success, 0=Failure) &$ABCD$   &   $BDAC$   &   $CADB$   &   $DCBA$  \\
				\hline
				
				(0,0,0,0)       &    1     &    0     &    1     &    1     \\
				(0,0,0,1)       &    0     &    1     &    1     &    0     \\
				(0,0,1,0)       &    1     &    1     &    0     &    1     \\
				(0,0,1,1)       &    1     &    0     &    0     &    0     \\
				(0,1,0,0)       &    1     &    1     &    1     &    0     \\
				(0,1,0,1)       &    1     &    1     &    1     &    2     \\
				(0,1,1,0)       &    1     &    1     &    1     &    2     \\
				(0,1,1,1)       &    0     &    1     &    1     &    0     \\
				(1,0,0,0)       &    1     &    0     &    1     &    0     \\
				(1,0,0,1)       &    1     &    1     &    0     &    0     \\
				(1,0,1,0)       &    1     &    0     &    1     &    0     \\
				(1,0,1,1)       &    2     &    0     &    0     &    1     \\
				(1,1,0,0)       &    1     &    1     &    1     &    0     \\
				(1,1,0,1)       &    0     &    2     &    2     &    4     \\
				(1,1,1,0)       &    2     &    3     &    3     &    0     \\
				(1,1,1,1)       &    4     &    9     &    5     &    10    \\
				\hline
			\end{tabular}
		\end{center}
	\end{table}

	\medskip \noindent	We use the correlation matrices defined in Section~\ref{DiffCorrStr} and calculate the optimal proportions. As mentioned earlier for estimating parameters we have considered the baseline constraints as $\beta_{1} = \tau_{A} = \rho_{A} = 0$, so that the design matrix has full column rank and all other parameters are estimable. 
	
	\medskip\noindent \bl{Using these baseline constraints and {\tt glm} function in R we fit the model, which gives us parameter estimates for the given data. Then we use these parameter estimates to make a guess for values of unknown parameters. Our nominal guess for the parameter values is $\theta_2 =$ $[ 0.5$, $0.06$, $-0.53$, $-0.6$, $-0.35$, $0.025$, $-0.23$, $0.73$, $0.23$, $0.30 ]$. Now, we follow the same procedure as mentioned in above pseudo code and calculate the optimal designs for different correlation structures. We also calculate optimal proportions by considering parameter estimates that gives non-uniform designs i.e. $\theta_1 = $ $[-2$, $0.25$, $0$, $0.75$, $1$, $5$, $-1.5$, $-3.5$, $2.75$, $0.75]$. As seen from Table~\ref{Tab6:Opt_Prop}, for the Latin square design the optimal proportions that we obtain using  using $\theta_1$ are non-uniform and that using $\theta_2$ are nearly uniform.}
	
	\begin{table}[H]
		\centering \caption{Optimal proportions for different correlation matrices.}\label{Tab6:Opt_Prop}
		{\scriptsize	
		\begin{center}
			\begin{tabular}{ |p{1.5cm}|p{0.8cm}p{0.8cm}p{0.8cm}p{0.8cm}|p{0.8cm}p{0.8cm}p{0.8cm}p{0.8cm}|}
				\hline 
				&            \multicolumn{8}{c|}{}   \\ 
				Correlation &  \multicolumn{4}{c|}{$\theta_1$} &  \multicolumn{4}{c|}{$\theta_2$}\\
				\cline{2-9} 
				&&&&&&&&\\
				Structure &${ A B C D }$ & ${ B D A C }$ & ${ C A D B }$ & ${ D C B A }$ & ${ A B C D }$ & ${ B D A C }$ & ${ C A D B }$ & ${ D C B A }$ \\
				&&&&&&&&\\                    
				\hline                        
				&&&&&&&&\\                    
				$Corr(1)$  &    { 0.1725 } & { 0.2483 } & { 0.2223 } & { 0.3569 } & { 0.2463 }  & { 0.2493 }  & { 0.2504 }  & { 0.2540 } \\
				$Corr(2)$  &    { 0.1747 } & { 0.2490 } & { 0.2184 } & { 0.3579 } & { 0.2461 }  & { 0.2493 }  & { 0.2501 }  & { 0.2546 } \\  
				$Corr(3)$  & 	{ 0.1714 } & { 0.2480 } & { 0.2236 } & { 0.3570 } & { 0.2461 }  & { 0.2492 }  & { 0.2507 }  & { 0.2540 } \\  
				$Corr(4)$  &	{ 0.1788 } & { 0.2556 } & { 0.2163 } & { 0.3493 } & { 0.2478 }  & { 0.2634 }  & { 0.2334 }  & { 0.2554 } \\  
				$Corr(5)$  &	{ 0.1784 } & { 0.2465 } & { 0.2101 } & { 0.3650 } & { 0.2480 }  & { 0.2517 }  & { 0.2442 }  & { 0.2561 } \\  
				$Corr(6)$  &	{ 0.1752 } & { 0.2531 } & { 0.2170 } & { 0.3547 } & { 0.2470 }  & { 0.2656 }  & { 0.2320 }  & { 0.2554 } \\
				\hline
			\end{tabular}
		\end{center}
	}
	\end{table}
	
	\bl{\medskip \noindent We also calculate optimal design considering all 24 sequences. We consider $Corr(2)$ and calculate optimal proportions for different values of $\rho$. Please refer the Supplementary Materials for details. From the tables in the Supplementary Materials it can be noted that corresponding to $\theta_1$ we have non uniform allocations for the Latin Square design, and almost uniform allocation corresponding to $\theta_2$. In case of non-uniform allocations, although nothing is uniform, the optimal design corresponding to $\theta_1$ has more zeros. Also note that the allocations do not vary a lot as $\rho$ changes, particularly for the sequences where we have zero allocations.}
	
	\medskip\noindent
	%%%%%%%%%%%%%%%%%%%%%%%%%%%%%%%%%%%%%%%%%%%%%%%%%%%%%%%%%%%%%%%%%%%%%%%%%%%%%%%%%%%%%                                                          
	%                                                                                   %                        
	%                                                                                   %
	%                                                                                   %
	%                                                                                   %
	%              SUB-SECTION 2: Sensitivity Study and Relative $D$-efficiency         %
	%                                                                                   %
	%                                                                                   %
	%                                                                                   %
	%                                                                                   %
	%%%%%%%%%%%%%%%%%%%%%%%%%%%%%%%%%%%%%%%%%%%%%%%%%%%%%%%%%%%%%%%%%%%%%%%%%%%%%%%%%%%%%
	
	\subsection{Sensitivity Study and Relative $D$-efficiency}\label{RelD_eff}
	
	\bl{In this section, we study the performance of the proposed locally optimal designs via sensitivity study in terms of relative $D$-efficiencies. Let $\theta_{t}$ be true parameter values and $\theta_{c}$ be assumed parameter values. Then we have corresponding objective function for these two choice of parameter values i.e $det(var(\hat{\tau_{t}}))$ and $det(var(\hat{\tau_{c}}))$ respectively. Hence the relative loss of efficiency of choosing $\theta_{c}$ instead of $\theta_{t}$ can be formulated as
		\begin{eqnarray*}\label{Sen}
			S(\tau_{t},\tau_{c}) = \frac{det(var(\hat{\tau_{t}}))^{(-\frac{1}{k})} - det(var(\hat{\tau_{c}}))^{(-\frac{1}{k})}}{det(var(\hat{\tau_{t}}))^{(-\frac{1}{k})}},
		\end{eqnarray*}
		where $k$ is the dimension of $\tau$. Then the relative $D$-efficiency of the original design $\xi$ compared to the optimal design $\xi^{*}$ can be computed using the formula:
		\begin{eqnarray*}\label{Eff}	
			E_{\xi} = \left[ \frac{det(var(\hat{\tau_{c}}))_{\xi^{*}}}{det(var(\hat{\tau_{t}}))_{\xi}} \right]^{-\frac{1}{k}}.
		\end{eqnarray*}
		
		\medskip \noindent For the Latin square design example we consider following two cases of assumed values $\theta_{c}$ for model parameters. For each case the values of parameters are simulated from a uniform distribution. The range of uniform distribution is obtained by $\pm 1$ and $\pm 2$ from true parameter values $\theta_{t}$ for each case respectively. Here we consider $\theta_{t} =$ $[ 0.5$, $0.06$, $-0.53$, $-0.6$, $-0.35$, $0.025$, $-0.23$, $0.73$, $0.23$, $0.30 ]$. }
	
	\begin{table}[h]
		\caption{Assumed values for model parameters.}
		\centering
		\begin{tabular}{|c|l|l|}
			\hline 
			&&\\
			Parameters $\theta_{c}$ & Case 1 & Case 2 \\
			&&\\ 
			\hline
			&&\\
			$\lambda$ & U($-0.5,1.5$) & U($-1.5,2.5$) \\
			$\beta_{2}$ & U($-0.04,0.16$) & U($-0.14,0.26$) \\
			$\beta_{3}$ & U($-1.53,0.47$) & U($-2.53,1.47$) \\
			$\beta_{4}$ & U($-1.6,0.4$) & U($-2.6,1.4$) \\
			$\tau_{2}$  & U($-1.35,0.65$) & U($-2.35,1.65$) \\
			$\tau_{3}$  & U($-0.075,0.125$) & U($-0.175,0.225$) \\
			$\tau_{4}$  & U($-1.23,0.77$) & U($-2.23,1.77$) \\
			$\rho_{2}$  & U($-0.27,1.73$) & U($-1.27,2.73$) \\
			$\rho_{3}$  & U($-0.77,1.23$) & U($-1.77,2.23$) \\
			$\rho_{4}$  & U($-0.70,1.30$) & U($-1.70,2.30$) \\
			&&\\
			\hline			
		\end{tabular}
	\end{table}
	%\begin{landscape}
	\begin{center}
	\begin{figure}[h]
		\begin{tabular}{cc}
			\includegraphics[scale=.25, trim={1.5cm .75cm .5cm .05cm}]{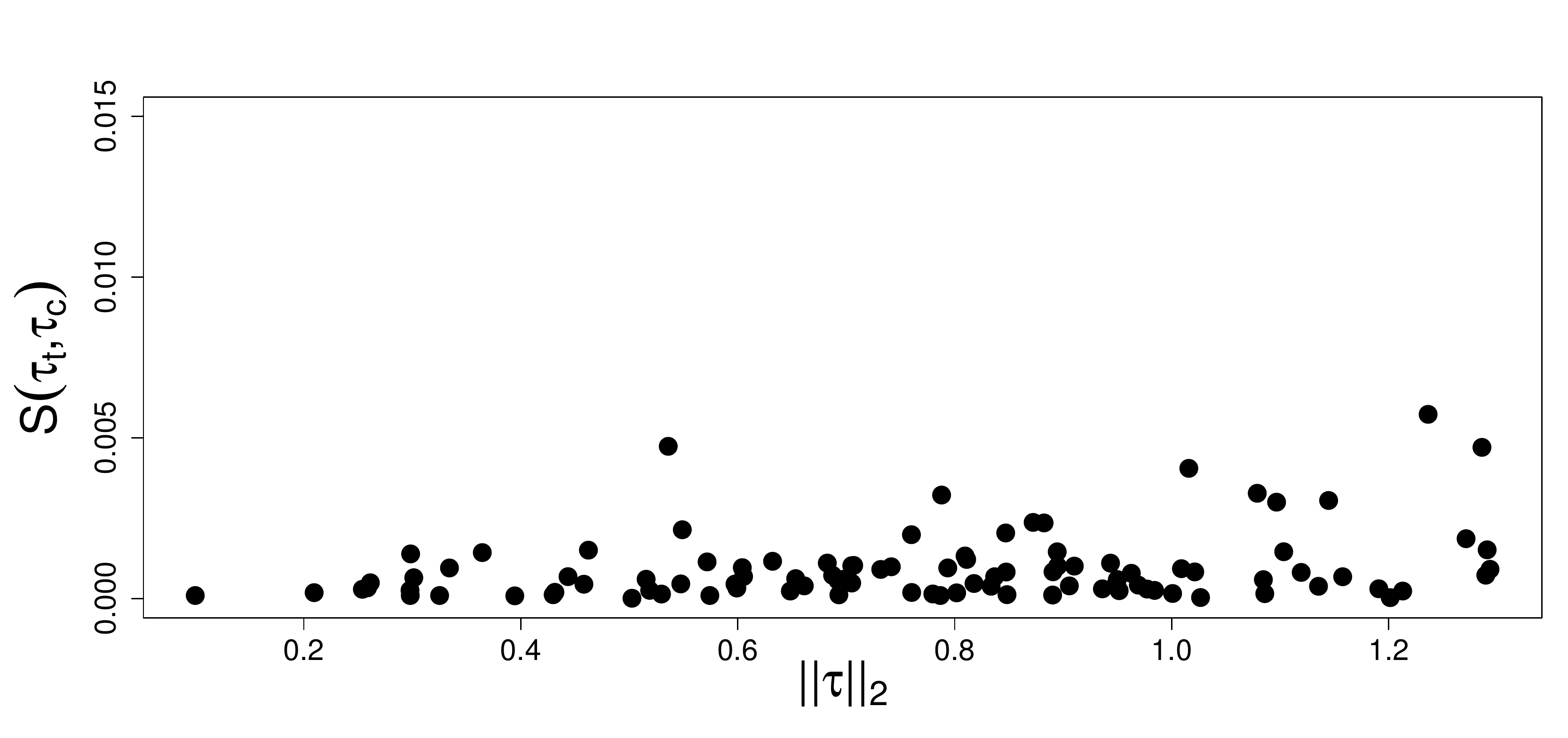} & \includegraphics[scale = .25, trim={0cm .75cm .5cm .05cm}]{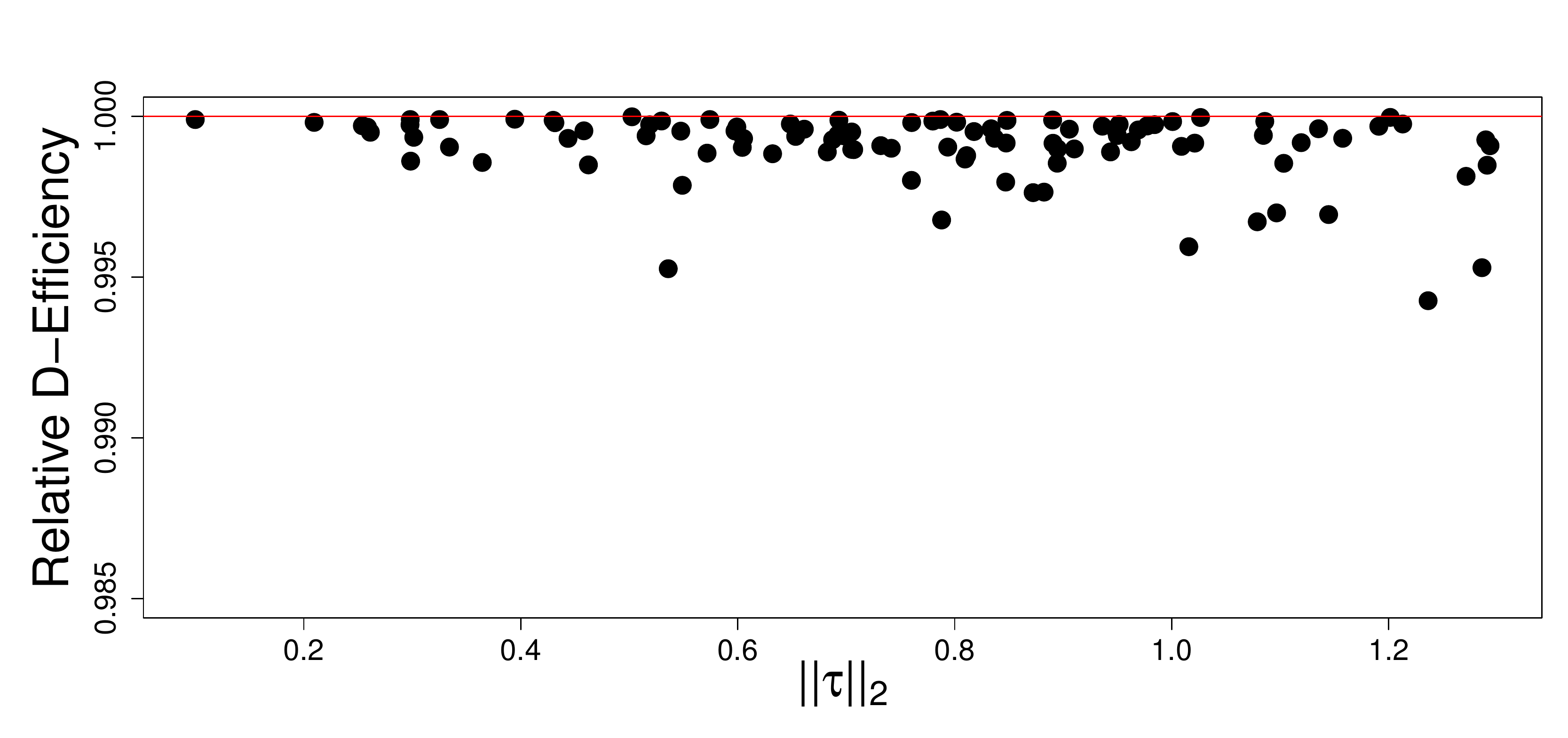}\\
			\text{Case 1: Relative loss of efficiency} & \text{Case 1: Relative $D$-efficicency}\\
			\includegraphics[scale=.25, trim={1.5cm .75cm .5cm .05cm}]{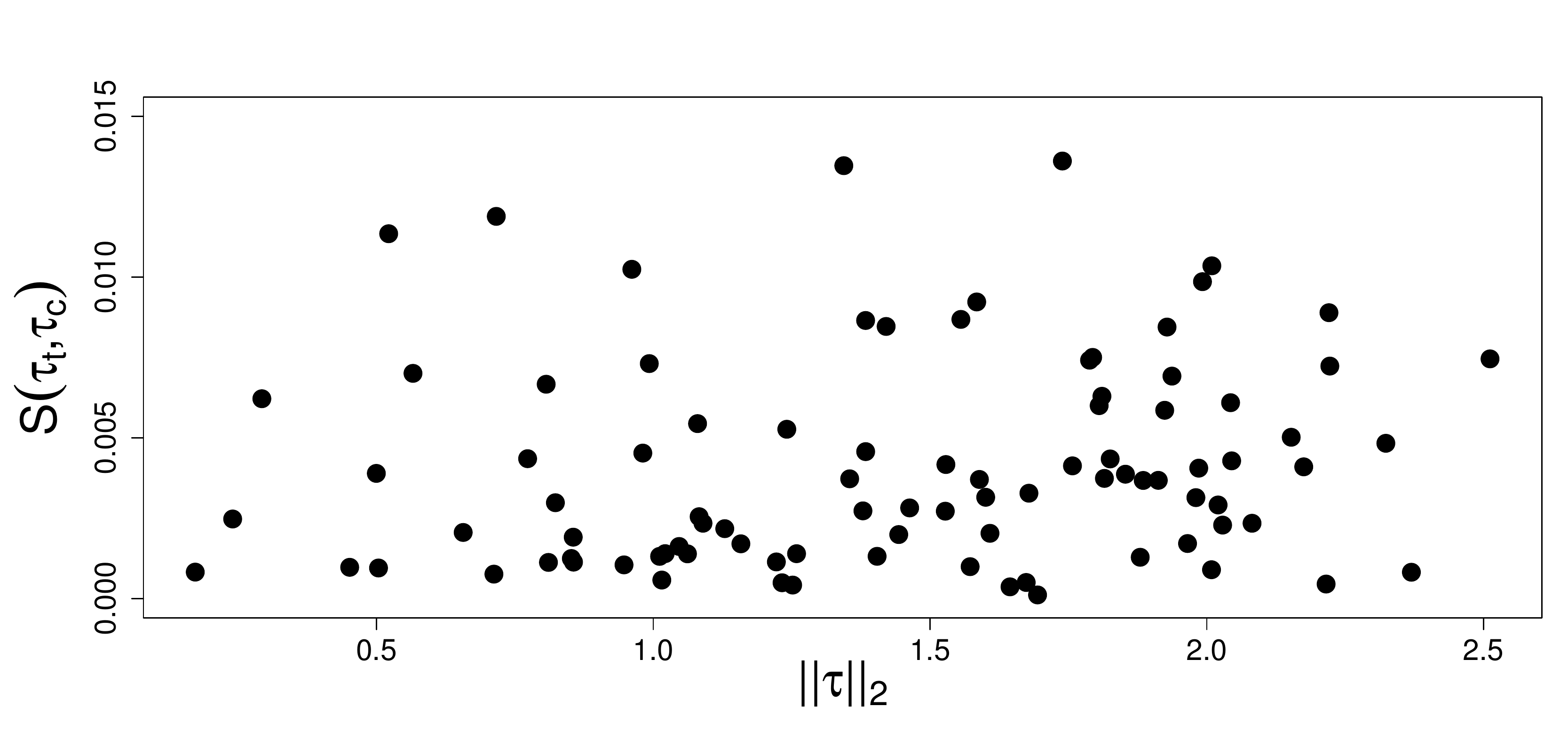} & \includegraphics[scale = .25, trim={0cm .75cm .5cm .05cm}]{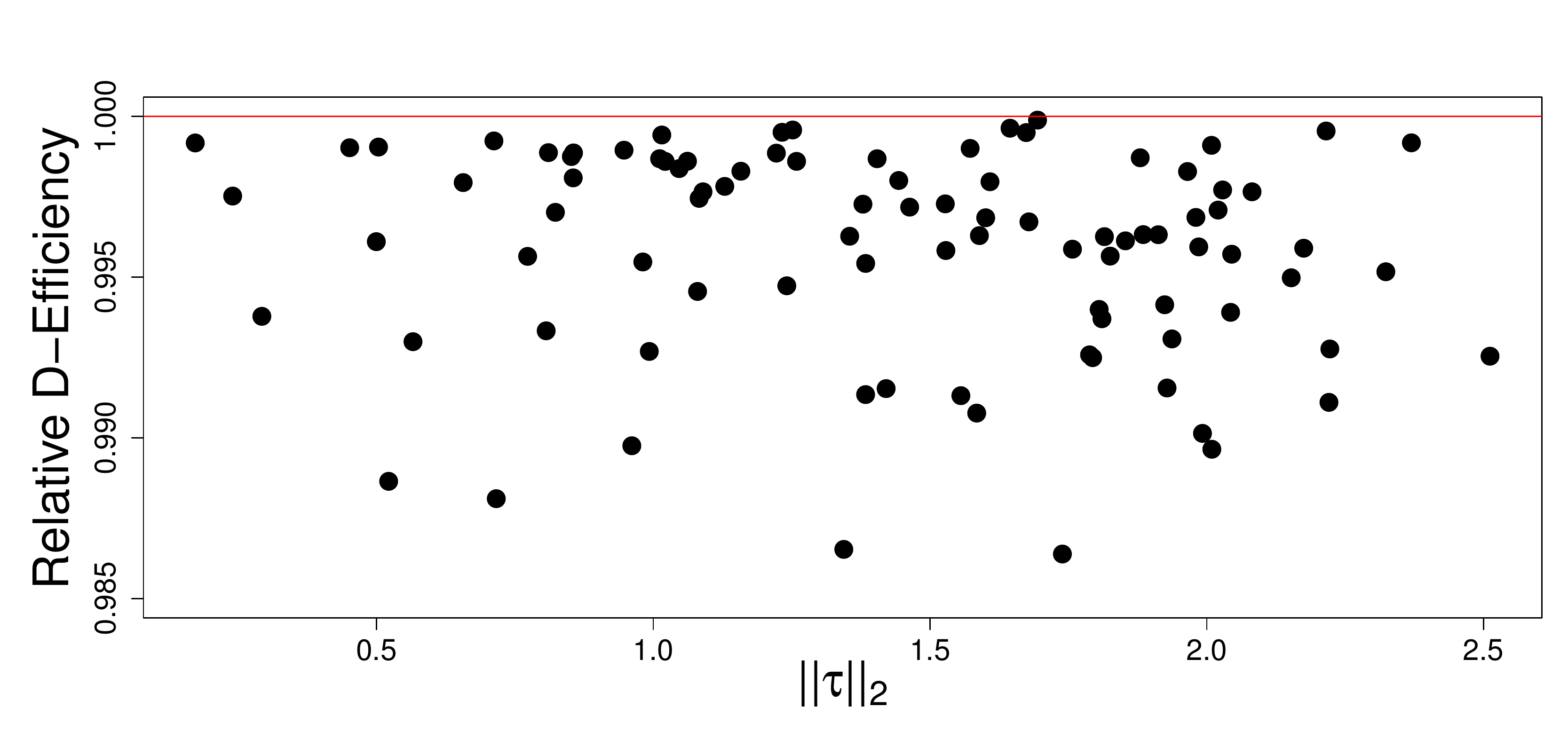}\\
			\text{Case 2: Relative loss of efficiency} & \text{Case 2: Relative $D$-efficicency}\\
		\end{tabular}
		\caption{Performance of the locally optimal designs.}
		\label{Eff_Sen}
	\end{figure}
	\end{center}
%\end{landscape}

	\clearpage
	%%%%%%%%%%%%%%%%%%%%%%%%%%%%%%%%%%%%%%%%%%%%%%%%%%%%%%%%%%%%%%%%%%%%%%%%%%%%%%%%%%%%%                             
	%                                                                                   %
	%                                                                                   %
	%                                                                                   %
	%                                                                                   %
	%                       SUB-SECTION 3: Simulation Studies                           %
	%                                                                                   %
	%                                                                                   %
	%                                                                                   %
	%                                                                                   %
	%%%%%%%%%%%%%%%%%%%%%%%%%%%%%%%%%%%%%%%%%%%%%%%%%%%%%%%%%%%%%%%%%%%%%%%%%%%%%%%%%%%%%

	\subsection{Simulation Studies with Two-Stage Designs}\label{SimStd}
	
	As stated earlier the main aim of this paper is to determine optimal and efficient crossover designs for experiments where the generalized linear model adequately describes the process under study. Crossover trials are repeated measurement designs, where these repeated measurements on the same subject have great advantages, but there are also many potential disadvantages associated with it. Nevertheless, the impact of these disadvantages can be minimized or reduced if we choose a proper design and analysis method.

	\noindent One of the major disadvantages of repeated measurement designs is that the effect of the treatment depends on the subject itself. Stronger subject effects cause more variation on estimated treatment effects.

	\medskip\noindent The simulation studies are motivated by the real-life example of Latin square design mentioned above. \bl{Since all the correlation structures mentioned in Section~\ref{DiffCorrStr} perform similarly in Table~\ref{Tab6:Opt_Prop}, we choose $Corr(2)$ for illustration purpose. Note that in $Corr(2)$, we have AR(1) structure, where the correlation between two responses decreases as the number of periods between responses increases, which makes good practical sense. For these simulation studies we are considering 400 observations and two different types of initial guess for $\theta$ values. In Case 1 we will use $\theta_2 =$ $[ 0.5$, $0.06$, $-0.53$, $-0.6$, $-0.35$, $0.025$, $-0.23$, $0.73$, $0.23$, $0.30 ]$ which is obtained from real data. This choice of $\theta_2$ gives optimal allocations as $ ( 0.2460, 0.2495, 0.2500, 0.2545 ) $, which is approximately uniform. For Case 2 we will use $\theta_1 = $ $[-2$, $0.25$, $0$, $0.75$, $1$, $5$, $-1.5$, $-3.5$, $2.75$, $0.75]$ and this guess of $\theta_1$ is such that optimal allocations are non-uniform. For example for $\rho = 0.1$ the optimal allocations are $ ( 0.172, 0.248, 0.222, 0.358 ) $. Optimal allocations are similar for other values of $\rho$. 
				
	\noindent The simulation process used here has two stages. First for a given parameter $\theta$, we define a design matrix corresponding to each treatment sequence along with correlation matrix.
	
		\begin{itemize}
			\item First Stage:
			\begin{enumerate}
				\item In this stage, we use {\tt rbin} function in R to simulate 30\% of observations uniformly over all four treatment sequence. These observations serve as our pilot study. Note that we use uniform design for pilot study.
				\item From these observations obtained in above step we estimate the correlation coefficient and regression parameters, which are used as the assumed parameter values for the second stage.
			\end{enumerate}
		\end{itemize}
		
		\begin{itemize}
			\item Second Stage:
			\begin{enumerate}
				\item Based on the assumed parameter values obtained in the first stage and the algorithm described in Section 2.4, we calculate the optimal allocation for the remaining 70\% of the subjects.  
				\item Using these optimal allocation we simulate observations for remaining 70\% of subjects according to the assumed parameter values.
				\item In case of uniform design, we simulate total number of observations uniformly over all treatment sequence i.e., one-fourth of the total observations correspond to each of the four treatment sequence.
			\end{enumerate}
		\end{itemize}

	 \noindent During this process we calculate the parameter estimates based on the simulated observations and calculate the corresponding Mean Square Error (MSE) from the true parameter values for each simulation. Above simulation procedure is repeated 100 times. Finally we take the average of those individual MSEs to calculate the overall MSE reported in Table~\ref{Tab5:SR}. We repeat the above simulation process for different correlation coefficients and for two different sets of initial $\theta$'s, $\theta_1$ and $\theta_2$. It is clear from Table~\ref{Tab5:SR} and Figure 10 that if the optimal allocations are non-uniform, then the proposed optimal design has a significant advantage over the traditional uniform designs, for all values of the correlation coefficients. It should be noted that those high values of MSEs for uniform designs are mostly due to a handful of ``bad'' datasets. In our experience, the proposed optimal designs never give rise to such data.}
	
	\begin{table}[H]
			\caption{Simulation results.}
			\medskip
			\label{Tab5:SR}
			\footnotesize	
			\centering
			\begin{tabular}{ |p{1.2cm}|p{1.2cm}p{1.2cm}|p{1.2cm}p{1.2cm}|}
				\hline
				&            \multicolumn{4}{c|}{}   \\ 
				Corr &  \multicolumn{4}{c|}{Mean Squared Errors} \\
				Corr(2) &  \multicolumn{2}{c|}{Case 1} &  \multicolumn{2}{c|}{Case 2} \\
				\cline{2-5} 
				&&&&\\
				&Uniform&Optimal &Uniform &Optimal\\
				\multicolumn{1}{|c|}{$\rho$} &Design& Design&Design&Design \\
				&            &            &        &    \\ 
				\cline{2-5} 
				%			0.0  & 0.120 & 0.122 & 0.983   &&&\\
				&	   &       &       &      \\ 
				0.1 \includegraphics[width=.2in,height = 0.02\textheight]{red.pdf}     & 0.109 & 0.108 & 2.834 & 0.393\\
				0.2 \includegraphics[width=.2in,height = 0.02\textheight]{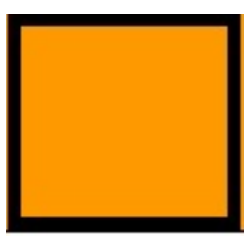}       & 0.103 & 0.100 & 2.718 & 0.659\\
				0.3 \includegraphics[width=.2in,height = 0.02\textheight]{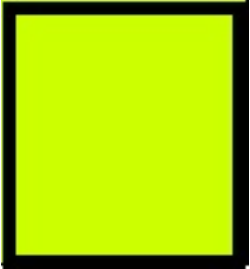}       & 0.101 & 0.140 & 4.925 & 0.490\\
				0.4 \includegraphics[width=.2in,height = 0.02\textheight]{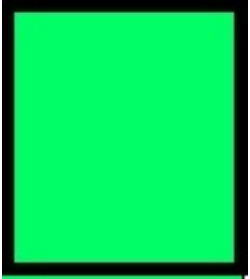}       & 0.094 & 0.127 & 4.896 & 0.484\\
				0.5 \includegraphics[width=.2in,height = 0.02\textheight]{6.pdf}       & 0.100 & 0.123 & 2.596 & 0.428\\
				0.6 \includegraphics[width=.2in,height = 0.02\textheight]{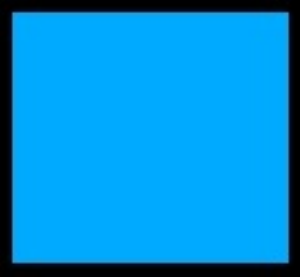} & 0.088 & 0.109 & 2.632 & 0.469\\
				0.7 \includegraphics[width=.2in,height = 0.02\textheight]{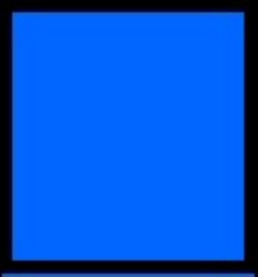}       & 0.086 & 0.095 & 5.110 & 0.458\\
				0.8 \includegraphics[width=.2in,height = 0.02\textheight]{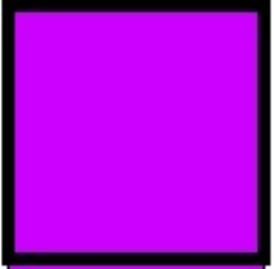}       & 0.066 & 0.077 & 2.705 & 0.586\\
				0.9 \includegraphics[width=.2in,height = 0.02\textheight]{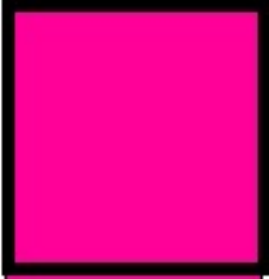}      & 0.050 & 0.051 & 2.761 & 0.559\\
				\hline
			\end{tabular}
		\end{table}

		\begin{center}
			\begin{figure}[h]
				\centering 
				\includegraphics[scale=.8,height = 0.4\textheight]{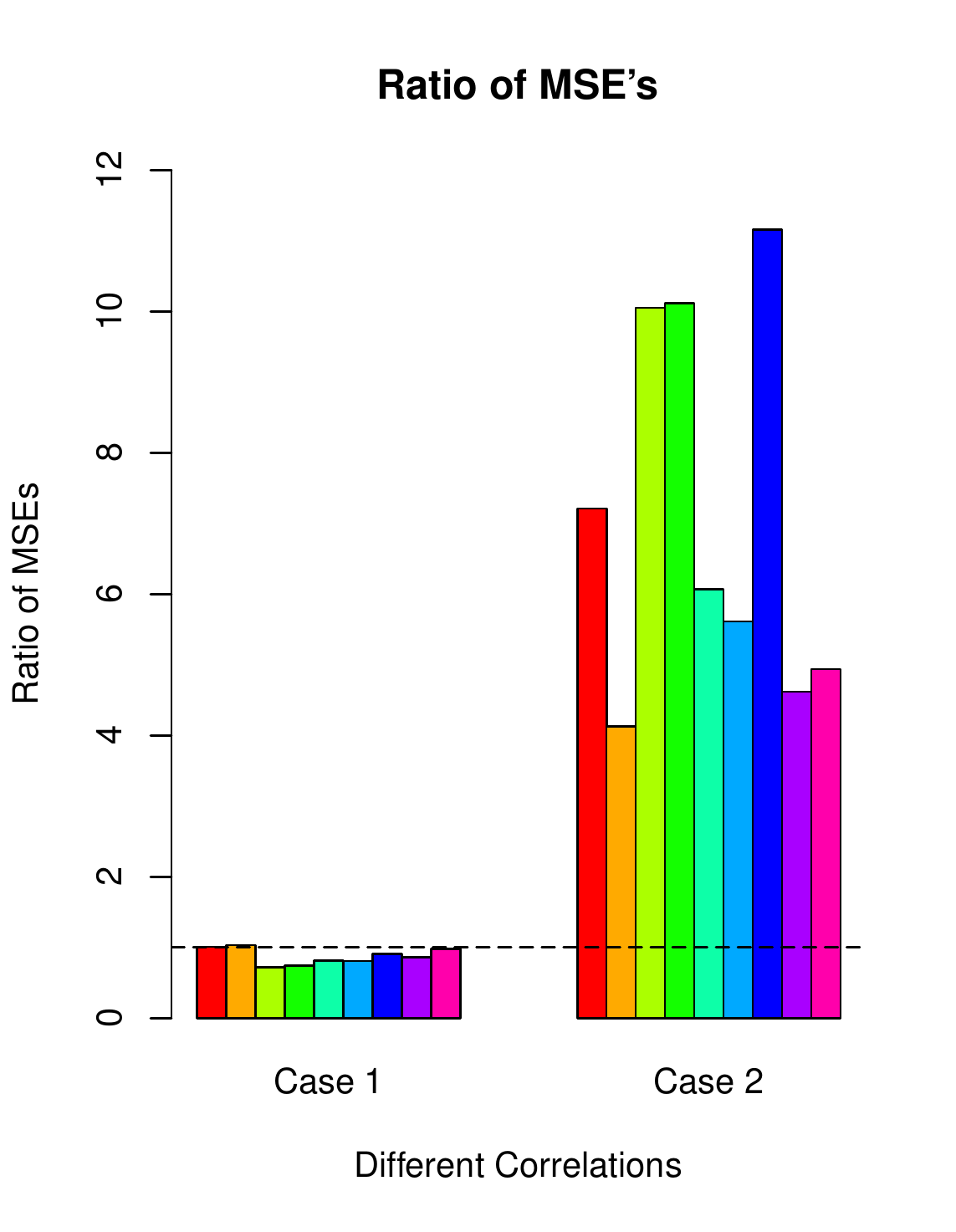}\\
				\caption{Simulation Results: Ratios of MSEs of the Uniform versus Optimal designs, for different values of $\rho$, for each of the two cases.}
				\label{Sim}
			\end{figure}
		\end{center}
	
	%%%%%%%%%%%%%%%%%%%%%%%%%%%%%%%%%%%%%%%%%%%%%%%%%%%%%%%%%%%%%%%%%%%%%%%%%%%%%%%%%%%%%
	%                                                                                   %
	%                                                                                   %
	%                                                                                   %
	%                                                                                   %
	%                                                                                   %
	%                     SECTION 5: Discussion and Future Work                         %
	%                                                                                   %
	%                                                                                   %
	%                                                                                   %
	%                                                                                   %
	%                                                                                   %
	%                                                                                   %
	%%%%%%%%%%%%%%%%%%%%%%%%%%%%%%%%%%%%%%%%%%%%%%%%%%%%%%%%%%%%%%%%%%%%%%%%%%%%%%%%%%%%%
	
	\section{Discussion}\label{DisFW}
	
	In practice, it is customary to use uniform designs where the same number of subjects are assigned to each treatment sequence. In the case of linear models, such uniform designs are optimal. However, optimal proportions obtained under generalized linear models are not uniform. We identified locally optimal designs under different correlation structures. Tables~\ref{Tab1:OP2Case} to \ref{Tab4:OP4Case} and graphs in Figures~\ref{Fig:OP2_1} to \ref{Fig:OP4_12} suggest that the optimal proportions do not vary much from one correlation structure to another. These results suggest that the identified designs are robust. Simulation studies and results in Table~\ref{Tab5:SR} and Figure~\ref{Sim} suggest that these designs are more efficient than uniform designs as well.
	
	\section*{Acknowledgement}
	The  authors  would  like  to  thank Dr. Pritam Ranjan for many helpful suggestions.
	
	%%%%%%%%%%%%%%%%%%%%%%%%%%%%%%%%%%%%%%%%%%%%%%%%%%%%%%%%%%%%%%%%%%%%%%%%%%%%%%%%%%%%%                                                    
	%                                                                                   %
	%                                   APPENDIX                                        %
	%                                                                                   %
	%%%%%%%%%%%%%%%%%%%%%%%%%%%%%%%%%%%%%%%%%%%%%%%%%%%%%%%%%%%%%%%%%%%%%%%%%%%%%%%%%%%%%

	\renewcommand{\thefigure}{A\arabic{figure}}
	\renewcommand{\thesection}{A\arabic{section}}
	\renewcommand{\theequation}{A\arabic{equation}}
	\setcounter{section}{0}	
	\section*{Appendix}\label{Appendix}
	
	%\setcounter{subsection}{0}	
	
	%%%%%%%%%%%%%%%%%%%%%%%%%%%%%%%%%%%%%%%%%%%%%%%%%%%%%%%%%%%%%%%%%%%%%%%%%%%%%%%%%%%%%                                                          
	%                                                                                   %                        
	%                                                                                   %
	%                                                                                   %
	%                                                                                   %
	%     SUB-SECTION 1: Effect of Misspecification of Working Correlation Structures   %
	%                                                                                   %
	%                                                                                   %
	%                                                                                   %
	%                                                                                   %
	%%%%%%%%%%%%%%%%%%%%%%%%%%%%%%%%%%%%%%%%%%%%%%%%%%%%%%%%%%%%%%%%%%%%%%%%%%%%%%%%%%%%%
	
	\subsection*{Effect of Misspecification of Working Correlation Structures}\label{Misspecification}
	
	\bl{The table below lists the locally optimal design when true correlation structure varies from working correlation structure. In the table first column represents true correlation structure and the corresponding optimal designs are calculated under $\theta_1$ and $\theta_2$ for each misspecified working correlation structure in second column. Also, relative $D$-efficiency is calculated under $\theta_1$ and $\theta_2$ for each design. It is clear from the relative $D$-efficiency values in the table that the effect of variance misspecification on the locally optimal design is minimal.} 
	
	\newpage	
	\begin{landscape}
		\begin{table}[h]
			\begin{center}
				\text{Table A1: Optimal design after variance misspecification.}
			\end{center}
			\medskip
			\begin{center}	
				\scriptsize	
				\begin{tabular}{ |p{1.5cm}|p{1.5cm}|cccc|cccc|c|c|}
					\hline 
					&&&&&&&&&& \multicolumn{2}{c|}{}\\
					True \hspace{.2in} Correlation & Working Correlation &  \multicolumn{4}{c|}{Optimal proportions for $\theta_1$} &  \multicolumn{4}{c|}{Optimal proportions for $\theta_2$} & \multicolumn{2}{c|}{Relative $D$-efficiency}\\
					\cline{3-10} 
					&&&&&&&&&&&\\
					Structure & Structure & ${ A B C D }$ & ${ B D A C }$ & ${ C A D B }$ & ${ D C B A }$ & ${ A B C D }$ & ${ B D A C }$ & ${ C A D B }$ & ${ D C B A }$ & under {$\theta_1$} & under {$\theta_2$} \\
					&&&&&&&&&&&\\
					\hline
					&&&&&&&&&&&\\
					& $Corr(2)$ & { 0.1723 } & { 0.2483 } & { 0.2222 } & { 0.3572 } & { 0.2463 } & { 0.2493 } & { 0.2504 } & { 0.2540 } & {0.9999} & {0.9999} \\
					& $Corr(3)$ & { 0.1726 } & { 0.2483 } & { 0.2223 } & { 0.3568 } & { 0.2463 } & { 0.2493 } & { 0.2504 } & { 0.2540 } & {0.9999} & {0.9999} \\ 
					$Corr(1)$  & $Corr(4)$   & { 0.1723 } & { 0.2513 } & { 0.2202 } & { 0.3562 } & { 0.2500 } & { 0.2500 } & { 0.2500 } & { 0.2500 } & {0.9997} & {0.9988} \\ 
					& $Corr(5)$ & { 0.2447 } & { 0.1713 } & { 0.2495 } & { 0.2223 } & { 0.3569 } & { 0.2475 } & { 0.2557 } & { 0.2521 } & {0.9994} & {0.9995} \\
					& $Corr(6)$ & { 0.2500 } & { 0.1724 } & { 0.2508 } & { 0.2197 } & { 0.3571 } & { 0.2500 } & { 0.2500 } & { 0.2500 } & {0.9999} & {0.9984} \\
					\hline
					&&&&&&&&&&&\\[-6pt]
					& $Corr(1)$ & { 0.1745 } & { 0.2489 } & { 0.2183 } & { 0.3583 } & { 0.2462 } & { 0.2493 } & { 0.2500 } & { 0.2545 } & {0.9999} & {0.9999} \\  
					& $Corr(3)$ & { 0.1744 } & { 0.2489 } & { 0.2182 } & { 0.3585 } & { 0.2462 } & { 0.2493 } & { 0.2500 } & { 0.2545 } & {0.9999} & {0.9999} \\   
					$Corr(2)$  & $Corr(4)$ & { 0.1745 } & { 0.2514 } & { 0.2177 } & { 0.3564 } & { 0.2500 } & { 0.2500 } & { 0.2500 } & { 0.2500 } & {0.9998} & {0.9987} \\  
					& $Corr(5)$ & { 0.1740 } & { 0.2503 } & { 0.2180 } & { 0.3577 } & { 0.2450 } & { 0.2480 } & { 0.2530 } & { 0.2540 } & {0.9997} & {0.9997} \\ 
					& $Corr(6)$ & { 0.1744 } & { 0.2512 } & { 0.2174 } & { 0.3570 } & { 0.2463 } & { 0.2497 } & { 0.2505 } & { 0.2535 } & {0.9999} & {0.9985} \\  
					\hline
					&&&&&&&&&&&\\[-6pt]
					& $Corr(1)$ & { 0.1714 } & { 0.2480 } & { 0.2236 } & { 0.3570 } & { 0.2461 } & { 0.2492 } & { 0.2507 } & { 0.2540 } & {0.9999} & {0.9999} \\ 
					& $Corr(2)$ & { 0.1711 } & { 0.2480 } & { 0.2235 } & { 0.3574 } & { 0.2462 } & { 0.2492 } & { 0.2506 } & { 0.2540 }  & {0.9999} & {0.9999} \\  
					$Corr(3)$  & $Corr(4)$ & { 0.1713 } & { 0.2516 } & { 0.2209 } & { 0.3562 } & { 0.2500 } & { 0.2500 } & { 0.2500 } & { 0.2500 }  & {0.9996} & {0.9987} \\  
					& $Corr(5)$ & { 0.1700 } & { 0.2463 } & { 0.2235 } & { 0.3572 } & { 0.2441 } & { 0.2476 } & { 0.2561 } & { 0.2522 } & {0.9992} & {0.9995} \\
					& $Corr(6)$ & { 0.1713 } & { 0.2510 } & { 0.2204 } & { 0.3573 } & { 0.2500 } & { 0.2500 } & { 0.2500 } & { 0.2500 } & {0.9999} & {0.9984} \\ 
					\hline
					&&&&&&&&&&&\\[-6pt]
					& $Corr(1)$ & { 0.1783 } & { 0.2585 } & { 0.2140 } & { 0.3492 } & { 0.2500 } & { 0.2637 } & { 0.2347 } & { 0.2516 } & {0.9994} & {0.9987} \\ 
					& $Corr(2)$ & { 0.1784 } & { 0.2580 } & { 0.2156 } & { 0.3480 } & { 0.2486 } & { 0.2640 } & { 0.2344 } & { 0.2530 } & {0.9996} & {0.9987} \\ 
					$Corr(4)$  & $Corr(3)$ & { 0.1782 } & { 0.2592 } & { 0.2131 } & { 0.3495 } & { 0.2498 } & { 0.2643 } & { 0.2342 } & { 0.2517 }  & {0.9992} & {0.9986} \\ 
					& $Corr(5)$ & { 0.1778 } & { 0.2579 } & { 0.2167 } & { 0.3476 } & { 0.2470 } & { 0.2650 } & { 0.2343 } & { 0.2537 } & {0.9992} & {0.9993} \\ 
					& $Corr(6)$ & { 0.1790 } & { 0.2555 } & { 0.2165 } & { 0.3490 } & { 0.2485 } & { 0.2631 } & { 0.2337 } & { 0.2547 } & {0.9999} & {0.9999} \\    
					\hline
					&&&&&&&&&&&\\[-6pt]
					& $Corr(1)$ & { 0.1774 } & { 0.2477 } & { 0.2092 } & { 0.3657 } & { 0.2466 } & { 0.2501 } & { 0.2486 } & { 0.2547 } & {0.9994} & {0.9999} \\  
					& $Corr(2)$ & { 0.1776 } & { 0.2476 } & { 0.2099 } & { 0.3649 } & { 0.2470 } & { 0.2506 } & { 0.2470 } & { 0.2554 } & {0.9997} & {0.9999} \\  
					$Corr(5)$  & $Corr(3)$ & { 0.1770 } & { 0.2477 } & { 0.2087 } & { 0.3666 } & { 0.2462 } & { 0.2503 } & { 0.2485 } & { 0.2550 } & {0.9992} & {0.9999} \\  
					& $Corr(4)$ & { 0.1776 } & { 0.2492 } & { 0.2108 } & { 0.3624 } & { 0.2472 } & { 0.2538 } & { 0.2450 } & { 0.2540 } & {0.9996} & {0.9994} \\  
					& $Corr(6)$ & { 0.1774 } & { 0.2496 } & { 0.2110 } & { 0.3620 } & { 0.2465 } & { 0.2535 } & { 0.2456 } & { 0.2544 } & {0.9998} & {0.9991} \\  
					\hline
					&&&&&&&&&&&\\[-6pt]
					& $Corr(1)$ & { 0.1748 } & { 0.2553 } & { 0.2142 } & { 0.3557 } & { 0.2482 } & { 0.2652 } & { 0.2332 } & { 0.2534 } & {0.9997} & {0.9985} \\  
					& $Corr(2)$ & { 0.1748 } & { 0.2551 } & { 0.2160 } & { 0.3541 } & { 0.2470 } & { 0.2657 } & { 0.2329 } & { 0.2544 } & {0.9999} & {0.9985} \\  
					$Corr(6)$  & $Corr(3)$ & { 0.1748 } & { 0.2558 } & { 0.2133 } & { 0.3561 } & { 0.2482 } & { 0.2660 } & { 0.2325 } & { 0.2533 } & {0.9996} & {0.9984} \\ 
					& $Corr(4)$ & { 0.1754 } & { 0.2530 } & { 0.2172 } & { 0.3544 } & { 0.2476 } & { 0.2652 } & { 0.2324 } & { 0.2548 } & {0.9999} & {0.9999} \\  
					& $Corr(5)$ & { 0.1741 } & { 0.2556 } & { 0.2180 } & { 0.3523 } & { 0.2452 } & { 0.2669 } & { 0.2339 } & { 0.2540 } & {0.9994} & {0.9991} \\  
					\hline
				\end{tabular}
			\end{center}
		\end{table}
	\end{landscape}

	%%%%%%%%%%%%%%%%%%%%%%%%%%%%%%%%%%%%%%%%%%%%%%%%%%%%%%%%%%%%%%%%%%%%%%%%%%%%%%%%%%%%%                                                    
	%                                                                                   %
	%                         SUPPLEMENTARY MATERIAL REF                                %
	%                                                                                   %
	%%%%%%%%%%%%%%%%%%%%%%%%%%%%%%%%%%%%%%%%%%%%%%%%%%%%%%%%%%%%%%%%%%%%%%%%%%%%%%%%%%%%%
	
	\renewcommand{\thefigure}{S\arabic{figure}}
	\renewcommand{\thesection}{S\arabic{section}}
	\renewcommand{\theequation}{S\arabic{equation}}
	\setcounter{section}{0}		
	\section{Supplementary Material}\label{SuppMatRef}	
	
	\setcounter{subsection}{0}
	
	\subsection{Optimal Design for Latin Square Example with 24 Sequences}
	
	In this section we present the optimal designs corresponding to 24 sequences under $\theta_1$ and $\theta_2$.
	
	\subsection{Explicit Expression of the Objective Function}
	
	In this section we workout the explicit expression for objective function and mention all the required steps.
	
	%%%%%%%%%%%%%%%%%%%%%%%%%%%%%%%%%%%%%%%%%%%%%%%%%%%%%%%%%%%%%%%%%%%%%%%%%%%%%%%%%%%%%                                                    
	%                                                                                   %
	%                                   REFRENCES                                       %
	%                                                                                   %
	%%%%%%%%%%%%%%%%%%%%%%%%%%%%%%%%%%%%%%%%%%%%%%%%%%%%%%%%%%%%%%%%%%%%%%%%%%%%%%%%%%%%%
	
	\section*{References}\label{Ref}
	\begin{enumerate}
		
		\item Anthony ,C. A and David, C. W. (2015). Designs for Generalized Linear Models. \emph{Handbook of Design and Analysis of Experiments} {\bf Chapter 13}. London: Chapman and Hall.
		
		\item Bose, M. and Dey, A. (2009). {\em Optimal Crossover Designs}. World Scientific. 
		
		\item Bose, M. and Dey, A. (2015). Crossover designs. In Handbook of Design and Analysis of Experiments (A. M. Dean, M. Morris, J. Stufken, D. Bingham, Eds.).
		
		\item Carriere, K. C. and R. Huang (2000). Crossover designs for two-treatment clinical trials. \emph{J. Statist. Plann. Inference} {\bf 87}, 125$-$134.
		
		\item Cheng, C. S. and C. F. Wu (1980). Balanced repeated measurements designs. \emph{Ann. Statist.} {\bf 8}, 1272$-$1283.
		
		\item Chernoff, H. (1953). Locally optimal designs for estimating parameters. \emph{Ann. Math. Statist.} {\bf 24} 586$-$602.
		
		\item Dey, A., V. K. Gupta and M. Singh (1983). Optimal change-over designs. \emph{Sankhya B45}, 233$-$239.
		
		\item Hedayat, A. and K. Afsarinejad (1975). Repeated measurements designs, I. In {\em A Survey of Statistical Designs and Linear Models (J. N. Srivastava,} Ed.). Amsterdam: North-Holland, pp. 229$-$242.
		
		\item Kenward, M. G. and Jones, B. (1992). Alternative approaches to the analysis of binary and categorical repeated measurements. \emph{Journal of Biopharmaceutical Statistics, 2:} 137$-$170. 
		
		\item Kenward, M. G. and Jones, B. (2014). {\em Design and Analysis of Cross-over Trials}, 3rd ed. London: Chapman and Hall. 
		
		\item Kershner, R. P. and W. T. Federer (1981). Two-treatment crossover designs for estimating a variety of effects. \emph{J. Amer. Statist. Assoc.} {\bf 76}, 612$-$619.
		
		\item Khuri, A. I., Mukherjee, B., Sinha, B. K. and Ghosh, M. (2006). Design issues for generalized linear models: A Review. \emph{Statistical Science}, {\bf 21}, 376$-$399.
		
		\item Kiefer, J. (1975). Construction and optimality of generalized Youden designs. \emph{In A Survey of Statistical Designs and Linear Models} (J. N. Srivastava, Ed.). Amsterdam: North-Holland, pp. 333$-$353.
		
		\item Kunert, J. (1983). Optimal design and refinement of the linear model with applications to repeated measurements designs. \emph{Ann. Statist.} {\bf 11}, 247$-$257.
		
		\item Kunert, J. (1984b). Optimality of balanced uniform repeated measurements designs. \emph{Ann. Statist.} {\bf 12}, 1006$-$1017.
		
		\item Kushner, H. B. (1997b). Optimal repeated measurements designs: the linear optimality equations. \emph{Ann. Statist.} {\bf 25}, 2328$-$2344.
		
		\item Laska, E. and M. Meisner (1985). A variational approach to optimal two treatment crossover designs: application to carryover effect models. \emph{J. Amer. Statist. Assoc.} {\bf 80}, 704$-$710.
		
		\item Layard, M.W and Arvesen, J.N (1978). Analysis of Poisson data in crossover experimental designs. \emph{Biometrics}{\bf 34}, 421$-$428.
		
		\item Liang, K. Y. and Zeger, S. L. (1986). Longitudinal data analysis using generalized linear models. Biometrika {\bf 73}, 13$-$22.
		
		\item Liang, K. Y. and Zeger, S. L. and Albert, P. S. (1988). Models for longitudinal data: A generalized estimating equation approach. Biometrics {\bf 44}, 1049$-$1060.
		
		\item Matthews, J. N. S. (1987). Recent developments in crossover designs. \emph{Internat. Statist. Rev.} {\bf 56}, 117$-$127.
		
		\item McCullagh, P. and J. A. Nelder (1989). {\em Generalized Linear Models} 2nd Edition. London: Chapman and Hall.
		
		\item Pukelsheim, F. (1993). Optimal Design of Experiments. New York: Wiley.
		
		\item Senn, S. (2003). {\em Cross-over Trials in Clinical Research,} 2nd ed. Chichester, England: Wiley.
		
		\item Singh, S. P. and Mukhopadhyay, S. (2016). Bayesian crossover design for generalized linear models. \emph{Computational Statistics and Data Analysis} {\bf 104}, 35$-$50.
		
		\item Stufken, J. (1996). Optimal crossover designs. In {\it Handbook of Statistics} {\bf 13} (S. Ghosh and C. R. Rao, Eds.). Amsterdam: North-Holland, pp. 63$-$90.
		
		\item Stufken, J. and Yang, M. (2012). Optimal designs for generalized linear models. In {\it Design and Analysis of Experiments}, Volume 3: Special Designs and Applications, (Edited by K.Hinkelmann). Wiley, New York.
		
	\end{enumerate}

	%%%%%%%%%%%%%%%%%%%%%%%%%%%%%%%%%%%%%%%%%%%%%%%%%%%%%%%%%%%%%%%%%%%%%%%%%%%%%%%%%%%%%                                                          
	%                                                                                   %
	%                           SUPPLEMENTARY MATERIAL                                  %
	%                                                                                   %
	%%%%%%%%%%%%%%%%%%%%%%%%%%%%%%%%%%%%%%%%%%%%%%%%%%%%%%%%%%%%%%%%%%%%%%%%%%%%%%%%%%%%%

	\newpage
	\begin{center}
		\textbf{\huge Optimal Crossover Designs for Generalized Linear Models}
	\end{center}	
	
	\renewcommand{\thefigure}{S\arabic{figure}}
	\renewcommand{\thesection}{S\arabic{section}}
	\renewcommand{\theequation}{S\arabic{equation}}
	\setcounter{section}{0}		
	\section{Supplementary Material}\label{SuppMat}	
	
	\setcounter{subsection}{0}
	%%%%%%%%%%%%%%%%%%%%%%%%%%%%%%%%%%%%%%%%%%%%%%%%%%%%%%%%%%%%%%%%%%%%%%%%%%%%%%%%%%%%%                                                          
	%                                                                                   %                        
	%                                                                                   %
	%                                                                                   %
	%                                                                                   %
	%             Optimal Design for Latin Square Example with 24 Sequencesn            %
	%                                                                                   %
	%                                                                                   %
	%                                                                                   %
	%                                                                                   %
	%%%%%%%%%%%%%%%%%%%%%%%%%%%%%%%%%%%%%%%%%%%%%%%%%%%%%%%%%%%%%%%%%%%%%%%%%%%%%%%%%%%%%

	\subsection{Optimal Design for Latin Square Example with 24 Sequences.}\label{24Seq}
	
	\bl{Following tables represent optimal designs for Latin square example with 24 sequences under $\theta_1$ and $\theta_2$.

		\begin{center}
			\text{Table S1.1: Optimal design considering 24 sequences under $\theta_1$}
		\end{center}		
		\begin{center}
			{\scriptsize 
			\begin{tabular}{ |p{2cm}|cccccc|}
				\hline
				&&&&&&\\ 
				Treatment Sequence  &  \multicolumn{6}{c|}{Optimal Designs for $\theta_1$}\\ 
				&&&&&&\\
				& $\rho = 0.1$ & $\rho = 0.2$ & $\rho = 0.5$ & $\rho = 0.6$ & $\rho = 0.7$ & $\rho = 0.9$\\
				\hline   
				&&&&&&\\
				$ABCD$ & 0.0094 & 0.0071 & 0.0109 & 0.0119 & 0.0125 & 0.0122 \\
				$ABDC$ &        &        &        &        &        &        \\
				$ACBD$ & 0.0716 & 0.1037 & 0.1148 & 0.1156 & 0.1153 & 0.1115 \\
				$ADBC$ & 0.1096 & 0.0820 & 0.0753 & 0.0795 & 0.0859 & 0.1003 \\
				$ACDB$ &        &        &        &        &        &        \\
				$ADCB$ &        &        &        &        &        &        \\
				$BACD$ & 0.0513 & 0.0537 & 0.0459 & 0.0417 & 0.0362 & 0.0250 \\
				$BADC$ &        &        &        &        &        &        \\
				$CABD$ & 0.1254 & 0.1162 & 0.1042 & 0.1007 & 0.0972 & 0.0878 \\
				$DABC$ & 0.0200 & 0.0447 & 0.0469 & 0.0421 & 0.0356 & 0.0194 \\
				$CADB$ &        &        &        &        &        &        \\
				$DACB$ & 0.0122 &        &        &        &        &        \\
				$BCAD$ &        &        &        &        &        &        \\
				$BDAC$ & 0.1735 & 0.1993 & 0.2055 & 0.2045 & 0.2031 & 0.2019 \\
				$CBAD$ &        &        &        &        &        &        \\
				$DBAC$ & 0.1667 & 0.1404 & 0.1374 & 0.1461 & 0.1588 & 0.1924 \\
				$CDAB$ & 0.1265 & 0.1426 & 0.1483 & 0.1473 & 0.1448 & 0.1358 \\
				$DCAB$ & 0.1114 & 0.1082 & 0.1108 & 0.1107 & 0.1106 & 0.1120 \\
				$BCDA$ &        &        &        &        &        &        \\
				$BDCA$ & 0.0224 & 0.0003 &        &        &        &        \\
				$CBDA$ &        &        &        &        &        &        \\
				$DBCA$ &        &        &        &        &        &        \\
				$CDBA$ &        &        &        &        &        &        \\
				$DCBA$ &        &        &        &        &        &        \\
				&&&&&&\\
				\hline
				
			\end{tabular}
		}
		\end{center}
		
		\clearpage	
		\begin{center}
			\text{Table S1.2: Optimal design considering 24 sequences under $\theta_2$.}
		\end{center}
		\begin{center}
			{\scriptsize 
			\begin{tabular}{ |p{2cm}|cccccc|}
				\hline
				&&&&&&\\ 
				Treatment Sequence  &  \multicolumn{6}{c|}{Optimal Designs for $\theta_2$}\\ 
				&&&&&&\\
				& $\rho = 0.1$ & $\rho = 0.2$ & $\rho = 0.5$ & $\rho = 0.6$ & $\rho = 0.7$ & $\rho = 0.9$\\
				\hline   
				&&&&&&\\
				$ABCD$ & 0.1105 & 0.1107 & 0.0875 & 0.0870 & 0.0876 & 0.0846 \\
				$ABDC$ &        &        &        &        &        & 0.0112 \\
				$ACBD$ & 0.0488 & 0.0525 & 0.0615 & 0.0624 & 0.0625 & 0.0522 \\
				$ADBC$ & 0.0347 & 0.0329 & 0.0516 & 0.0561 & 0.0618 & 0.0807 \\
				$ACDB$ & 0.0402 & 0.0348 & 0.0126 & 0.0128 & 0.0135 & 0.0247 \\
				$ADCB$ & 0.0370 & 0.0417 & 0.0645 & 0.0625 & 0.0587 & 0.0383 \\
				$BACD$ &        &        &        &        &        & 0.0052 \\
				$BADC$ & 0.1125 & 0.1109 & 0.0903 & 0.0855 & 0.0801 & 0.0545 \\
				$CABD$ & 0.0467 & 0.0419 & 0.0127 & 0.0087 & 0.0054 & 0.0125 \\
				$DABC$ &        & 0.0041 & 0.0213 & 0.0192 & 0.0152 &        \\
				$CADB$ & 0.0611 & 0.0619 & 0.0729 & 0.0733 & 0.0737 & 0.0674 \\
				$DACB$ &        &        & 0.0136 & 0.0198 & 0.0272 & 0.0537 \\
				$BCAD$ & 0.0363 & 0.0371 & 0.0472 & 0.0441 & 0.0392 & 0.0141 \\
				$BDAC$ & 0.0034 &        & 0.0003 & 0.0008 & 0.0027 & 0.0224 \\
				$CBAD$ &        & 0.0004 & 0.0360 & 0.0427 & 0.0503 & 0.0744 \\
				$DBAC$ & 0.1034 & 0.1056 & 0.0854 & 0.0859 & 0.0858 & 0.0728 \\
				$CDAB$ &        &        &        &        &        & 0.0055 \\
				$DCAB$ & 0.1157 & 0.1163 & 0.0946 & 0.0915 & 0.0888 & 0.0780 \\
				$BCDA$ &        &        & 0.0241 & 0.0294 & 0.0361 & 0.0617 \\
				$BDCA$ & 0.0882 & 0.0901 & 0.0719 & 0.0728 & 0.0733 & 0.0678 \\
				$CBDA$ & 0.0239 & 0.0297 & 0.0369 & 0.0356 & 0.0326 & 0.0166 \\
				$DBCA$ & 0.0276 & 0.0201 & 0.0238 & 0.0192 & 0.0153 & 0.0109 \\
				$CDBA$ & 0.1100 & 0.1093 & 0.0913 & 0.0907 & 0.0902 & 0.0802 \\
				$DCBA$ &        &        &        &        &        & 0.0106 \\
				&&&&&&\\
				\hline
				
			\end{tabular}
		}
		\end{center}	
	}
	
	\setcounter{subsection}{1}
	%%%%%%%%%%%%%%%%%%%%%%%%%%%%%%%%%%%%%%%%%%%%%%%%%%%%%%%%%%%%%%%%%%%%%%%%%%%%%%%%%%%%%                                                          
	%                                                                                   %                        
	%                                                                                   %
	%                                                                                   %
	%                                                                                   %
	%           SUB-SECTION 2: Explicit Expression of the Objective Function            %
	%                                                                                   %
	%                                                                                   %
	%                                                                                   %
	%                                                                                   %
	%%%%%%%%%%%%%%%%%%%%%%%%%%%%%%%%%%%%%%%%%%%%%%%%%%%%%%%%%%%%%%%%%%%%%%%%%%%%%%%%%%%%%

	\subsection{Explicit Expression of the Objective Function}\label{ObjFn}

	\bl{We give an example of Latin square design to illustrate how we obtain the objective function. We use $Corr(1)$ correlation structure with $\rho = 0.2$.
		
		\medskip\noindent First we look at the design matrix for each of the subject. Design matrix is obtained by using expression of $X_{j}$ mentioned in Section~\ref{PreSetup}.
		
		\bigskip\noindent
		
		\[
		X_1 = X_{ABCD} = \left(\begin{array}{rrrrrrrrrrrrr}
		1 & 1 & 0 & 0 & 0 & 1 & 0 & 0 & 0 & 0 & 0 & 0 & 0 \\ 
		1 & 0 & 1 & 0 & 0 & 0 & 1 & 0 & 0 & 1 & 0 & 0 & 0 \\ 
		1 & 0 & 0 & 1 & 0 & 0 & 0 & 1 & 0 & 0 & 1 & 0 & 0 \\ 
		1 & 0 & 0 & 0 & 1 & 0 & 0 & 0 & 1 & 0 & 0 & 1 & 0 \\ 
		\end{array}\right),
		\]
		
		\bigskip\noindent
		
		\[
		X_2 = X_{CADB} = \left(\begin{array}{rrrrrrrrrrrrr}
		1 & 1 & 0 & 0 & 0 & 0 & 1 & 0 & 0 & 0 & 0 & 0 & 0 \\ 
		1 & 0 & 1 & 0 & 0 & 0 & 0 & 0 & 1 & 0 & 1 & 0 & 0 \\ 
		1 & 0 & 0 & 1 & 0 & 1 & 0 & 0 & 0 & 0 & 0 & 0 & 1 \\ 
		1 & 0 & 0 & 0 & 1 & 0 & 0 & 1 & 0 & 1 & 0 & 0 & 0 \\ 
		\end{array}\right),
		\]
		
		\bigskip\noindent
		
		\[
		X_3 = X_{BDAC} = \left(\begin{array}{rrrrrrrrrrrrr}
		1 & 1 & 0 & 0 & 0 & 0 & 0 & 1 & 0 & 0 & 0 & 0 & 0 \\ 
		1 & 0 & 1 & 0 & 0 & 1 & 0 & 0 & 0 & 0 & 0 & 1 & 0 \\ 
		1 & 0 & 0 & 1 & 0 & 0 & 0 & 0 & 1 & 1 & 0 & 0 & 0 \\ 
		1 & 0 & 0 & 0 & 1 & 0 & 1 & 0 & 0 & 0 & 0 & 0 & 1 \\ 
		\end{array}\right),
		\]
		
		\bigskip\noindent
		
		\[
		X_4 = X_{DCBA} = \left(\begin{array}{rrrrrrrrrrrrr}
		1 & 1 & 0 & 0 & 0 & 0 & 0 & 0 & 1 & 0 & 0 & 0 & 0 \\ 
		1 & 0 & 1 & 0 & 0 & 0 & 0 & 1 & 0 & 0 & 0 & 0 & 1 \\ 
		1 & 0 & 0 & 1 & 0 & 0 & 1 & 0 & 0 & 0 & 0 & 1 & 0 \\ 
		1 & 0 & 0 & 0 & 1 & 1 & 0 & 0 & 0 & 0 & 1 & 0 & 0 \\ 
		\end{array}\right).
		\]
		
		\medskip\noindent Now, using the above design matrix for each subject and estimates of parameter values, we consider $\hat{\theta} =$ $[ 0.5$, $0.06$, $-0.53$, $-0.6$, $-0.35$, $0.025$, $-0.23$, $0.73$, $0.23]$. Then the values of $\eta_{j} = X_j \hat{\theta} $ for each subject can be obtained as follows:
		
		\bigskip
		\begin{minipage}{0.4\textwidth}
			
			\[
			\eta_{1} = X_1 \hat{\theta} = \left(\begin{array}{r}
			0.534 \\
			0.278 \\
			0.761 \\
			-0.070 \\
			\end{array}\right),
			\]
			
		\end{minipage}
		\hspace{0.05\textwidth}
		\begin{minipage}{0.4\textwidth}
			
			\[
			\eta_{2} = X_2 \hat{\theta} = \left(\begin{array}{r}
			0.185 \\
			1.131 \\
			0.307 \\
			-0.050 \\
			\end{array}\right),
			\]
			
		\end{minipage}
		
		\medskip\noindent
		
		\begin{minipage}{0.4\textwidth}
			
			\[
			\eta_{3} = X_3 \hat{\theta} = \left(\begin{array}{r}
			0.557 \\
			0.857 \\
			-0.220 \\
			-0.122 \\
			\end{array}\right),
			\]
			
		\end{minipage}
		\hspace{0.05\textwidth}
		\begin{minipage}{0.4\textwidth}
			
			\[
			\eta_{4} = X_4 \hat{\theta} = \left(\begin{array}{r}
			0.307 \\
			0.950 \\
			-0.112 \\
			0.658 \\
			\end{array}\right).
			\]
			
		\end{minipage}
		
		\medskip\noindent Hence, using model~(\ref{logitmodel}) mentioned in Section~\ref{PreSetup}, we can get corresponding $\mu_{j} = \frac{\exp\{\eta_j\}}{1+\exp\{\eta_j\}}$, they are as follows:
		
		\bigskip
		\begin{minipage}{0.4\textwidth}
			
			\[
			\mu_{1} = \frac{\exp\{\eta_1\}}{1+\exp\{\eta_1\}} = \left(\begin{array}{r}
			0.6304156 \\
			0.5690558 \\
			0.6815708 \\
			0.4825071 \\
			\end{array}\right),
			\]
			
		\end{minipage}
		\hspace{0.05\textwidth}
		\begin{minipage}{0.4\textwidth}
			
			\[
			\mu_{2} = \frac{\exp\{\eta_2\}}{1+\exp\{\eta_2\}} = \left(\begin{array}{r}
			0.5461185 \\
			0.7560234 \\
			0.5761528 \\
			0.4875026 \\
			\end{array}\right),
			\]
			
		\end{minipage}
		
		\medskip
		
		\begin{minipage}{0.4\textwidth}
			
			\[
			\mu_{3} = \frac{\exp\{\eta_3\}}{1+\exp\{\eta_3\}} = \left(\begin{array}{r}
			0.6357581 \\
			0.7020335 \\
			0.4452208 \\
			0.4695378 \\
			\end{array}\right),
			\]
			
		\end{minipage}
		\hspace{0.05\textwidth}
		\begin{minipage}{0.4\textwidth}
			
			\[
			\mu_{4} = \frac{\exp\{\eta_4\}}{1+\exp\{\eta_4\}} = \left(\begin{array}{r}
			0.5761528 \\
			0.7211152 \\
			0.4720292 \\
			0.6588110 \\ 
			\end{array}\right).
			\]
			
		\end{minipage}
		
		\medskip\noindent
		We are using compound symmetric correlation structure $Corr(1)$ with $\rho = 0.2$. Hence we have $C(\alpha) = Corr(1)$ as true correlation matrix. 
		
		\medskip\noindent Correlation matrix $C(\alpha)$ and matrix $H$ can be written down as follows:
		
		\bigskip
		\begin{minipage}{0.35\textwidth}
			
			\[
			C(\alpha) = \left(\begin{array}{rrrr}
			1 	& 0.2 & 0.2 & 0.2 \\ 
			0.2 & 1   & 0.2 & 0.2 \\ 
			0.2 & 0.2 & 1   & 0.2 \\ 
			0.2 & 0.2 & 0.2 & 1   \\ 
			\end{array}\right),
			\]
			
		\end{minipage}
		\hspace{0.032\textwidth}
		\begin{minipage}{0.35\textwidth}
			
			\[
			H = \left(\begin{array}{rrrrrrrrrrrrr}
			0 & 0 & 0 & 0 & 0 & 1 & 0 & 0 & 0 & 0 & 0 & 0 & 0 \\ 
			0 & 0 & 0 & 0 & 0 & 0 & 1 & 0 & 0 & 0 & 0 & 0 & 0 \\ 
			0 & 0 & 0 & 0 & 0 & 0 & 0 & 1 & 0 & 0 & 0 & 0 & 0 \\ 
			0 & 0 & 0 & 0 & 0 & 0 & 0 & 0 & 1 & 0 & 0 & 0 & 0 \\ 
			\end{array}\right).
			\]
			
		\end{minipage}	
		
		\medskip \noindent Using the expression for $Cov[Y_{j}]$ mentioned below we compute covariance matrix for each subject. We denote this covariance matrix by $W_{j}$ for each subject $j$:
		
		\indent
		\begin{eqnarray*}
			Cov[Y_{j}] = W_{j} = D_{j}^{1/2}Corr_{1}D_{j}^{1/2}
		\end{eqnarray*}
		
		where $D_{j}$ in above equation is $ diag\Big(\mu_{1j}(1-\mu_{1j}),\ldots,\mu_{pj}(1-\mu_{pj})\Big)$ and $p$ is number of periods. 
		
		\bigskip\noindent Hence corresponding $D_{j}$ for Latin square example are as follows:
		
		\bigskip\noindent
		\begin{minipage}{0.4\textwidth}
			
			\[
			D_{1} = \left(\begin{array}{rrrr}
			0.23 & 0    & 0    & 0 \\ 
			0    & 0.24 & 0    & 0 \\ 
			0    & 0    & 0.22 & 0 \\ 
			0    & 0    & 0    & 0.25 \\ 
			\end{array}\right),
			\]
			
		\end{minipage}
		\hspace{0.05\textwidth}
		\begin{minipage}{0.5\textwidth}
			
			\[
			D_{2} = \left(\begin{array}{rrrr}
			0.25 & 0    & 0    & 0 \\ 
			0    & 0.18 & 0    & 0 \\ 
			0    & 0    & 0.24 & 0 \\ 
			0    & 0    & 0    & 0.25 \\ 
			\end{array}\right),
			\]
			
		\end{minipage}
		
		\bigskip\noindent
		\begin{minipage}{0.4\textwidth}
			
			\[
			D_{3} = \left(\begin{array}{rrrr}
			0.23 & 0    & 0    & 0 \\ 
			0    & 0.21 & 0    & 0 \\ 
			0    & 0    & 0.25 & 0 \\ 
			0    & 0    & 0    & 0.25 \\ 
			\end{array}\right),
			\]
			
		\end{minipage}
		\hspace{0.05\textwidth}
		\begin{minipage}{0.5\textwidth}
			
			\[
			D_{4} = \left(\begin{array}{rrrr}
			0.24 & 0    & 0    & 0 \\ 
			0    & 0.20 & 0    & 0 \\ 
			0    & 0    & 0.25 & 0 \\ 
			0    & 0    & 0    & 0.22 \\ 
			\end{array}\right).
			\]
			
		\end{minipage}
		
		\bigskip\noindent Calculating matrix $D_{j}^{1/2}$ and using the above formula for $W_{j}$ , we have inverse of $W_{j}$ matrices as follows:
		
		\bigskip\noindent
		\begin{minipage}{0.45\textwidth}
			
			\[
			W^{-1}_{1} = \left(\begin{array}{rrrr}
			4.69 & -0.65 & -0.69 & -0.65 \\ 
			-0.65 &  4.46 & -0.68 & -0.63 \\ 
			-0.69 & -0.68 &  5.04 & -0.67 \\ 
			-0.65 & -0.63 & -0.67 &  4.38 \\  
			\end{array}\right),
			\]
			
		\end{minipage}
		\hspace{0.05\textwidth}
		\begin{minipage}{0.45\textwidth}
			
			\[
			W^{-1}_{2} = \left(\begin{array}{rrrr}
			4.41 & -0.73 & -0.64 & -0.63 \\ 
			-0.73 &  5.93 & -0.74 & -0.73 \\ 
			-0.64 & -0.74 &  4.48 & -0.63 \\ 
			-0.63 & -0.73 & -0.63 &  4.38 \\ 
			\end{array}\right),
			\]
			
		\end{minipage}
		
		\bigskip
		
		\noindent \begin{minipage}{0.45\textwidth}
			
			\[
			W^{-1}_{3} = \left(\begin{array}{rrrr}
			4.72 & -0.71 & -0.65 & -0.65 \\ 
			-0.71 &  5.23 & -0.69 & -0.68 \\ 
			-0.65 & -0.69 &  4.43 & -0.63 \\ 
			-0.65 & -0.68 & -0.63 &  4.39 \\
			\end{array}\right),
			\]
			
		\end{minipage}
		\hspace{0.05\textwidth}
		\begin{minipage}{0.45\textwidth}
			
			\[
			W^{-1}_{4} = \left(\begin{array}{rrrr}
			4.48 & -0.70 & -0.63 & -0.67 \\ 
			-0.70 &  5.44 & -0.70 & -0.73 \\ 
			-0.63 & -0.70 &  4.39 & -0.66 \\ 
			-0.67 & -0.73 & -0.66 &  4.87 \\
			\end{array}\right).
			\]
			
		\end{minipage}
		
		\medskip\noindent Note that $D_{\omega} = D_{j}$ and $W_{\omega} = W_{j}$.
		
		\bigskip\noindent
		The variance of parameter estimate $ {\rm Var} (\hat{\theta}) = \left[\sum_{\omega \epsilon \Omega} np_{\omega} \frac{\partial \mu_{\omega}^{\prime}}{\partial \theta} W_{\omega}^{-1} \frac{\partial \mu_{\omega}}{\partial \theta}\right]^{-1} $ has another component which is $ \frac{\partial \mu_{\omega}}{\partial \theta} $ and the $ith$ row of $\frac{\partial \mu_{\omega}}{\partial \theta}$ is $ x^{\prime}_{i}d_{i},$ where $x_{i}$ is the $i$th row of design matrix $X_{\omega}$ and $d_{i}$ corresponds to $i$th diagonal entry of matrix $D_{j}$.
		
		Hence, $ \frac{\partial \mu_{\omega}}{\partial \theta} $ matrix for each subject are as follows:
		
		\bigskip\noindent
		
		\[
		\frac{\partial \mu_{1}}{\partial \theta} = \left(\begin{array}{rrrrrrrrrrrrr}
		0.23 & 0.23 & 0    & 0    & 0    & 0.23 & 0    & 0    & 0    & 0    & 0    & 0    & 0 \\ 
		0.24 & 0    & 0.24 & 0    & 0    & 0    & 0.24 & 0    & 0    & 0.24 & 0    & 0    & 0 \\ 
		0.22 & 0    & 0    & 0.22 & 0    & 0    & 0    & 0.22 & 0    & 0    & 0.22 & 0    & 0 \\ 
		0.25 & 0    & 0    & 0    & 0.25 & 0    & 0    & 0    & 0.25 & 0    & 0    & 0.25 & 0 \\ 
		\end{array}\right),
		\]
		
		\bigskip\noindent
		
		\[
		\frac{\partial \mu_{2}}{\partial \theta} = \left(\begin{array}{rrrrrrrrrrrrr}
		0.25 & 0.25 & 0    & 0    & 0    & 0    & 0.25 & 0    & 0    & 0    & 0    & 0 & 0    \\ 
		0.18 & 0    & 0.18 & 0    & 0    & 0    & 0    & 0    & 0.18 & 0    & 0.18 & 0 & 0    \\ 
		0.24 & 0    & 0    & 0.24 & 0    & 0.24 & 0    & 0    & 0    & 0    & 0    & 0 & 0.24 \\ 
		0.25 & 0    & 0    & 0    & 0.25 & 0    & 0    & 0.25 & 0    & 0.25 & 0    & 0 & 0    \\ 
		\end{array}\right),
		\]
		
		\bigskip\noindent
		
		\[
		\frac{\partial \mu_{3}}{\partial \theta} = \left(\begin{array}{rrrrrrrrrrrrr}
		0.23 & 0.23 & 0    & 0    & 0    & 0    & 0    & 0.23 & 0    & 0    & 0 & 0    & 0    \\ 
		0.21 & 0    & 0.21 & 0    & 0    & 0.21 & 0    & 0    & 0    & 0    & 0 & 0.21 & 0    \\ 
		0.25 & 0    & 0    & 0.25 & 0    & 0    & 0    & 0    & 0.25 & 0.25 & 0 & 0    & 0    \\ 
		0.25 & 0    & 0    & 0    & 0.25 & 0    & 0.25 & 0    & 0    & 0    & 0 & 0    & 0.25 \\ 
		\end{array}\right),
		\]
		
		\bigskip\noindent
		
		\[
		\frac{\partial \mu_{4}}{\partial \theta} = \left(\begin{array}{rrrrrrrrrrrrr}
		0.24 & 0.24 & 0    & 0    & 0    & 0    & 0    & 0    & 0.24 & 0 & 0    & 0    & 0    \\ 
		0.20 & 0    & 0.20 & 0    & 0    & 0    & 0    & 0.20 & 0    & 0 & 0    & 0    & 0.20 \\ 
		0.25 & 0    & 0    & 0.25 & 0    & 0    & 0.25 & 0    & 0    & 0 & 0    & 0.25 & 0    \\ 
		0.22 & 0    & 0    & 0    & 0.22 & 0.22 & 0    & 0    & 0    & 0 & 0.22 & 0    & 0    \\ 
		\end{array}\right).
		\]
		
		\bigskip\noindent Using above calculated inverse of each $W_{\omega}$ matrix,  and the corresponding calculated $ \frac{\partial \mu_{\omega}}{\partial \theta} $ matrices we can calculate required $13\times13$ matrices $ \frac{\partial \mu_{\omega}^{\prime}}{\partial \theta} W_{\omega}^{-1} \frac{\partial \mu_{\omega}}{\partial \theta} $ for each $\omega$.
		
		\bigskip\noindent Further, inverse of $ \left[\sum_{\omega \epsilon \Omega} np_{\omega} \frac{\partial \mu_{\omega}^{\prime}}{\partial \theta} W_{\omega}^{-1} \frac{\partial \mu_{\omega}}{\partial \theta}\right] $ i.e ${\rm Var} (\hat{\theta})$ is found numerically and calculate objective function $ {\rm Var}(\hat\tau) = H{\rm Var}(\hat\theta)H^{\prime} $, where we try to minimize $ {\rm Var}(\hat\tau) $ w.r.t $ p_{\omega}.$ This values of $ p_{\omega}$ which minimizes objective function are optimal proportions we are looking for.}
	
\end{document}